%% file: Five_Class_S_Conjectures_v4.tex
\documentclass[aps,prl,reprint,amsmath,amssymb,nofootinbib]{revtex4-2}
\usepackage[T1]{fontenc}
\usepackage{array}[=2016-10-06]
\usepackage{graphicx,booktabs,tabularx,xcolor}
\usepackage{tikz}
\usepackage[hidelinks]{hyperref}
\input{figures/Family_Diagrams}

\input{Scan_Counts}
\renewcommand{\thesection}{\arabic{section}}
\renewcommand{\thesubsection}{\thesection.\arabic{subsection}}
\makeatletter
\let\p@subsection\@empty
\makeatother
\renewcommand{\thetable}{\arabic{table}}
\newcommand{\Spin}{\operatorname{Spin}}
\newcommand{\OO}{\mathcal{O}}
\newcommand{\Ia}{\mathcal{I}}
\newcommand{\Sp}{\operatorname{Sp}}
\newcommand{\PE}{\operatorname{PE}}
\newcommand{\youngrow}[4]{%
  \multiput(#1,#2)(1,0){#3}{%
    \makebox(0,0)[bl]{\color{#4}\rule{\unitlength}{\unitlength}}%
    \framebox(1,1){}%
  }%
}
\hypersetup{pdftitle={Isomorphisms of Class \texorpdfstring{$\mathcal{S}$}{S} Theories with IndexLab},
  pdfauthor={Zhihao Duan, Zhenghao Zhong}}

\begin{document}
\raggedbottom
\title{Isomorphisms of Class \texorpdfstring{$\mathcal{S}$}{S} Theories with \texttt{IndexLab}}
\author{Zhihao Duan}
\email{xduanz@gmail.com}
\affiliation{Shanghai Institute for Mathematics and Interdisciplinary Sciences (SIMIS),
  657 Songhu Road, Shanghai 200433, China}
\author{Zhenghao Zhong}
\email{zhong@simis.cn}
\affiliation{Shanghai Institute for Mathematics and Interdisciplinary Sciences (SIMIS),
  657 Songhu Road, Shanghai 200433, China}

\begin{abstract}
We use IndexLab to scan untwisted and twisted class $\mathcal{S}$ constructions based on ADE Lie algebras of rank up to seven, as well as $A_8$ and $E_8$. We consider surfaces of genus at most three with up to five punctures, covering 167,244,647 constructions in total. To find candidate pairs of isomorphic theories, we compare Schur and Hall-Littlewood indices, Weyl anomalies, Coulomb branch spectra, and flavor data. We set aside pairs explained by known mechanisms, leaving a catalogue of 5,278 candidate pairs. Recurring changes in rank, puncture labels, and twists suggest families of candidate isomorphisms. We group 99.58\% of the pairs into 25 families and 6 finite families. IndexLab is available at \href{https://superconformal.com}{superconformal.com}.
\end{abstract}
\maketitle

\section{Introduction}
\label{sec:introduction}

Class $\mathcal{S}$ theories are four dimensional $\mathcal N=2$ superconformal field theories (SCFTs) obtained by compactifying a six dimensional $(2,0)$ theory on a punctured surface~\cite{Gaiotto,GMN}. To specify a construction, we choose the $ADE$ parent, the surface, the puncture labels, and any twists. In many classical examples, the punctures are labeled by Young diagrams. These diagrams determine the flavor symmetries visible at the punctures and contribute to the spectrum of Coulomb branch operators~\cite{CDTypeA,CDTypeD,CDTDefects}.

Dualities give us different descriptions of the same theory, often making it easier to study. They play a central role in string theory and supersymmetric field theory. Here we focus on isomorphisms between SCFTs that arise from distinct class $\mathcal{S}$ constructions. Such equivalences have been studied since the early development of class $\mathcal{S}$~\cite{Gaiotto,CDTypeD}. The simplest examples come from isomorphisms of the parent Lie algebras. For instance, $A_3\simeq D_3$ lets us write the same construction using two puncture conventions~\cite[Sec.~2.4.1]{CDTypeD}. More subtle equivalences can arise when we keep the parent fixed but change the punctures. For the $E_7$ parent, three punctured spheres with different puncture labels have been shown to give isomorphic SCFTs~\cite{DEKLIsomorphisms}. Some of these isomorphisms can be established by realizing both theories as torus compactifications of the same six-dimensional $(1,0)$ SCFT~\cite{OSTY,BKL,DEKLIsomorphisms}. Fixture classifications have also led to conjectured isomorphisms supported by matching superconformal indices and other protected data~\cite{CDTTypeE6,CDTTwistedD}. One family we study relates twisted $D$-type and untwisted $A$-type constructions. It also gives us a way to check proposed twisted $D$-type full indices.

We search systematically for new isomorphic pairs of class $\mathcal{S}$ constructions by comparing protected observables and intrinsic SCFT data. Our main tool is the superconformal index~\cite{RomelsbergerIndex,KinneyIndex}, including its Schur, Hall--Littlewood, and Macdonald limits~\cite{GRRYSchur,GRRYMacdonald,RRReview}. These indices capture protected information about the four dimensional theory. We also compare intrinsic SCFT data: the Weyl anomaly coefficients $a$ and $c$, the graded spectrum of Coulomb branch operators, and the available flavor symmetry data~\cite{ShapereTachikawa,CDTDefects}.

We first find constructions whose protected data agree, then test them with complementary observables and higher order index expansions. For selected pairs, we also keep the flavor fugacities. Distinct theories can agree on individual protected observables~\cite{DESDistinguishing,SongADHL}, so our proposed isomorphisms rely on the combined evidence. Among the surviving pairs, we find recurring changes in ranks and punctures. These patterns suggest families of conjectures and let us predict further pairs, trios, quartets, and larger sets of isomorphic constructions.

To carry out these calculations, we introduce \texttt{IndexLab}, a standalone browser tool for studying four dimensional $\mathcal N=2$ theories. It supports regular untwisted and twisted class $\mathcal{S}$ constructions~\cite{LPRTypeD,BPTwisted}, Argyres--Douglas theories~\cite{BNADSchur,MSMinimalAD}, and Lagrangian gauge theories. Using exact arithmetic, it evaluates finite order expansions of the Schur, Hall--Littlewood, and Macdonald indices. For supported theories, it also computes finite order full $(p,q,t)$ indices using exact calculations, conjectural elliptic wavefunctions, or reconstructions from protected data. For class $\mathcal{S}$ theories, IndexLab computes anomalies and Coulomb branch spectra from puncture data and displays the available flavor information. Each calculation records its truncation order and any assumptions or conjectural identifications it uses. Supplemental Sec.~\ref{app:index-calculations} describes the methods and gives validation examples.

Our dataset begins with an exhaustive census of untwisted and twisted $ADE$ constructions with parent algebras of rank up to seven, as well as $A_8$ and $E_8$, on surfaces of genus at most three with up to five punctures. We discuss \MainFamilies{} representative families in the main text and give a five-way example of an isolated SCFT. The Supplemental Material gives the simplest recorded pair for each family and finite set of related examples. One family, for example, relates constructions from the $D_n$ and $D_{n+1}$ six dimensional parents. We give explicit rules for changing their puncture Young diagrams and use a $\Spin(8)$ character comparison to predict higher rank isomorphisms. Other families relate untwisted $A$-type spheres to twisted $D$-type spheres, or connect $E_6$, $E_7$, and $E_8$ constructions by changing both the punctures and the parent algebra. In the catalogue, \UmbrellaFamilies{} families and \FiniteRuleGroups{} finite sets of related examples account for \OrganizedCandidatePairs{} of the \CandidateDualities{} candidate pairs, or \OrganizedCandidatePercent{}\%. We list the \IsolatedComparisons{} pairs without a family assignment in Supplemental Sec.~\ref{supp:isolated-comparisons}.

\section{Method}
\label{sec:method}

\textit{The Scan.}
We specify each construction by its parent algebra, genus, puncture labels, and twist data, including very even colors where needed. Several constructions can describe the same four dimensional SCFT. Our scan covers 29 combinations of parent algebra and outer automorphism twist sector, all with regular punctures. There are 15 untwisted cases ($A_1$ through $A_8$, $D_4$ through $D_7$, and $E_6$, $E_7$, $E_8$), 12 order two twisted cases ($A_2$ through $A_8$, $D_4$ through $D_7$, and $E_6$), and two further $D_4$ cases with $\mathbb Z_3$ or $S_3$ twists. The systematic census contains \RegularCensus{} regular constructions on surfaces with $0\le g\le3$ and at most five punctures. We include all explicitly checked members of the rank dependent families in the same dataset and classification, with their full rank and puncture data.

\textit{Screening and candidate pairs.}
For each construction, we compute the unrefined Schur index by setting all flavor fugacities to one and writing $q=\tau^2$. The class $\mathcal{S}$ representation sum includes every representation that can contribute through the chosen order~\cite{GRRYSchur,GRRYMacdonald,LPRTypeD,BPTwisted}. We compare the plethystic logarithms coefficient by coefficient, first through $\tau^{12}$ and then through $\tau^{20}$. At each stage, we remove pairs whose data differ and count each distinct construction once if it still has a partner. Next, we compare Hall--Littlewood coefficients through $x^{20}$, with $x=\sqrt t$. The surviving pairs undergo the higher order index tests listed in Supplemental Table~\ref{tab:landscape-scan}.

We also require agreement of the Weyl anomalies $a,c$ and the Coulomb branch spectra. We use the effective numbers of vector multiplets and hypermultiplets, $n_v=4(2a-c)$ and $n_h=4(5c-4a)$, with $n_h-n_v=24(c-a)$~\cite[Sec.~3.3]{CDTDefects}. We apply the additional comparison of spinor contributions described in Supplemental Table~\ref{tab:landscape-scan} and set aside pairs with an established mismatch in the full flavor symmetry algebra.

Supplemental Fig.~\ref{fig:landscape-scan} and Table~\ref{tab:landscape-scan} show how the construction counts change at each screening step. For every pair, we record the index orders reached, the invariant checks, and the family assignment.

\textit{Candidate selection.}
We first identify pairs whose unrefined Macdonald sums agree term by term. Their puncture contributions and denominator powers coincide after allowing puncture permutations and compatible Dynkin diagram symmetries. The TQFT formula therefore directly explains their index agreement for generic $(q,t)$ to all orders~\cite{GRRYMacdonald,LPRTypeD}. Their graded Coulomb branch spectra, Weyl anomalies $a,c$, and available flavor data also agree independently. Taken together, these matches recorded in the IndexLab database provide strong evidence that the paired constructions describe isomorphic SCFTs. We conjecture these isomorphisms and therefore expect the full superconformal indices to agree as well. An explicit equivalence between the underlying compactifications, including all puncture and twist data, would establish these isomorphisms. We set aside this group and focus on candidate dualities whose index agreement still needs an explanation.

The superconformal index is invariant under exactly marginal deformations, so it cannot distinguish different points on a conformal manifold~\cite{RRReview}. We set aside pairs whose puncture resolutions identify a common conformal family, even when their original constructions select different coupling loci. Resolving an atypical puncture makes an exactly marginal coupling explicit; colliding the resulting punctures can select a specific coupling locus~\cite{CDTTwistedA,CDTTwistedD}. The IndexLab database records \CollisionPairs{} such pairs, including $D_4$ cases with $S_3$ holonomies~\cite{DESD4}.
 For \CollisionSharedPairs{} pairs (\CollisionSharedPercent), resolving the punctures directly or taking another common collision identifies a locus at finite coupling where both presentations give the same local SCFT. This is a separate check from identifying a common conformal family. For each of the remaining \CollisionNoSharedPairs{} pairs, we conjecture that the two presentations also give isomorphic local SCFTs at some finite coupling. 

Setting aside all pairs identified in a common conformal family reduces the number of surviving constructions from about \AfterIdentical{} to about \AfterCollisions{} distinct constructions. Further common conformal families may remain unidentified among the surviving candidates. We set aside comparisons explained by known Higgsings, gaugings, or gluing constructions and require both sides to have zero decoupled free hypermultiplets. Combining the surviving census pairs with checked family extensions gives \CandidateDualities{} candidate pairs in the IndexLab database.

These families and the surviving candidates are the main focus of this paper.

\section{Dualities}
\label{sec:dualities}

Among the matches that survive the scan, the same puncture changes often work in several examples. We use these patterns to pair constructions across ranks, parent types, or twists. Below, we present representative examples and the checks supporting them.

The IndexLab database lists all \SubfamilyCount{} documented subfamilies across the \UmbrellaFamilies{} families. Supplemental Sec.~\ref{supp:simplest-representatives} gives the simplest recorded pair for each family and finite set of related examples.

We write $G\{P,Q,R\}$ for a sphere of parent type $G$ with punctures $P,Q,R$. A subscript $t$ marks a twisted puncture, as in $[5^2]_t$. The superscript $\mathrm{tw}$ labels the whole construction, which can contain untwisted punctures too.

\subsection{Eight-box partitions and \texorpdfstring{$\Spin(8)$}{Spin(8)}}
\label{sec:spin8}

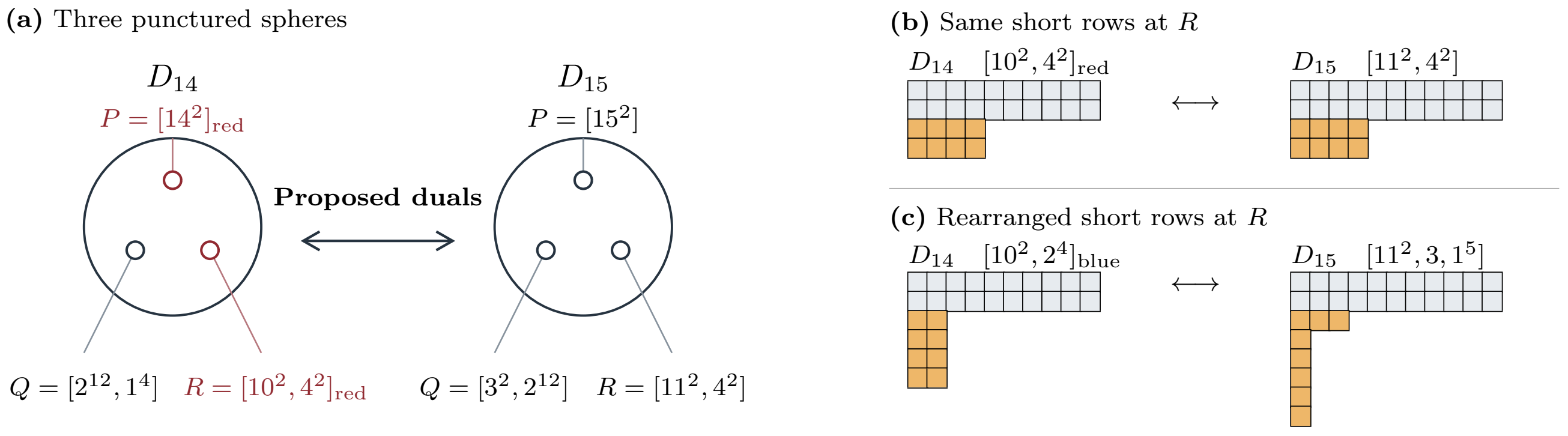
\begin{figure*}[t]
\centering
\begingroup
\footnotesize
\begin{minipage}[t]{0.48\textwidth}
\vspace{0pt}
\centering
\setlength{\unitlength}{1pt}
\definecolor{spinred}{RGB}{145,41,48}
\begin{picture}(234,134)
  \put(0,129){\makebox(0,0)[l]{\textbf{(a)} Three punctured spheres}}
  \put(0,0){\includegraphics[width=234pt,height=120pt]{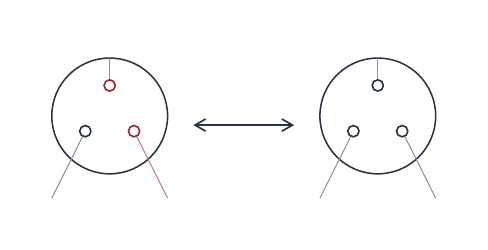}}
  \put(52.7,111){\makebox(0,0){{\normalsize $D_{14}$}}}
  \put(181.4,111){\makebox(0,0){{\normalsize $D_{15}$}}}
  \put(52.7,98){\makebox(0,0){\color{spinred}$P=[14^2]_{\mathrm{red}}$}}
  \put(181.4,98){\makebox(0,0){$P=[15^2]$}}
  \put(25,14){\makebox(0,0){$Q=[2^{12},1^4]$}}
  \put(85,14){\makebox(0,0){\color{spinred}$R=[10^2,4^2]_{\mathrm{red}}$}}
  \put(153.5,14){\makebox(0,0){$Q=[3^2,2^{12}]$}}
  \put(209,14){\makebox(0,0){$R=[11^2,4^2]$}}
  \put(117,73){\makebox(0,0){\textbf{Proposed duals}}}
\end{picture}
\end{minipage}\hfill
\begin{minipage}[t]{0.48\textwidth}
\vspace{0pt}
\centering
\setlength{\unitlength}{6pt}
\linethickness{0.32pt}
\definecolor{spinlong}{RGB}{231,235,239}
\definecolor{spinresidual}{RGB}{238,183,105}
\begin{picture}(35,22.3)(0,3.2)
  \put(0,24.5){\makebox(0,0)[l]{\textbf{(b)} Same short rows at $R$}}
  \put(1,22.5){\makebox(0,0)[l]{$D_{14}\quad [10^2,4^2]_{\mathrm{red}}$}}
  \put(21,22.5){\makebox(0,0)[l]{$D_{15}\quad [11^2,4^2]$}}
  \put(16,20.3){\makebox(0,0){$\longleftrightarrow$}}
  \youngrow{1}{20.5}{10}{spinlong}
  \youngrow{1}{19.5}{10}{spinlong}
  \youngrow{1}{18.5}{4}{spinresidual}
  \youngrow{1}{17.5}{4}{spinresidual}
  \youngrow{21}{20.5}{11}{spinlong}
  \youngrow{21}{19.5}{11}{spinlong}
  \youngrow{21}{18.5}{4}{spinresidual}
  \youngrow{21}{17.5}{4}{spinresidual}
  \put(0,15.9){\color{black!35}\line(1,0){35}}
  \put(0,14.3){\makebox(0,0)[l]{\textbf{(c)} Rearranged short rows at $R$}}
  \put(1,12.4){\makebox(0,0)[l]{$D_{14}\quad [10^2,2^4]_{\mathrm{blue}}$}}
  \put(21,12.4){\makebox(0,0)[l]{$D_{15}\quad [11^2,3,1^5]$}}
  \put(16,10.8){\makebox(0,0){$\longleftrightarrow$}}
  \youngrow{1}{10.5}{10}{spinlong}
  \youngrow{1}{9.5}{10}{spinlong}
  \youngrow{1}{8.5}{2}{spinresidual}
  \youngrow{1}{7.5}{2}{spinresidual}
  \youngrow{1}{6.5}{2}{spinresidual}
  \youngrow{1}{5.5}{2}{spinresidual}
  \youngrow{21}{10.5}{11}{spinlong}
  \youngrow{21}{9.5}{11}{spinlong}
  \youngrow{21}{8.5}{3}{spinresidual}
  \youngrow{21}{7.5}{1}{spinresidual}
  \youngrow{21}{6.5}{1}{spinresidual}
  \youngrow{21}{5.5}{1}{spinresidual}
  \youngrow{21}{4.5}{1}{spinresidual}
  \youngrow{21}{3.5}{1}{spinresidual}
\end{picture}
\end{minipage}
\endgroup
\caption{Two examples of the adjacent $D_n/D_{n+1}$ family at $n=14$ and $i=j=0$. (a) A pair of three punctured spheres. (b) The $R$ punctures of (a), with the short partition $[4^2]$ unchanged. (c) The $R$ punctures of a second pair, obtained from (a) by changing only $R$ on each sphere to $[10^2,2^4]_{\mathrm{blue}}$ for $D_{14}$ and $[11^2,3,1^5]$ for $D_{15}$. In (b,c), each gray long row grows by one box, while the eight gold boxes show the short partitions. Red and blue subscripts distinguish the very even punctures. Both examples are special cases of Fig.~\ref{fig:spin8-general}.}
\label{fig:spin8-spheres}
\label{fig:spin8-young}
\end{figure*}

\begin{figure*}[t]
\centering
\begingroup
\footnotesize
\definecolor{spinfamilybase}{RGB}{231,235,239}
\definecolor{spinfamilygold}{RGB}{238,183,105}
\definecolor{spinfamilyadd}{RGB}{150,207,184}
\definecolor{spinfamilyink}{RGB}{32,48,67}
\begin{minipage}[t]{0.48\textwidth}
\vspace{0pt}
\centering
\setlength{\unitlength}{1pt}
\newcommand{\spingeneralsphere}[4]{%
  \begin{picture}(0,0)
    \put(0,163){\makebox(0,0){\normalsize $#1$}}
    \put(0,146){\makebox(0,0){$#2$}}
    \put(0,53){\makebox(0,0){$#3$}}
    \put(0,78){\makebox(0,0){$#4$}}
  \end{picture}%
}
\begin{picture}(232,193)
  \put(0,0){\includegraphics[width=232pt]{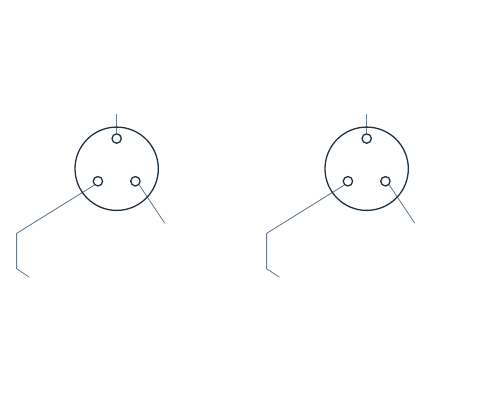}}
  \put(0,187){\makebox(0,0)[l]{\textbf{(a)} Three punctured spheres}}
  \put(56,0){\spingeneralsphere{D_n}
    {P_0=[(n-i)^2]\cup\lambda_0}
    {Q_0=[3^j,2^{2(h-j)},1^{j+4+2\epsilon}]}
    {R_0=[(n-4+i)^2]\cup\mu_0}}
  \put(176,0){\spingeneralsphere{D_{n+1}}
    {P_1=[(n+1-i)^2]\cup\lambda_1}
    {Q_1=[3^{j+2},2^{2(h-j)},1^{j+2\epsilon}]}
    {R_1=[(n-3+i)^2]\cup\mu_1}}
  \put(116,123){\makebox(0,0){\color{spinfamilyink}Proposed duals}}
  \put(116,112){\makebox(0,0){\color{spinfamilyink}$\longleftrightarrow$}}
  \put(116,27){\makebox(0,0){$h=\lfloor(n-2)/2\rfloor,\quad \epsilon=n\bmod 2$}}
\end{picture}
\end{minipage}\hfill
\begin{minipage}[t]{0.48\textwidth}
\vspace{0pt}
\centering
\setlength{\unitlength}{5.5pt}
\setlength{\fboxsep}{1.65pt}
\linethickness{0.32pt}
\newcommand{\spinfamilylongpair}[2]{%
  \begin{picture}(0,0)
    \put(3.35,3){\makebox(0,0){$#1$}}
    \youngrow{0}{1}{2}{spinfamilybase}
    \youngrow{0}{0}{2}{spinfamilybase}
    \put(3.1,1.5){\makebox(0,0){$\cdots$}}
    \put(3.1,0.5){\makebox(0,0){$\cdots$}}
    \youngrow{4.2}{1}{2}{spinfamilybase}
    \youngrow{4.2}{0}{2}{spinfamilybase}
    \multiput(0.5,-1.55)(0,0.55){3}{\makebox(0,0){\rule{0.8pt}{0.8pt}}}
    \put(0,-4.3){\makebox(0,0)[bl]{\colorbox{spinfamilygold}{\strut$#2$}}}
  \end{picture}%
}
\newcommand{\spinfamilyrowgroup}[2]{%
  \begin{picture}(0,0)
    \youngrow{0}{0}{#1}{spinfamilybase}
    \put(3.65,0.5){\makebox(0,0)[l]{$\times #2$}}
  \end{picture}%
}
\begin{picture}(42,35.1)(0,3.2)
  \put(0,37.3){\makebox(0,0)[l]{\textbf{(b)} Young diagrams}}
  \put(9,34.35){\makebox(0,0){$D_n$}}
  \put(28.2,34.35){\makebox(0,0){$D_{n+1}$}}
  \put(1,30.1){\makebox(0,0){$P$}}
  \put(17.6,30.1){\makebox(0,0){$\longrightarrow$}}
  \put(5.6,29.1){\spinfamilylongpair{n-i}{\lambda_0}}
  \put(24.7,29.1){\spinfamilylongpair{n+1-i}{\lambda_1}}
  \youngrow{30.9}{30.1}{1}{spinfamilyadd}
  \youngrow{30.9}{29.1}{1}{spinfamilyadd}
  \put(0,24.25){\color{black!22}\line(1,0){42}}
  \put(1,21.3){\makebox(0,0){$R$}}
  \put(17.6,21.3){\makebox(0,0){$\longrightarrow$}}
  \put(5.6,20.3){\spinfamilylongpair{n-4+i}{\mu_0}}
  \put(24.7,20.3){\spinfamilylongpair{n-3+i}{\mu_1}}
  \youngrow{30.9}{21.3}{1}{spinfamilyadd}
  \youngrow{30.9}{20.3}{1}{spinfamilyadd}
  \put(0,15.45){\color{black!22}\line(1,0){42}}
  \put(1,12.6){\makebox(0,0){$Q$}}
  \put(17.6,12.6){\makebox(0,0){$\longrightarrow$}}
  \put(5.6,13.7){\spinfamilyrowgroup{3}{j}}
  \put(24.7,13.7){\spinfamilyrowgroup{3}{j}}
  \youngrow{24.7}{12.7}{3}{spinfamilygold}
  \youngrow{24.7}{11.7}{3}{spinfamilygold}
  \put(5.6,10.3){\spinfamilyrowgroup{2}{2(h-j)}}
  \put(24.7,10.3){\spinfamilyrowgroup{2}{2(h-j)}}
  \put(5.6,8.9){\spinfamilyrowgroup{1}{j+2\epsilon}}
  \put(24.7,8.9){\spinfamilyrowgroup{1}{j+2\epsilon}}
  \youngrow{5.6}{7.9}{1}{spinfamilygold}
  \youngrow{5.6}{6.9}{1}{spinfamilygold}
  \youngrow{5.6}{5.9}{1}{spinfamilygold}
  \youngrow{5.6}{4.9}{1}{spinfamilygold}
\end{picture}
\end{minipage}
\endgroup
\caption{The general $D_n/D_{n+1}$ pair from Eq.~\eqref{eq:letter-punctures}. At $P,R$, horizontal dots abbreviate the long rows and green marks the added boxes. Below the vertical dots, the gold labels stand for the short partitions. At $Q$, each gray row is shown once with its multiplicity, and the gold rows show the four rows replaced by two. Empty partitions and row groups of multiplicity zero contribute no boxes.}
\label{fig:spin8-general}
\end{figure*}
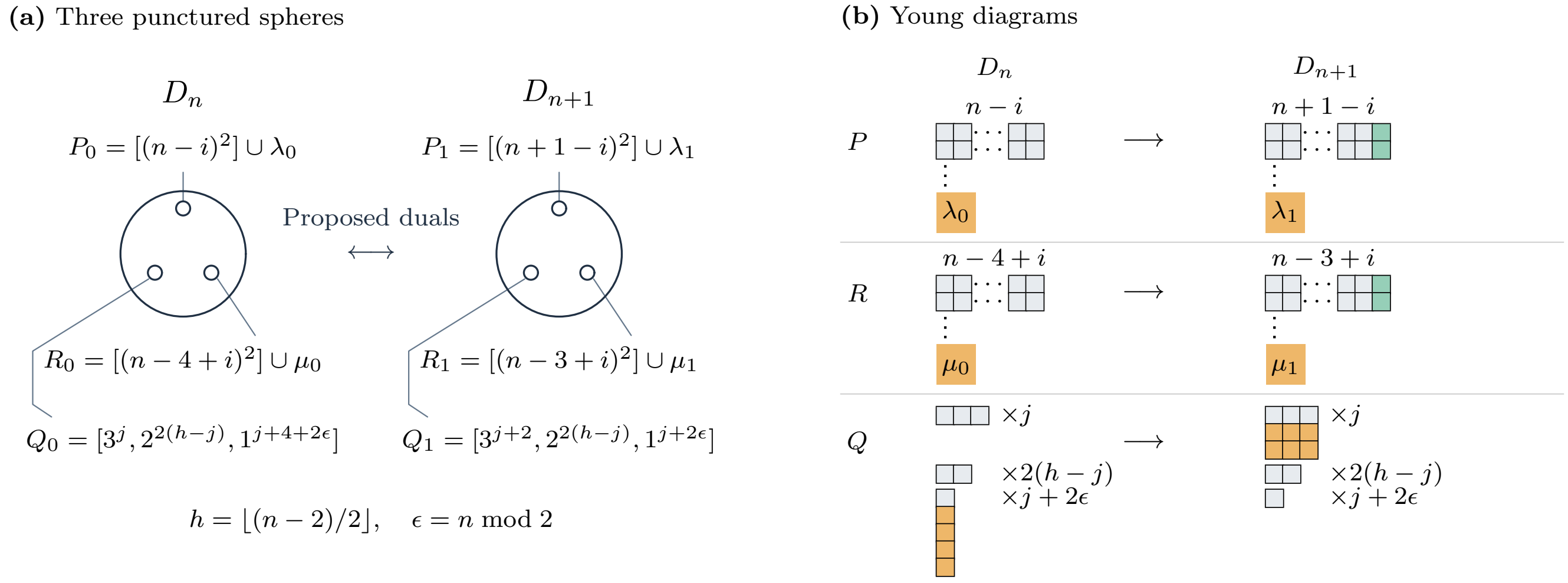

We checked $D_4/D_5$, the smallest adjacent pair of type $D$ parents, as well as the two $D_{14}/D_{15}$ examples in Fig.~\ref{fig:spin8-spheres}, with the results recorded in the IndexLab database. The larger examples make the puncture changes easier to see. Figure~\ref{fig:spin8-general} shows the general family that contains these examples.

Raising the rank from $n$ to $n+1$ adds two boxes at every puncture~\cite{CDTypeD}. At $P$ and $R$ we extend the two longest rows by one box each. At $Q$ we replace four rows of length one by two rows of length three.

In Fig.~\ref{fig:spin8-general}, $r=n+s$ with $s=0,1$ and $i=0,2$. Here $\epsilon=n\bmod 2$ and $0\le j\le h=\lfloor(n-2)/2\rfloor$.
\begin{align}
 P_s&=[(r-i)^2]\cup\lambda_s,\qquad
 R_s=[(r-4+i)^2]\cup\mu_s,\nonumber\\
 Q_s&=[3^{j+2s},2^{2(h-j)},1^{j+4+2\epsilon-4s}].
\label{eq:letter-punctures}
\end{align}
The union $\cup$ combines the rows and sorts them, keeping repeated rows. For example, $[3,1]\cup[2,2]=[3,2,2,1]=[3,2^2,1]$, where $2^2$ means two rows of length two. A $D$-partition is a partition in which each even row length occurs an even number of times~\cite[Sec.~2.1]{CDTypeD}. The row lengths in $\lambda_s$ and $\mu_s$ sum to $2i$ and $8-2i$, respectively. Together, these two short partitions contain eight boxes. The choices $i=0,2$ split these boxes as $0+8$ or $4+4$, with $\lambda_s=\varnothing$ for $i=0$. The short rows may stay unchanged. At $i=0$, the allowed exchanges are $[3,1^5]\leftrightarrow[2^4]$ and $[5,1^3]\leftrightarrow[4^2]$. At $i=2$, we allow $([1^4],[3,1])\leftrightarrow([2^2],[2^2])$, up to interchanging $P$ and $R$.

Each full label $P_s,Q_s,R_s$ must be a valid $D_r$ partition of $2r$. We retain spheres with nonnegative graded Coulomb branch dimensions and no decoupled free hypermultiplets. Every row of $\lambda_s$ must have length at most $r-i$, and every row of $\mu_s$ must have length at most $r-4+i$. These bounds ensure that the two equal rows displayed explicitly in $P_s$ and $R_s$ are among the longest rows of those punctures.

Both examples in Fig.~\ref{fig:spin8-spheres} have $n=14$, $i=j=0$, and $\lambda_0=\lambda_1=\varnothing$. Panels (a,b) take $\mu_0=\mu_1=[4^2]$. Panel (c) instead takes $\mu_0=[2^4]$ and $\mu_1=[3,1^5]$, keeping $P,Q$ fixed. The general family also allows $i=2$, with four short boxes at each of $P$ and $R$.

For a very even $D$ partition, whose parts are all even and occur with even multiplicity, the same Young diagram labels two distinct nilpotent orbits. Specifying the puncture therefore requires an additional choice, denoted red or blue~\cite[Sec.~2.1]{CDTypeD}. In the $D_{14}$ examples of Fig.~\ref{fig:spin8-spheres}, both $P$ and $R$ require this choice, while $Q$ does not.

With the puncture partitions held fixed, simultaneously exchanging red and blue at $P$ and $R$ is induced by an outer automorphism and gives an isomorphic SCFT~\cite[Sec.~2.1]{DESDistinguishing}. Thus the assignments $(\mathrm{red},\mathrm{red})$ and $(\mathrm{blue},\mathrm{blue})$ are equivalent, as are $(\mathrm{red},\mathrm{blue})$ and $(\mathrm{blue},\mathrm{red})$, with the entries ordered as $(P,R)$. We can therefore fix $P$ to red, leaving two choices: $R$ red (same color) or $R$ blue (opposite colors). We refer to this choice as the \emph{relative color} of $P$ and $R$ on the same $D_{14}$ sphere. In Fig.~\ref{fig:spin8-spheres}(a,b), the $D_{14}$ punctures $P=[14^2]_{\mathrm{red}}$ and $R=[10^2,4^2]_{\mathrm{red}}$ are both red. In panel (c), $P$ remains red while $R=[10^2,2^4]_{\mathrm{blue}}$ is blue.

The combined short partition $\lambda_s\cup\mu_s$ specifies an $\mathfrak{su}(2)$ embedding in $\mathfrak{so}(8)$. Each row of length $d$ gives an irreducible summand in the vector representation, with weights $d-1,d-3,\ldots,1-d$~\cite{CDTDefects}. On each side, sort the positive weights in decreasing order and pad with zeros to obtain $u_1,\ldots,u_4$. On the odd rank side, we take the vector weights $\pm u_a$. On the even rank side, we take the half sums $\tfrac12(\pm u_1\pm u_2\pm u_3\pm u_4)$ with an odd number of minus signs. We then compare the two sets of weights. When the combined short partition on the even rank side is very even, negate $u_4$ for opposite $P,R$ colors at even $n$ or equal colors at odd $n$~\cite{CDTypeD}. When the two characters agree, we conjecture that the SCFTs are equivalent for the stated short row choices. For example, the short partition $[4^2]$ in Fig.~\ref{fig:spin8-spheres}(b) gives two four dimensional $\mathfrak{su}(2)$ representations. With the displayed colors, the selected spinor has the same decomposition, so
\begin{equation}
 \chi_{\mathrm{vec}}(z)=\chi_{\mathrm{spin}}(z)
 =2\bigl(z^3+z+z^{-1}+z^{-3}\bigr),
\label{eq:spin8-character-example}
\end{equation}
where $z$ records the $\mathfrak{su}(2)$ weight $2J_3$. This is the character match underlying this example. A related $D_6$ example in the literature distinguishes gauged theories by their $\Spin(7)$ embedding in a puncture's $\Spin(8)$ flavor group~\cite[Sec.~5.1]{DEKLIsomorphisms}.

For $j<h$, the $2(h-j)$ rows of length two at $Q$ give a flavor factor $\Sp(h-j)$, with $\Sp(m)=\mathrm{USp}(2m)$. The four dimensional flavor central charge $k$ is also called the flavor level. It fixes the normalization of the two point function of the flavor currents. We use the convention in which a half hypermultiplet in the fundamental contributes $k=1$. If $b_1,b_2$ are the heights of the first two columns in our row length convention for the partition, then $k=b_1+b_2$~\cite[Sec.~2.3]{CDTypeD}. In Fig.~\ref{fig:spin8-spheres}, both $Q$ punctures have twelve rows of length two and hence a $\Sp(6)$ factor. Their column heights change from $(16,12)$ to $(14,14)$, preserving $k=28$. The displayed move also preserves $24(c-a)$.

To isolate the effect of this color choice, keep all puncture partitions in Fig.~\ref{fig:spin8-spheres}(a) fixed and change $R=[10^2,4^2]_{\mathrm{red}}$ to $[10^2,4^2]_{\mathrm{blue}}$, keeping $P$ red. This changes the relative color and the selected spinor character. Compared with the unchanged $D_{15}$ theory, the Schur index difference $\Ia_{14}-\Ia_{15}$ computed in IndexLab starts at $-2\tau^7$. This sensitivity to color lets us distinguish theories with the same anomalies and Coulomb branch spectra~\cite{DESDistinguishing}. A similar dependence on the choice of very even orbit occurs in six dimensional conformal matter~\cite{DKLDistinguishing6d}. Since the Schur index is a specialization of the full index, this mismatch also rules out equality of the full indices.

\subsection{\texorpdfstring{$A/D$}{A/D} comparisons with rectangular punctures}
\label{sec:rectangular-ad}

This family relates untwisted $A_{N-1}$ spheres to $D_M$ spheres with a $\mathbb Z_2$ twist, with either four or five punctures on each side, as shown in Fig.~\ref{fig:rectangular-ad-versions}. Here $N$ is the number of boxes in each $A$ puncture partition, $M$ is the rank of the $D$ parent algebra, and $k$ counts the rows of length two at the first $A$ puncture.

\begin{figure*}[t]
\centering
\begin{minipage}[t]{0.48\textwidth}
\vspace{0pt}
\centering\footnotesize
\textbf{(a) Four-punctured spheres}\par\smallskip
\AtlasGeneral{3}
\end{minipage}\hfill
\begin{minipage}[t]{0.48\textwidth}
\vspace{0pt}
\centering\footnotesize
\textbf{(b) Five-punctured spheres}\par\smallskip
\AtlasPair{A_{N-1}}{5}{%
  \AtlasKey{1}{[2^k,1^{N-2k}]}%
  \AtlasKey{2}{[N-n,n]}%
  \AtlasKey{3}{[N-n,n]}%
  \AtlasKey{4}{[N-1,1]}%
  \AtlasKey{5}{[N-1,1]}%
}{D_M^{\mathrm{tw}}}{5}{%
  \AtlasKey{1}{[2^{2d},1^{2M-4d}]}%
  \AtlasKey{2}{[(M-1)^2,1^2]}%
  \AtlasKey{3}{[M^2]}%
  \AtlasKey{4}{[2M-2]_t}%
  \AtlasKey{5}{[2M-2]_t}%
}
\end{minipage}
\caption{Four- and five-punctured spheres in the $A/D$ family with rectangular punctures. Numbered dots identify the punctures listed below each sphere. Both panels use $M=N-k$, $n=\lfloor N/2\rfloor$, and $d=M-n-1$. The choice $\eta=[1^2]$ or $[2]$ enters panel (a).}
\label{fig:rectangular-ad-versions}
\end{figure*}

Set $n=\lfloor N/2\rfloor$ and $d=M-n-1$, and define
\begin{gather}
 P=[2^k,1^{N-2k}],\qquad R=[N-n,n], \nonumber\\
 T=[2^{2d},1^{2M-4d}].
\end{gather}
The two copies of $R$ are rectangular for even $N$ and have rows differing by one box for odd $N$. For the four-punctured constructions, the same partition of two boxes $\eta=[1^2]$ or $[2]$ enters the $A$ and twisted $D$ puncture lists:
\begin{align}
 &A_{N-1}\{P,R,R,[N-2]\cup\eta\}\nonumber\\
 &\quad\longleftrightarrow D_M^{\mathrm{tw}}
 \{T, ([(M-2)^2]\cup\eta)_t,[M^2],[2M-2]_t\}.
\label{eq:rectangular-ad-first}
\end{align}

On the $A_{N-1}$ side, the five-punctured construction contains two simple punctures $[N-1,1]$~\cite[Sec.~2.1]{CDTypeA}:
\begin{align}
 &A_{N-1}\{P,R,R,[N-1,1],[N-1,1]\}\nonumber\\
 &\quad\longleftrightarrow D_M^{\mathrm{tw}}
 \{T,[(M-1)^2,1^2],[M^2],([2M-2]_t)^2\}.
\label{eq:rectangular-ad-second}
\end{align}
When $M$ is even, $[M^2]$ is very even. The recorded checks use the blue label for that puncture. The four-punctured spheres have one exactly marginal coupling on each side, and the five-punctured spheres have two~\cite{Gaiotto,CDTypeA,CDTTwistedD}.

For the four-punctured constructions, the sphere and puncture anomaly contributions~\cite[Sec.~3.3]{CDTDefects} give
\begin{equation}
 n_v^{(A)}-n_v^{(D)}=(N-k)^2-M^2.
\end{equation}
Matching $n_v$ fixes $M=N-k$. With this relation between the ranks, the difference in the effective hypermultiplet numbers is $n_h^{(A)}-n_h^{(D)}=k(N\bmod 2)$. This difference vanishes for even $N$. For odd $N$, matching both anomalies requires $k=0$, so $M=N$. 

For both the four- and five-punctured cases without decoupled free hypermultiplets, even $N$ allows $\max(4,n+1)\le M\le N$, or equivalently $0\le k\le\min(N-4,N/2-1)$. The upper bound follows from $k=N-M\ge0$, while $d=M-n-1\ge0$ gives $M\ge n+1$. The remaining bound $M\ge4$ is the lower rank cutoff for the $D$ parents considered here. For example, $P=[2,1^4]$ has $N=6$ and $k=1$, giving $M=5$. With $\eta=[1^2]$, the proposed duality between four-punctured spheres is
\begin{align*}
 &A_5\{[2,1^4],[3^2],[3^2],[4,1^2]\}\\
 &\quad\longleftrightarrow D_5^{\mathrm{tw}}
 \{[2^2,1^6],[3^2,1^2]_t,[5^2],[8]_t\}.
\end{align*}
Both constructions have $(n_v,n_h)=(42,66)$ and Coulomb branch scaling dimensions $2,3,4,4,5,6$. The five-punctured constructions at the same ranks have $(n_v,n_h)=(45,68)$ and one additional Coulomb branch generator of scaling dimension two on each side. 

\subsection{Exceptional puncture replacements}
\AtlasMain{eq:exceptional-puncture-diagram}{\AtlasGeneral{6}}

This family includes matches between $E_6$ and $E_7$ and between $E_7$ and $E_8$. Exceptional punctures are specified by Bala--Carter labels of nilpotent orbits. Thus a label such as $A_2$ or $D_4(a_1)$ inside the braces names a puncture, while the symbol outside the braces names the six dimensional parent~\cite[App.~A]{CDTTypeE6}.

Let $d_\Delta$ count the independent Coulomb branch generators of scaling dimension $\Delta$ in the complete SCFT. For a puncture $P$ of parent $G$, define
\begin{equation*}
 \mathbf v_G(P)=\bigl(n_v^{G}(P),n_h^{G}(P),\{d_\Delta^{G}(P)\}\bigr).
\end{equation*}
The first two entries are the puncture contributions to the effective numbers of vector multiplets and hypermultiplets. The remaining entries give its contributions to the graded Coulomb branch dimensions, including the effects of local constraints~\cite{CDTTypeE6,CDTZTypeE7,CDTZTypeE8}. For a sphere with three punctures, adding the bulk and puncture contributions gives the complete theory data:
\begin{equation*}
 \bigl(n_v,n_h,\{d_\Delta\}\bigr)
 =\mathbf b_G+\sum_{i=1}^{3}\mathbf v_G(P_i).
\end{equation*}
Here $\mathbf b_G$ is the bulk contribution fixed by the parent on a sphere~\cite[Sec.~3.3]{CDTDefects}. We use the same list of scaling dimensions for both parents, inserting zero components where necessary. For three replacements $P_i\mapsto O_i$, from parent $G$ to parent $H$, define
\begin{equation}
 \boldsymbol\delta_i=\mathbf v_H(O_i)-\mathbf v_G(P_i),
 \qquad i=1,2,3.
\end{equation}
Equality of the anomalies and graded Coulomb branch spectra requires
\begin{equation}
 \sum_{i=1}^3\boldsymbol\delta_i=\mathbf b_G-\mathbf b_H.
 \label{eq:exceptional-puncture-balance}
\end{equation}
The puncture changes must therefore offset the change of parent in every component. We use this condition to select candidates, then test whether their indices agree.

The comparison
\begin{equation}
 E_6\{A_2,A_2,E_6(a_3)\}
 \longleftrightarrow E_7\{D_4(a_1),A_4,E_7(a_3)\}
\end{equation}
gives the differences in puncture contributions listed below. In each row, $P_i$ is a puncture of the $E_6$ construction and $O_i$ is its replacement in the $E_7$ construction. The last two columns are the differences $n_v^{E_7}(O_i)-n_v^{E_6}(P_i)$ and $n_h^{E_7}(O_i)-n_h^{E_6}(P_i)$, respectively:
\begin{equation}
\begin{array}{@{}llrr@{}}
 P_i & O_i & (\boldsymbol\delta_i)_{n_v} & (\boldsymbol\delta_i)_{n_h}\\
 A_2 & D_4(a_1) & 795 & 796\\
 A_2 & A_4 & 718 & 716\\
 E_6(a_3) & E_7(a_3) & 432 & 432
\end{array}
\end{equation}
For example, the first row gives $1316-521=795$ and $1332-536=796$~\cite{CDTTypeE6,CDTZTypeE7}. The column sums $(1945,1944)$ equal the $(n_v,n_h)$ components of $\mathbf b_{E_6}-\mathbf b_{E_7}$. This opposite order for the bulk difference follows from requiring the total theory data to agree in Eq.~\eqref{eq:exceptional-puncture-balance}. Both constructions have $(n_v,n_h)=(113,152)$ and the same Coulomb branch spectrum. Sending both $A_2$ punctures to $D_4(a_1)$ instead gives $(n_v,n_h)=(190,232)$ on the $E_7$ side and fails the anomaly test. The checks cover 156 $E_6/E_7$, 61 $E_7/E_8$, and 36 $D_5/E_6$, plus 19 $D_6/E_6$ entries. In twelve $D_6/E_6$ tests, changing the relative colors preserves the anomalies and Coulomb branch spectrum but changes the first Schur coefficient at $\tau^3$, as predicted by the leading spinor term.

The $E_7/E_8$ example illustrates the Coulomb components of Eq.~\eqref{eq:exceptional-puncture-balance}:
\begin{align*}
 &E_7\{D_4(a_1),D_4(a_1),E_7(a_3)\}\\
 &\qquad\longleftrightarrow E_8\{E_8(a_7),A_6,E_8(a_3)\},
\end{align*}
On a sphere, each parent Casimir of degree $\Delta$ contributes $-(2\Delta-1)$ to the bulk component $d_\Delta$. The degrees $20,24,30$ occur among the $E_8$ Casimirs but not among those of $E_7$. At these degrees, changing the parent from $E_7$ to $E_8$ therefore changes the bulk contributions by $(-39,-47,-59)$. The selected puncture changes contribute $(39,47,59)$, cancelling these differences. Neither construction has Coulomb branch generators at those three degrees, and the full spectra agree degree by degree. Both sides have Coulomb branch rank 12, meaning twelve independent Coulomb branch generators, and $(a,c)=(197/4,51)$. 

\subsection{\texorpdfstring{From tori to spheres in type \(A\)}{From tori to spheres in type A}}
\label{sec:torus-sphere}
\AtlasMain{eq:torus-sphere-opened}{%
  \AtlasTorusPair{A_{n-1}^{g=1}}{2}{%
    \AtlasKey{1}{[n-2,2]}%
    \AtlasKey{2}{[n-2,2]}%
  }{A_{2n-1}^{\mathrm{tw}}}{5}{%
    \AtlasKey{1}{[2n-3,3,1]_t}%
    \AtlasKey{2}{[2n-1,1]}%
    \AtlasKey{3}{[2n+1]_t}%
    \AtlasKey{4}{[2n+1]_t}%
    \AtlasKey{5}{[2n+1]_t}%
  }%
}

Equation~\eqref{eq:torus-sphere-opened} gives a candidate identification between a torus with two punctures and a twisted sphere with five punctures for \(n\ge4\). Both sides have two exactly marginal couplings. The torus punctures give \((n_v,n_h)=(4n^2-20,4n^2-16)\), two Coulomb generators of each dimension \(2\) and \(3\), and four of each dimension \(4,\dots,n\). For all \(n\ge4\), the sphere reproduces these anomalies and the Coulomb branch spectrum. 

Cutting the torus gives two copies of the trinion \(\widetilde R_{0,n}=A_{n-1}\{[1^n],[1^n],[n-2,2]\}\), joined by \(SU(n)\times SU(n)\) gauging~\cite[Sec.~2.2]{AMSE7}. In a bubbling degeneration, a cluster of approaching punctures lies on a separate sphere joined to the rest of the surface by a tube that pinches in the limit. On the sphere, bubbling off \([2n-1,1]\) and two \([2n+1]_t\) punctures gives an empty component attached by an \(SU(n)\times SU(n)\) cylinder to the twisted trinion \(A_{2n-1}^{\mathrm{tw}}\{[1^{2n}],[2n-3,3,1]_t,[2n+1]_t\}\)~\cite[Sec.~4.2.1]{CDTTwistedA}.

The interesting observation is the connection with the twisted trinion \(A_{2n-1}^{\mathrm{tw}}\{[1^{2n}],[2n-3,1^4]_t,[2n+1]_t\}\) which describes a product of two \(R_{0,n}\) sectors~\cite[Sec.~5.1.1]{CDTTwistedA}. Each \(R_{0,n}\) has an \(SU(2)\) flavor symmetry at its \([n-2,1^2]\) puncture. Higgsing that symmetry changes the puncture to \([n-2,2]\), giving \(\widetilde R_{0,n}\)~\cite[Sec.~2.2]{AMSE7}. On the twisted side the two Higgsings act diagonally on the four unit rows, with \(\mathbf{2}\otimes\mathbf{2}=\mathbf{3}\oplus\mathbf{1}\), so \([2n-3,1^4]_t\) becomes \([2n-3,3,1]_t\). This recovers the twisted trinion above and suggests that it describes two \(\widetilde R_{0,n}\) sectors, supporting the proposed torus--sphere identification. 

\subsection{Five constructions of an isolated theory}
\label{sec:five-way}

The five spheres in Fig.~\ref{fig:five-way} are candidate realizations of one isolated SCFT. 

\input{figures/Five_Way_Duality}

All five have $(a,c)=(41/3,46/3)$, Coulomb branch generators of scaling dimensions $(4,6,6,10)$, and no decoupled free hypermultiplets. The proposed common SCFT is \emph{isolated}, meaning that it has no exactly marginal deformation preserving $\mathcal N=2$ supersymmetry. Such a deformation would be a supersymmetry descendant of a Coulomb branch operator of scaling dimension two, which is absent here~\cite[Table~22]{CDIDeformations}.

The relative color remains essential. In Fig.~\ref{fig:five-way}(b), change $[6^2]$ from red to blue while keeping $[2^6]$ blue. This changes the two very even punctures from opposite to equal colors. The anomalies and Coulomb branch spectrum are unchanged, but the original and color flipped Schur indices begin
\begin{align}
 \Ia_{\mathrm{original}}&=1+56\tau^2+\OO(\tau^3),\nonumber\\
 \Ia_{\mathrm{flip}}&=1+2\tau+51\tau^2+\OO(\tau^3).
\end{align}
The comparison in IndexLab gives $\Ia_{\mathrm{flip}}-\Ia_{\mathrm{original}}=2\tau-5\tau^2+\OO(\tau^3)$. One full free hypermultiplet has unrefined Schur index $\Ia_{\mathrm{hyper}}=1+2\tau+3\tau^2+\OO(\tau^3)$~\cite{GRRYSchur,GRRYMacdonald}. Tensoring the original SCFT with it would give $\Ia_{\mathrm{original}}\Ia_{\mathrm{hyper}}=1+2\tau+59\tau^2+\OO(\tau^3)$. The color flipped theory therefore cannot be obtained merely by adding a free hypermultiplet to the original SCFT.

\section{Conclusion}
\label{sec:conclusion}

Our census organizes candidate class $\mathcal{S}$ isomorphisms by recurring changes in ranks, punctures, and twists. 

The conjectured rules still need a derivation. The coupling maps for families with exactly marginal deformations also remain to be determined. Our conjectures concern equivalence of the local SCFTs; extending them to line operators requires matching the global data~\cite{TachikawaGlobal}.

From the surviving candidates, all 1{,}569 tested pairs (within the scope of IndexLab) agree in the full $(p,q,t)$ index with flavor fugacities set to one. The comparisons reach common orders 5, 6, or 7 in $s$, with $p=s^3y$, $q=s^3/y$, and $t=s^4/v$. The IndexLab database records 12 comparisons using published formulas, 1{,}314 using conjectural elliptic wavefunctions, and 243 using reconstructions from protected data. To our knowledge, explicit perturbative expansions of the full $p,q,t$ index have yet to appear for many class $\mathcal{S}$ theories in our catalogue. Candidate dualities can help us develop and check these calculations. For example, in the $A_5/D_5$ comparisons of Sec.~\ref{sec:rectangular-ad}, we use the untwisted $A_5$ index to check the proposed elliptic wavefunctions for twisted $D_5$. Similar comparisons helped guide the full index constructions in Supplemental Sec.~\ref{sec:full-index}.

\section*{ACKNOWLEDGMENTS}
Z.D.\ and Z.Z.\ thank Hongliang Jiang for useful conversations. Z.D.\ and Z.Z.\ are supported by a SIMIS starting grant. We acknowledge the use of generative AI in large scale scanning of examples and polishing of this paper. The authors take full responsibility for the content of this paper.

\let\standardbibfont\bibfont
\renewcommand{\bibfont}{\standardbibfont\setlength{\baselineskip}{9pt}}
% References are supplied directly below; skip REVTeX automatic BibTeX insertion.
\makeatletter
\let\auto@bib@innerbib\@empty
\makeatother

\input{Class_S_References}
\let\bibfont\standardbibfont

\onecolumngrid
\clearpage
\input{Supplemental_Material}

\end{document}

%% file: figures/Family_Diagrams.tex
\definecolor{atlasink}{RGB}{32,48,67}
\definecolor{atlascell}{RGB}{231,235,239}
\definecolor{atlastail}{RGB}{248,222,182}

\newcommand{\AtlasNumber}[1]{%
  \textcolor{atlasink}{\ensuremath{\scriptstyle #1}}%
}
\newcommand{\AtlasDot}[3]{%
  \put(#1,#2){\makebox(0,0){\textcolor{black!15}{\scalebox{2.1}{$\bullet$}}}}%
  \put(#1,#2){\makebox(0,0){\AtlasNumber{#3}}}%
}
\newcommand{\AtlasSphere}[1]{%
  \begingroup
  \setlength{\unitlength}{1pt}%
  \linethickness{0.50pt}%
  \begin{picture}(38,38)
    {\color{atlasink}%
    \qbezier(19,2)(26.04,2)(31.02,6.98)
    \qbezier(31.02,6.98)(36,11.96)(36,19)
    \qbezier(36,19)(36,26.04)(31.02,31.02)
    \qbezier(31.02,31.02)(26.04,36)(19,36)
    \qbezier(19,36)(11.96,36)(6.98,31.02)
    \qbezier(6.98,31.02)(2,26.04)(2,19)
    \qbezier(2,19)(2,11.96)(6.98,6.98)
    \qbezier(6.98,6.98)(11.96,2)(19,2)}
    \ifcase#1\relax\or\or\or
      \AtlasDot{19}{27}{1}\AtlasDot{27}{14}{2}\AtlasDot{11}{14}{3}%
    \or
      \AtlasDot{12}{25}{1}\AtlasDot{26}{25}{2}%
      \AtlasDot{26}{13}{3}\AtlasDot{12}{13}{4}%
    \or
      \AtlasDot{19}{28}{1}\AtlasDot{27}{22}{2}\AtlasDot{24}{12}{3}%
      \AtlasDot{14}{12}{4}\AtlasDot{11}{22}{5}%
    \fi
  \end{picture}%
  \endgroup
}
\newcommand{\AtlasKey}[2]{\AtlasNumber{#1}&$#2$\\}
\newcommand{\AtlasSide}[3]{%
  \begin{minipage}[t]{.46\linewidth}
  \centering
  $#1$\par
  \AtlasSphere{#2}\par
  \begin{tabular}{@{}r@{\hspace{4pt}}l@{}}#3\end{tabular}
  \end{minipage}%
}
\newcommand{\AtlasPair}[6]{%
  \par\noindent
  \begingroup\footnotesize
  \AtlasSide{#1}{#2}{#3}%
  \hfill\raisebox{-22pt}[0pt][0pt]{$\leftrightarrow$}\hfill
  \AtlasSide{#4}{#5}{#6}\par
  \endgroup
}
\newcommand{\AtlasNote}[1]{%
  \par\smallskip{\centering\footnotesize #1\par}%
}

\makeatletter
\newcommand{\AtlasBoxes}[1]{%
  \begingroup
  \fboxsep=0pt \fboxrule=.22pt
  \@tempcnta=#1\relax
  \loop\ifnum\@tempcnta>0
    \fcolorbox{atlasink}{atlascell}{\rule{0pt}{2.7pt}\hspace{2.7pt}}%
    \advance\@tempcnta by -1
  \repeat
  \endgroup
}
\newcommand{\AtlasGrid}[1]{%
  $\vcenter{\offinterlineskip
    \@for\AtlasNext:=#1\do{\hbox{\AtlasBoxes{\AtlasNext}}}%
  }$%
}
\newcommand{\AtlasRows}[1]{$\vcenter{\offinterlineskip #1}$}
\newcommand{\AtlasRow}[2]{%
  \hbox{\AtlasBoxes{#1}\kern 3pt$\scriptstyle\times #2$}\kern 2pt%
}
\newcommand{\AtlasLong}[2]{%
  \hbox{\AtlasBoxes{1}$\scriptstyle\cdots$\AtlasBoxes{1}%
    \kern 3pt$\scriptstyle (#1)\times #2$}\kern 2pt%
}
\newcommand{\AtlasTail}[1]{\hbox{\AtlasBadge{#1}}\kern 1pt}
\newcommand{\AtlasBadge}[1]{%
  \begingroup\setlength{\fboxsep}{2pt}\colorbox{atlastail}{$#1$}\endgroup
}
\newcommand{\AtlasCell}[2]{%
  \AtlasNumber{#1}&
  \raisebox{\dimexpr 2pt+(\depth-\height)/2\relax}{#2}\\
  \noalign{\vskip 3pt}%
}
\newcommand{\AtlasBoardSide}[2]{%
  \begin{minipage}[t]{.49\linewidth}
  \centering
  $#1$\par\smallskip
  \begin{tabular}{@{}r@{\hspace{4pt}}l@{}}#2\end{tabular}
  \end{minipage}%
}
\newcommand{\AtlasBoards}[4]{%
  \par\noindent
  \begingroup\footnotesize
  \AtlasBoardSide{#1}{#2}\hfill\AtlasBoardSide{#3}{#4}\par
  \endgroup
}
\newcommand{\AtlasDefine}[5]{%
  \@namedef{AtlasGeneral@#1}{#2}%
  \@namedef{AtlasShapes@#1}{#3}%
  \@namedef{AtlasWorked@#1}{#4}%
  \@namedef{AtlasGrids@#1}{#5}%
}
\newcommand{\AtlasGeneral}[1]{\@nameuse{AtlasGeneral@#1}}
\newcommand{\AtlasPlateRow}[4]{%
  \noindent
  \begin{minipage}[t]{.48\linewidth}
    \centering\footnotesize\textbf{#1}\par\smallskip
    #2%
  \end{minipage}\hfill
  \begin{minipage}[t]{.48\linewidth}
    \centering\footnotesize\textbf{#3}\par\smallskip
    #4%
  \end{minipage}\par
}
\newcommand{\AtlasPlate}[1]{%
  \par\nobreak\smallskip
  \AtlasPlateRow{General punctures}{\AtlasGeneral{#1}}%
    {General shapes}{\@nameuse{AtlasShapes@#1}}%
  \medskip
  \AtlasPlateRow{Worked example (\thesubsection)}{\@nameuse{AtlasWorked@#1}}%
    {Worked shapes}{\@nameuse{AtlasGrids@#1}}%
  \medskip
}
\newcommand{\AtlasMain}[2]{%
  \begin{equation}
    \label{#1}%
    \text{\begin{minipage}{.86\linewidth}#2\end{minipage}}%
  \end{equation}%
}
\makeatother

\input{figures/Family_Diagrams_Orbits}
\input{figures/Family_Diagrams_Main}
\input{figures/Family_Diagrams_Supp}

\input{figures/Family_Diagrams_Additional}

%% file: figures/Family_Diagrams_Main.tex
\AtlasDefine{1}{%
  \AtlasPair{D_n}{3}{%
    \AtlasKey{1}{[(n-i)^2]\cup\lambda_0}%
    \AtlasKey{2}{[3^j,2^{2(h-j)},1^{j+4+2\epsilon}]}%
    \AtlasKey{3}{[(n-4+i)^2]\cup\mu_0}%
  }{D_{n+1}}{3}{%
    \AtlasKey{1}{[(n+1-i)^2]\cup\lambda_1}%
    \AtlasKey{2}{[3^{j+2},2^{2(h-j)},1^{j+2\epsilon}]}%
    \AtlasKey{3}{[(n-3+i)^2]\cup\mu_1}%
  }%
  \AtlasNote{$h=\lfloor(n-2)/2\rfloor,\quad \epsilon=n\bmod 2.$}%
}{%
  \AtlasBoards{D_n}{%
    \AtlasCell{1}{\AtlasRows{\AtlasLong{n-i}{2}\AtlasTail{\lambda_0}}}%
    \AtlasCell{2}{\AtlasRows{\AtlasRow{3}{j}\AtlasRow{2}{2(h-j)}\AtlasRow{1}{j+4+2\epsilon}}}%
    \AtlasCell{3}{\AtlasRows{\AtlasLong{n-4+i}{2}\AtlasTail{\mu_0}}}%
  }{D_{n+1}}{%
    \AtlasCell{1}{\AtlasRows{\AtlasLong{n+1-i}{2}\AtlasTail{\lambda_1}}}%
    \AtlasCell{2}{\AtlasRows{\AtlasRow{3}{j+2}\AtlasRow{2}{2(h-j)}\AtlasRow{1}{j+2\epsilon}}}%
    \AtlasCell{3}{\AtlasRows{\AtlasLong{n-3+i}{2}\AtlasTail{\mu_1}}}%
  }%
}{%
  \AtlasPair{D_{14}}{3}{%
    \AtlasKey{1}{[14^2]_{\mathrm{red}}}%
    \AtlasKey{2}{[2^{12},1^4]}%
    \AtlasKey{3}{[10^2,4^2]_{\mathrm{red}}}%
  }{D_{15}}{3}{%
    \AtlasKey{1}{[15^2]}%
    \AtlasKey{2}{[3^2,2^{12}]}%
    \AtlasKey{3}{[11^2,4^2]}%
  }%
}{%
  \AtlasBoards{D_{14}}{%
    \AtlasCell{1}{\AtlasGrid{14,14}}%
    \AtlasCell{2}{\AtlasGrid{2,2,2,2,2,2,2,2,2,2,2,2,1,1,1,1}}%
    \AtlasCell{3}{\AtlasGrid{10,10,4,4}}%
  }{D_{15}}{%
    \AtlasCell{1}{\AtlasGrid{15,15}}%
    \AtlasCell{2}{\AtlasGrid{3,3,2,2,2,2,2,2,2,2,2,2,2,2}}%
    \AtlasCell{3}{\AtlasGrid{11,11,4,4}}%
  }%
}

\AtlasDefine{2}{%
  \AtlasPair{D_N^{\mathrm{tw}}}{4}{%
    \AtlasKey{1}{[3^j,2^{2(k-j)},1^{2N-4k+j}]}%
    \AtlasKey{2}{[N^2]}%
    \AtlasKey{3}{[(N-1)^2]_t}%
    \AtlasKey{4}{([2N-4]\cup\rho)_t}%
  }{D_M^{\mathrm{tw}}}{4}{%
    \AtlasKey{1}{[3^{j+2d},2^{2(k-j)},1^{j+2\epsilon}]}%
    \AtlasKey{2}{[M^2]}%
    \AtlasKey{3}{[(M-1)^2]_t}%
    \AtlasKey{4}{([2M-4]\cup\rho)_t}%
  }%
  \AtlasNote{$d=M-N,\quad k=\lfloor N/2\rfloor-d,\quad \epsilon=N\bmod 2,\quad
  \rho\in\{[1^2],[2]\}.$}%
}{%
  \AtlasBoards{D_N^{\mathrm{tw}}}{%
    \AtlasCell{1}{\AtlasRows{\AtlasRow{3}{j}\AtlasRow{2}{2(k-j)}\AtlasRow{1}{2N-4k+j}}}%
    \AtlasCell{2}{\AtlasRows{\AtlasLong{N}{2}}}%
    \AtlasCell{3}{\AtlasRows{\AtlasLong{N-1}{2}}}%
    \AtlasCell{4}{\AtlasRows{\AtlasLong{2N-4}{1}\AtlasTail{\rho}}}%
  }{D_M^{\mathrm{tw}}}{%
    \AtlasCell{1}{\AtlasRows{\AtlasRow{3}{j+2d}\AtlasRow{2}{2(k-j)}\AtlasRow{1}{j+2\epsilon}}}%
    \AtlasCell{2}{\AtlasRows{\AtlasLong{M}{2}}}%
    \AtlasCell{3}{\AtlasRows{\AtlasLong{M-1}{2}}}%
    \AtlasCell{4}{\AtlasRows{\AtlasLong{2M-4}{1}\AtlasTail{\rho}}}%
  }%
}{%
  \AtlasPair{D_6^{\mathrm{tw}}}{4}{%
    \AtlasKey{1}{[3,2^2,1^5]}%
    \AtlasKey{2}{[6^2]}%
    \AtlasKey{3}{[5^2]_t}%
    \AtlasKey{4}{[8,1^2]_t}%
  }{D_7^{\mathrm{tw}}}{4}{%
    \AtlasKey{1}{[3^3,2^2,1]}%
    \AtlasKey{2}{[7^2]}%
    \AtlasKey{3}{[6^2]_t}%
    \AtlasKey{4}{[10,1^2]_t}%
  }%
}{%
  \AtlasBoards{D_6^{\mathrm{tw}}}{%
    \AtlasCell{1}{\AtlasGrid{3,2,2,1,1,1,1,1}}%
    \AtlasCell{2}{\AtlasGrid{6,6}}%
    \AtlasCell{3}{\AtlasGrid{5,5}}%
    \AtlasCell{4}{\AtlasGrid{8,1,1}}%
  }{D_7^{\mathrm{tw}}}{%
    \AtlasCell{1}{\AtlasGrid{3,3,3,2,2,1}}%
    \AtlasCell{2}{\AtlasGrid{7,7}}%
    \AtlasCell{3}{\AtlasGrid{6,6}}%
    \AtlasCell{4}{\AtlasGrid{10,1,1}}%
  }%
}

\AtlasDefine{3}{%
  \AtlasPair{A_{N-1}}{4}{%
    \AtlasKey{1}{[2^k,1^{N-2k}]}%
    \AtlasKey{2}{[N-n,n]}%
    \AtlasKey{3}{[N-n,n]}%
    \AtlasKey{4}{[N-2]\cup\eta}%
  }{D_M^{\mathrm{tw}}}{4}{%
    \AtlasKey{1}{[2^{2d},1^{2M-4d}]}%
    \AtlasKey{2}{([(M-2)^2]\cup\eta)_t}%
    \AtlasKey{3}{[M^2]}%
    \AtlasKey{4}{[2M-2]_t}%
  }%
  \AtlasNote{$M=N-k,\quad n=\lfloor N/2\rfloor,\quad d=M-n-1,\quad
  \eta\in\{[1^2],[2]\}.$}%
}{%
  \AtlasBoards{A_{N-1}}{%
    \AtlasCell{1}{\AtlasRows{\AtlasRow{2}{k}\AtlasRow{1}{N-2k}}}%
    \AtlasCell{2}{\AtlasRows{\AtlasLong{N-n}{1}\AtlasLong{n}{1}}}%
    \AtlasCell{3}{\AtlasRows{\AtlasLong{N-n}{1}\AtlasLong{n}{1}}}%
    \AtlasCell{4}{\AtlasRows{\AtlasLong{N-2}{1}\AtlasTail{\eta}}}%
  }{D_M^{\mathrm{tw}}}{%
    \AtlasCell{1}{\AtlasRows{\AtlasRow{2}{2d}\AtlasRow{1}{2M-4d}}}%
    \AtlasCell{2}{\AtlasRows{\AtlasLong{M-2}{2}\AtlasTail{\eta}}}%
    \AtlasCell{3}{\AtlasRows{\AtlasLong{M}{2}}}%
    \AtlasCell{4}{\AtlasRows{\AtlasLong{2M-2}{1}}}%
  }%
}{%
  \AtlasPair{A_5}{4}{%
    \AtlasKey{1}{[2,1^4]}%
    \AtlasKey{2}{[3^2]}%
    \AtlasKey{3}{[3^2]}%
    \AtlasKey{4}{[4,1^2]}%
  }{D_5^{\mathrm{tw}}}{4}{%
    \AtlasKey{1}{[2^2,1^6]}%
    \AtlasKey{2}{[3^2,1^2]_t}%
    \AtlasKey{3}{[5^2]}%
    \AtlasKey{4}{[8]_t}%
  }%
}{%
  \AtlasBoards{A_5}{%
    \AtlasCell{1}{\AtlasGrid{2,1,1,1,1}}%
    \AtlasCell{2}{\AtlasGrid{3,3}}%
    \AtlasCell{3}{\AtlasGrid{3,3}}%
    \AtlasCell{4}{\AtlasGrid{4,1,1}}%
  }{D_5^{\mathrm{tw}}}{%
    \AtlasCell{1}{\AtlasGrid{2,2,1,1,1,1,1,1}}%
    \AtlasCell{2}{\AtlasGrid{3,3,1,1}}%
    \AtlasCell{3}{\AtlasGrid{5,5}}%
    \AtlasCell{4}{\AtlasGrid{8}}%
  }%
}

\AtlasDefine{4}{%
  \AtlasPair{D_N}{3}{%
    \AtlasKey{1}{[2^{k_1},1^{2N-2k_1}]}%
    \AtlasKey{2}{[2^{k_2},1^{2N-2k_2}]}%
    \AtlasKey{3}{[2N-5]\cup\xi}%
  }{D_M^{\mathrm{tw}}}{3}{%
    \AtlasKey{1}{[M^2]}%
    \AtlasKey{2}{[2^{2(M-N)},1^{4N-2M-2}]_t}%
    \AtlasKey{3}{([(M-3)^2]\cup\zeta)_t}%
  }%
  \AtlasNote{$s=2N-1-M$ and
  $(k_1,k_2)=(2\lfloor s/2\rfloor,2\lceil s/2\rceil)$.}%
}{%
  \AtlasBoards{D_N}{%
    \AtlasCell{1}{\AtlasRows{\AtlasRow{2}{k_1}\AtlasRow{1}{2N-2k_1}}}%
    \AtlasCell{2}{\AtlasRows{\AtlasRow{2}{k_2}\AtlasRow{1}{2N-2k_2}}}%
    \AtlasCell{3}{\AtlasRows{\AtlasLong{2N-5}{1}\AtlasTail{\xi}}}%
  }{D_M^{\mathrm{tw}}}{%
    \AtlasCell{1}{\AtlasRows{\AtlasLong{M}{2}}}%
    \AtlasCell{2}{\AtlasRows{\AtlasRow{2}{2(M-N)}\AtlasRow{1}{4N-2M-2}}}%
    \AtlasCell{3}{\AtlasRows{\AtlasLong{M-3}{2}\AtlasTail{\zeta}}}%
  }%
}{%
  \AtlasPair{D_5}{3}{%
    \AtlasKey{1}{[1^{10}]}%
    \AtlasKey{2}{[2^2,1^6]}%
    \AtlasKey{3}{[5^2]}%
  }{D_8^{\mathrm{tw}}}{3}{%
    \AtlasKey{1}{[8^2]_{\mathrm{red}}}%
    \AtlasKey{2}{[2^6,1^2]_t}%
    \AtlasKey{3}{[5^2,4]_t}%
  }%
}{%
  \AtlasBoards{D_5}{%
    \AtlasCell{1}{\AtlasGrid{1,1,1,1,1,1,1,1,1,1}}%
    \AtlasCell{2}{\AtlasGrid{2,2,1,1,1,1,1,1}}%
    \AtlasCell{3}{\AtlasGrid{5,5}}%
  }{D_8^{\mathrm{tw}}}{%
    \AtlasCell{1}{\AtlasGrid{8,8}}%
    \AtlasCell{2}{\AtlasGrid{2,2,2,2,2,2,1,1}}%
    \AtlasCell{3}{\AtlasGrid{5,5,4}}%
  }%
}

\AtlasDefine{5}{%
  \AtlasPair{A_{m-1}}{3}{%
    \AtlasKey{1}{[2^{m-n},1^{2n-m}]}%
    \AtlasKey{2}{[2^{m-n},1^{2n-m}]}%
    \AtlasKey{3}{[m-4]\cup\rho}%
  }{A_{2n-1}^{\mathrm{tw}}}{3}{%
    \AtlasKey{1}{[2^{2n-m},1^{2m-2n}]}%
    \AtlasKey{2}{([2n-5]\cup\sigma)_t}%
    \AtlasKey{3}{[2n+1]_t}%
  }%
}{%
  \AtlasBoards{A_{m-1}}{%
    \AtlasCell{1}{\AtlasRows{\AtlasRow{2}{m-n}\AtlasRow{1}{2n-m}}}%
    \AtlasCell{2}{\AtlasRows{\AtlasRow{2}{m-n}\AtlasRow{1}{2n-m}}}%
    \AtlasCell{3}{\AtlasRows{\AtlasLong{m-4}{1}\AtlasTail{\rho}}}%
  }{A_{2n-1}^{\mathrm{tw}}}{%
    \AtlasCell{1}{\AtlasRows{\AtlasRow{2}{2n-m}\AtlasRow{1}{2m-2n}}}%
    \AtlasCell{2}{\AtlasRows{\AtlasLong{2n-5}{1}\AtlasTail{\sigma}}}%
    \AtlasCell{3}{\AtlasRows{\AtlasLong{2n+1}{1}}}%
  }%
}{%
  \AtlasPair{A_4}{3}{%
    \AtlasKey{1}{[2^2,1]}%
    \AtlasKey{2}{[2^2,1]}%
    \AtlasKey{3}{[1^5]}%
  }{A_5^{\mathrm{tw}}}{3}{%
    \AtlasKey{1}{[2,1^4]}%
    \AtlasKey{2}{[1^7]_t}%
    \AtlasKey{3}{[7]_t}%
  }%
}{%
  \AtlasBoards{A_4}{%
    \AtlasCell{1}{\AtlasGrid{2,2,1}}%
    \AtlasCell{2}{\AtlasGrid{2,2,1}}%
    \AtlasCell{3}{\AtlasGrid{1,1,1,1,1}}%
  }{A_5^{\mathrm{tw}}}{%
    \AtlasCell{1}{\AtlasGrid{2,1,1,1,1}}%
    \AtlasCell{2}{\AtlasGrid{1,1,1,1,1,1,1}}%
    \AtlasCell{3}{\AtlasGrid{7}}%
  }%
}

\AtlasDefine{6}{%
  \AtlasPair{G}{3}{%
    \AtlasKey{1}{P_1}\AtlasKey{2}{P_2}\AtlasKey{3}{P_3}%
  }{H}{3}{%
    \AtlasKey{1}{O_1}\AtlasKey{2}{O_2}\AtlasKey{3}{O_3}%
  }%
  \AtlasNote{Three puncture branch with $P_i\mapsto O_i$ and
  $\boldsymbol\delta_1+\boldsymbol\delta_2+\boldsymbol\delta_3=\mathbf b_G-\mathbf b_H$.}%
}{%
  \AtlasBoards{G}{%
    \AtlasCell{1}{\AtlasBadge{P_1}}%
    \AtlasCell{2}{\AtlasBadge{P_2}}%
    \AtlasCell{3}{\AtlasBadge{P_3}}%
  }{H}{%
    \AtlasCell{1}{\AtlasBadge{O_1}}%
    \AtlasCell{2}{\AtlasBadge{O_2}}%
    \AtlasCell{3}{\AtlasBadge{O_3}}%
  }%
  \AtlasNote{$\boldsymbol\delta_i$ is the change in puncture contributions to the protected data.}%
}{%
  \AtlasPair{E_6}{3}{%
    \AtlasKey{1}{A_2}\AtlasKey{2}{A_2}\AtlasKey{3}{E_6(a_3)}%
  }{E_7}{3}{%
    \AtlasKey{1}{D_4(a_1)}\AtlasKey{2}{A_4}\AtlasKey{3}{E_7(a_3)}%
  }%
  \AtlasNote{The puncture contributions compensate the change of parent.}%
}{%
  \AtlasBoards{E_6}{%
    \AtlasCell{1}{\AtlasOrbit{A_2}{E6}{000002}}%
    \AtlasCell{2}{\AtlasOrbit{A_2}{E6}{000002}}%
    \AtlasCell{3}{\AtlasOrbit{E_6(a_3)}{E6}{202020}}%
  }{E_7}{%
    \AtlasCell{1}{\AtlasOrbit{D_4(a_1)}{E7}{0200000}}%
    \AtlasCell{2}{\AtlasOrbit{A_4}{E7}{2000200}}%
    \AtlasCell{3}{\AtlasOrbit{E_7(a_3)}{E7}{2020220}}%
  }%
}

\AtlasDefine{7}{%
  \AtlasPair{D_N^{\mathrm{tw}}}{3}{%
    \AtlasKey{1}{Q}\AtlasKey{2}{B_e}\AtlasKey{3}{B_f}%
  }{E_6^{\mathrm{tw}}}{3}{%
    \AtlasKey{1}{D_5}\AtlasKey{2}{I}\AtlasKey{3}{X_{e+f}}%
  }%
  \AtlasNote{$b=N-5\in\{0,1\},\quad
  B_e=[3^{2b},2^{3-2b+e},1^{2-2e}]_t,\quad e,f\in\{0,1\}$.}%
  \AtlasNote{$(X_0,X_1,X_2)=((B_2)_t,(C_3(a_1))_t,(F_4(a_3))_t)$.}%
}{%
  \AtlasBoards{D_N^{\mathrm{tw}}}{%
    \AtlasCell{1}{\AtlasRows{\AtlasTail{Q}}}%
    \AtlasCell{2}{\AtlasRows{\AtlasRow{3}{2b}\AtlasRow{2}{3-2b+e}\AtlasRow{1}{2-2e}}}%
    \AtlasCell{3}{\AtlasRows{\AtlasRow{3}{2b}\AtlasRow{2}{3-2b+f}\AtlasRow{1}{2-2f}}}%
  }{E_6^{\mathrm{tw}}}{%
    \AtlasCell{1}{\AtlasOrbit{D_5}{E6}{202022}}%
    \AtlasCell{2}{\AtlasBadge{I}}%
    \AtlasCell{3}{\AtlasBadge{X_{e+f}}}%
  }%
}{%
  \AtlasPair{D_5^{\mathrm{tw}}}{3}{%
    \AtlasKey{1}{[5,3,1^2]}%
    \AtlasKey{2}{[2^3,1^2]_t}%
    \AtlasKey{3}{[2^4]_t}%
  }{E_6^{\mathrm{tw}}}{3}{%
    \AtlasKey{1}{D_5}%
    \AtlasKey{2}{(A_2)_t}%
    \AtlasKey{3}{(C_3(a_1))_t}%
  }%
}{%
  \AtlasBoards{D_5^{\mathrm{tw}}}{%
    \AtlasCell{1}{\AtlasGrid{5,3,1,1}}%
    \AtlasCell{2}{\AtlasGrid{2,2,2,1,1}}%
    \AtlasCell{3}{\AtlasGrid{2,2,2,2}}%
  }{E_6^{\mathrm{tw}}}{%
    \AtlasCell{1}{\AtlasOrbit{D_5}{E6}{202022}}%
    \AtlasCell{2}{\AtlasOrbit{(A_2)_t}{F4}{2000}}%
    \AtlasCell{3}{\AtlasOrbit{(C_3(a_1))_t}{F4}{1010}}%
  }%
}

\AtlasDefine{8}{%
  \AtlasNote{Published $G_2$ resolution.}%
  \AtlasPair{D_4^{S_3}}{4}{%
    \AtlasKey{1}{[1^6]_{t,\gamma}}%
    \AtlasKey{2}{[1^6]_{t,\delta}}%
    \AtlasKey{3}{(G_2)_{t,h}}%
    \AtlasKey{4}{[5,3]^{\times k}}%
  }{D_4^{S_3}}{5}{%
    \AtlasKey{1}{[1^6]_{t,\gamma}}%
    \AtlasKey{2}{[1^6]_{t,\delta}}%
    \AtlasKey{3}{[6]_{t,\alpha}}%
    \AtlasKey{4}{[6]_{t,\beta}}%
    \AtlasKey{5}{[5,3]^{\times k}}%
  }%
  \AtlasNote{$h=\alpha\beta,\quad\gamma\delta h=1$.
  The last marker groups $k$ untwisted punctures and is omitted for $k=0$.}%
}{%
  \AtlasBoards{D_4^{S_3}}{%
    \AtlasCell{1}{\AtlasGrid{1,1,1,1,1,1}}%
    \AtlasCell{2}{\AtlasGrid{1,1,1,1,1,1}}%
    \AtlasCell{3}{\AtlasOrbit{(G_2)_{t,h}}{G2}{22}}%
    \AtlasCell{4}{\AtlasGrid{5,3}}%
  }{D_4^{S_3}}{%
    \AtlasCell{1}{\AtlasGrid{1,1,1,1,1,1}}%
    \AtlasCell{2}{\AtlasGrid{1,1,1,1,1,1}}%
    \AtlasCell{3}{\AtlasGrid{6}}%
    \AtlasCell{4}{\AtlasGrid{6}}%
    \AtlasCell{5}{\AtlasGrid{5,3}}%
  }%
  \AtlasNote{The last shape is repeated $k$ times on each side.}%
}{%
  \AtlasNote{Checked $D_4/D_5$ comparison.}%
  \AtlasPair{D_4^{S_3}}{4}{%
    \AtlasKey{1}{[1^6]_{t,\alpha}}%
    \AtlasKey{2}{[1^6]_{t,\alpha}}%
    \AtlasKey{3}{[6]_{t,\beta}}%
    \AtlasKey{4}{[6]_{t,\beta}}%
  }{D_5}{4}{%
    \AtlasKey{1}{[3^2,1^4]}%
    \AtlasKey{2}{[3^3,1]}%
    \AtlasKey{3}{[5^2]}%
    \AtlasKey{4}{[7,3]}%
  }%
}{%
  \AtlasBoards{D_4^{S_3}}{%
    \AtlasCell{1}{\AtlasGrid{1,1,1,1,1,1}}%
    \AtlasCell{2}{\AtlasGrid{1,1,1,1,1,1}}%
    \AtlasCell{3}{\AtlasGrid{6}}%
    \AtlasCell{4}{\AtlasGrid{6}}%
  }{D_5}{%
    \AtlasCell{1}{\AtlasGrid{3,3,1,1,1,1}}%
    \AtlasCell{2}{\AtlasGrid{3,3,3,1}}%
    \AtlasCell{3}{\AtlasGrid{5,5}}%
    \AtlasCell{4}{\AtlasGrid{7,3}}%
  }%
}

%% file: figures/Family_Diagrams_Supp.tex
\AtlasDefine{9}{%
  \AtlasPair{D_N}{4}{%
    \AtlasKey{1}{U_{k_1}}\AtlasKey{2}{U_{k_2}}\AtlasKey{3}{S_e}\AtlasKey{4}{S_f}%
  }{D_M^{\mathrm{tw}}}{3}{%
    \AtlasKey{1}{[(M-2)^2]\cup\eta_{e+f}}%
    \AtlasKey{2}{[2^{2(M-N)},1^{4N-2M-2}]_t}%
    \AtlasKey{3}{[(M-1)^2]_t}%
  }%
  \AtlasNote{With $s=2N-1-M$, take $k_1=2\lfloor s/2\rfloor$ and $k_2=2\lceil s/2\rceil$. Here $U_k=[2^k,1^{2N-2k}]$, $S_a=[2N-3]\cup\sigma_a$, $(\sigma_0,\sigma_1)=([1^3],[3])$, and $(\eta_0,\eta_1,\eta_2)=([1^4],[2^2],[3,1])$.}%
}{%
  \AtlasBoards{D_N}{%
    \AtlasCell{1}{\AtlasRows{\AtlasRow{2}{k_1}\AtlasRow{1}{2N-2k_1}}}%
    \AtlasCell{2}{\AtlasRows{\AtlasRow{2}{k_2}\AtlasRow{1}{2N-2k_2}}}%
    \AtlasCell{3}{\AtlasRows{\AtlasLong{2N-3}{1}\AtlasTail{\sigma_e}}}%
    \AtlasCell{4}{\AtlasRows{\AtlasLong{2N-3}{1}\AtlasTail{\sigma_f}}}%
  }{D_M^{\mathrm{tw}}}{%
    \AtlasCell{1}{\AtlasRows{\AtlasLong{M-2}{2}\AtlasTail{\eta_{e+f}}}}%
    \AtlasCell{2}{\AtlasRows{\AtlasRow{2}{2(M-N)}\AtlasRow{1}{4N-2M-2}}}%
    \AtlasCell{3}{\AtlasRows{\AtlasLong{M-1}{2}}}%
  }%
}{%
  \AtlasPair{D_4}{4}{%
    \AtlasKey{1}{[2^2,1^4]}\AtlasKey{2}{[2^2,1^4]}\AtlasKey{3}{[5,3]}\AtlasKey{4}{[5,3]}%
  }{D_5^{\mathrm{tw}}}{3}{%
    \AtlasKey{1}{[3^3,1]}\AtlasKey{2}{[2^2,1^4]_t}\AtlasKey{3}{[4^2]_t}%
  }%
  \AtlasNote{$N=4,\ M=5,\ e=f=1$.}%
}{%
  \AtlasBoards{D_4}{%
    \AtlasCell{1}{\AtlasGrid{2,2,1,1,1,1}}\AtlasCell{2}{\AtlasGrid{2,2,1,1,1,1}}%
    \AtlasCell{3}{\AtlasGrid{5,3}}\AtlasCell{4}{\AtlasGrid{5,3}}%
  }{D_5^{\mathrm{tw}}}{%
    \AtlasCell{1}{\AtlasGrid{3,3,3,1}}\AtlasCell{2}{\AtlasGrid{2,2,1,1,1,1}}%
    \AtlasCell{3}{\AtlasGrid{4,4}}%
  }%
}

\AtlasDefine{10}{%
  \AtlasPair{A_{2n-1}^{\mathrm{tw}}}{4}{%
    \AtlasKey{1}{[n^2]}\AtlasKey{2}{[2n-2]\cup\rho_e}%
    \AtlasKey{3}{X_f}\AtlasKey{4}{[2n+1]_t}%
  }{D_M^{\mathrm{tw}}}{4}{%
    \AtlasKey{1}{Y_f}\AtlasKey{2}{[(M-1)^2,1^2]}%
    \AtlasKey{3}{([(M-2)^2]\cup\rho_e)_t}\AtlasKey{4}{[2M-2]_t}%
  }%
  \AtlasNote{$X_f=[3^{a-f},2^{2f},1^{2n+1-3a-f}]_t$, $Y_f=[3^{b-f},2^{2f},1^{2M-3b-f}]$, $a=n+2-M$, $b=M-n+1$, and $(\rho_0,\rho_1)=([1^2],[2])$.}%
}{%
  \AtlasBoards{A_{2n-1}^{\mathrm{tw}}}{%
    \AtlasCell{1}{\AtlasRows{\AtlasLong{n}{2}}}%
    \AtlasCell{2}{\AtlasRows{\AtlasLong{2n-2}{1}\AtlasTail{\rho_e}}}%
    \AtlasCell{3}{\AtlasRows{\AtlasRow{3}{a-f}\AtlasRow{2}{2f}\AtlasRow{1}{2n+1-3a-f}}}%
    \AtlasCell{4}{\AtlasRows{\AtlasLong{2n+1}{1}}}%
  }{D_M^{\mathrm{tw}}}{%
    \AtlasCell{1}{\AtlasRows{\AtlasRow{3}{b-f}\AtlasRow{2}{2f}\AtlasRow{1}{2M-3b-f}}}%
    \AtlasCell{2}{\AtlasRows{\AtlasLong{M-1}{2}\AtlasRow{1}{2}}}%
    \AtlasCell{3}{\AtlasRows{\AtlasLong{M-2}{2}\AtlasTail{\rho_e}}}%
    \AtlasCell{4}{\AtlasRows{\AtlasLong{2M-2}{1}}}%
  }%
}{%
  \AtlasPair{A_5^{\mathrm{tw}}}{4}{%
    \AtlasKey{1}{[3^2]}\AtlasKey{2}{[4,2]}\AtlasKey{3}{[3,1^4]_t}\AtlasKey{4}{[7]_t}%
  }{D_4^{\mathrm{tw}}}{4}{%
    \AtlasKey{1}{[3^2,1^2]}\AtlasKey{2}{[3^2,1^2]}\AtlasKey{3}{[2^3]_t}\AtlasKey{4}{[6]_t}%
  }%
  \AtlasNote{$n=3,\ M=4,\ e=1,\ f=0$.}%
}{%
  \AtlasBoards{A_5^{\mathrm{tw}}}{%
    \AtlasCell{1}{\AtlasGrid{3,3}}\AtlasCell{2}{\AtlasGrid{4,2}}%
    \AtlasCell{3}{\AtlasGrid{3,1,1,1,1}}\AtlasCell{4}{\AtlasGrid{7}}%
  }{D_4^{\mathrm{tw}}}{%
    \AtlasCell{1}{\AtlasGrid{3,3,1,1}}\AtlasCell{2}{\AtlasGrid{3,3,1,1}}%
    \AtlasCell{3}{\AtlasGrid{2,2,2}}\AtlasCell{4}{\AtlasGrid{6}}%
  }%
}

\AtlasDefine{11}{%
  \AtlasPair{A_{2n-1}^{\mathrm{tw}}}{3}{%
    \AtlasKey{1}{[2n-2]\cup\rho}\AtlasKey{2}{(B_i)_t}\AtlasKey{3}{(B_j)_t}%
  }{D_M^{\mathrm{tw}}}{3}{%
    \AtlasKey{1}{[(M-1)^2,1^2]}\AtlasKey{2}{[2^{M-1-K},1^{2K}]_t}%
    \AtlasKey{3}{([(M-2)^2]\cup\rho)_t}%
  }%
  \AtlasNote{$B_s=[2^{2s},1^{2n+1-4s}]$, $i=\lfloor K/2\rfloor$, $j=\lceil K/2\rceil$, $M=2n-K$, and $\rho=[1^2]$ or $[2]$.}%
}{%
  \AtlasBoards{A_{2n-1}^{\mathrm{tw}}}{%
    \AtlasCell{1}{\AtlasRows{\AtlasLong{2n-2}{1}\AtlasTail{\rho}}}%
    \AtlasCell{2}{\AtlasRows{\AtlasRow{2}{2i}\AtlasRow{1}{2n+1-4i}}}%
    \AtlasCell{3}{\AtlasRows{\AtlasRow{2}{2j}\AtlasRow{1}{2n+1-4j}}}%
  }{D_M^{\mathrm{tw}}}{%
    \AtlasCell{1}{\AtlasRows{\AtlasLong{M-1}{2}\AtlasRow{1}{2}}}%
    \AtlasCell{2}{\AtlasRows{\AtlasRow{2}{M-1-K}\AtlasRow{1}{2K}}}%
    \AtlasCell{3}{\AtlasRows{\AtlasLong{M-2}{2}\AtlasTail{\rho}}}%
  }%
}{%
  \AtlasPair{A_7^{\mathrm{tw}}}{3}{%
    \AtlasKey{1}{[6,1^2]}\AtlasKey{2}{[2^2,1^5]_t}\AtlasKey{3}{[2^4,1]_t}%
  }{D_5^{\mathrm{tw}}}{3}{%
    \AtlasKey{1}{[4^2,1^2]}\AtlasKey{2}{[2,1^6]_t}\AtlasKey{3}{[3^2,1^2]_t}%
  }%
  \AtlasNote{$n=4,\ K=3,\ M=5,\ \rho=[1^2]$.}%
}{%
  \AtlasBoards{A_7^{\mathrm{tw}}}{%
    \AtlasCell{1}{\AtlasGrid{6,1,1}}\AtlasCell{2}{\AtlasGrid{2,2,1,1,1,1,1}}%
    \AtlasCell{3}{\AtlasGrid{2,2,2,2,1}}%
  }{D_5^{\mathrm{tw}}}{%
    \AtlasCell{1}{\AtlasGrid{4,4,1,1}}\AtlasCell{2}{\AtlasGrid{2,1,1,1,1,1,1}}%
    \AtlasCell{3}{\AtlasGrid{3,3,1,1}}%
  }%
}

\AtlasDefine{12}{%
  \AtlasPair{A_{N-1}}{3}{%
    \AtlasKey{1}{P}\AtlasKey{2}{P}\AtlasKey{3}{X_e}%
  }{D_N^{\mathrm{tw}}}{3}{%
    \AtlasKey{1}{[N^2]}\AtlasKey{2}{T}\AtlasKey{3}{Y_e}%
  }%
  \AtlasNote{Untwisted $A$ branch. $N=2n+\epsilon$ with $\epsilon\in\{0,1\}$. Here $P=[2^{n-1+\epsilon},1^{2-\epsilon}]$, $X_e=[3^\epsilon,2^{n-1-\epsilon+e},1^{2-2e}]$, $T=[3^{2+2\epsilon},2^{N-4-3\epsilon}]_t$, and $Y_e=[3^{2\epsilon},2^{N-2-3\epsilon+e},1^{2-2e}]_t$.}%
}{%
  \AtlasBoards{A_{N-1}}{%
    \AtlasCell{1}{\AtlasRows{\AtlasRow{2}{n-1+\epsilon}\AtlasRow{1}{2-\epsilon}}}%
    \AtlasCell{2}{\AtlasRows{\AtlasRow{2}{n-1+\epsilon}\AtlasRow{1}{2-\epsilon}}}%
    \AtlasCell{3}{\AtlasRows{\AtlasRow{3}{\epsilon}\AtlasRow{2}{n-1-\epsilon+e}\AtlasRow{1}{2-2e}}}%
  }{D_N^{\mathrm{tw}}}{%
    \AtlasCell{1}{\AtlasRows{\AtlasLong{N}{2}}}%
    \AtlasCell{2}{\AtlasRows{\AtlasRow{3}{2+2\epsilon}\AtlasRow{2}{N-4-3\epsilon}}}%
    \AtlasCell{3}{\AtlasRows{\AtlasRow{3}{2\epsilon}\AtlasRow{2}{N-2-3\epsilon+e}\AtlasRow{1}{2-2e}}}%
  }%
}{%
  \AtlasPair{A_5}{3}{%
    \AtlasKey{1}{[2^2,1^2]}\AtlasKey{2}{[2^2,1^2]}\AtlasKey{3}{[2^3]}%
  }{D_6^{\mathrm{tw}}}{3}{%
    \AtlasKey{1}{[6^2]_{\mathrm{blue}}}\AtlasKey{2}{[3^2,2^2]_t}\AtlasKey{3}{[2^5]_t}%
  }%
  \AtlasNote{$N=6,\ e=1$. The same $D_6$ sphere also matches $A_7^{\mathrm{tw}}\{[2^2,1^4],[3^2,1^3]_t,[9]_t\}$.}%
}{%
  \AtlasBoards{A_5}{%
    \AtlasCell{1}{\AtlasGrid{2,2,1,1}}\AtlasCell{2}{\AtlasGrid{2,2,1,1}}%
    \AtlasCell{3}{\AtlasGrid{2,2,2}}%
  }{D_6^{\mathrm{tw}}}{%
    \AtlasCell{1}{\AtlasGrid{6,6}}\AtlasCell{2}{\AtlasGrid{3,3,2,2}}%
    \AtlasCell{3}{\AtlasGrid{2,2,2,2,2}}%
  }%
}

\AtlasDefine{13}{%
  \AtlasPair{D_4}{3}{%
    \AtlasKey{1}{[1^8]}\AtlasKey{2}{[2^2,1^4]}\AtlasKey{3}{[2^2,1^4]}%
  }{D_6}{3}{%
    \AtlasKey{1}{[3^4]}\AtlasKey{2}{[4^2,1^4]}\AtlasKey{3}{[4^2,1^4]}%
  }%
  \AtlasNote{Untwisted trinion among the finite set of examples.}%
}{%
  \AtlasBoards{D_4}{%
    \AtlasCell{1}{\AtlasRows{\AtlasRow{1}{8}}}%
    \AtlasCell{2}{\AtlasRows{\AtlasRow{2}{2}\AtlasRow{1}{4}}}%
    \AtlasCell{3}{\AtlasRows{\AtlasRow{2}{2}\AtlasRow{1}{4}}}%
  }{D_6}{%
    \AtlasCell{1}{\AtlasRows{\AtlasRow{3}{4}}}%
    \AtlasCell{2}{\AtlasRows{\AtlasRow{4}{2}\AtlasRow{1}{4}}}%
    \AtlasCell{3}{\AtlasRows{\AtlasRow{4}{2}\AtlasRow{1}{4}}}%
  }%
}{%
  \AtlasPair{D_4}{3}{%
    \AtlasKey{1}{[1^8]}\AtlasKey{2}{[2^2,1^4]}\AtlasKey{3}{[2^2,1^4]}%
  }{D_6}{3}{%
    \AtlasKey{1}{[3^4]}\AtlasKey{2}{[4^2,1^4]}\AtlasKey{3}{[4^2,1^4]}%
  }%
  \AtlasNote{Checked $D_4/D_6$ example.}%
}{%
  \AtlasBoards{D_4}{%
    \AtlasCell{1}{\AtlasGrid{1,1,1,1,1,1,1,1}}%
    \AtlasCell{2}{\AtlasGrid{2,2,1,1,1,1}}\AtlasCell{3}{\AtlasGrid{2,2,1,1,1,1}}%
  }{D_6}{%
    \AtlasCell{1}{\AtlasGrid{3,3,3,3}}%
    \AtlasCell{2}{\AtlasGrid{4,4,1,1,1,1}}\AtlasCell{3}{\AtlasGrid{4,4,1,1,1,1}}%
  }%
}

\AtlasDefine{14}{%
  \AtlasPair{A_{N-1}}{4}{%
    \AtlasKey{1}{[1^N]}\AtlasKey{2}{[1^N]}\AtlasKey{3}{[N-1,1]}\AtlasKey{4}{[N-1,1]}%
  }{D_N^{\mathrm{tw}}}{3}{%
    \AtlasKey{1}{[(N-1)^2,1^2]}\AtlasKey{2}{[1^{2N-2}]_t}\AtlasKey{3}{[(N-1)^2]_t}%
  }%
  \AtlasNote{SQCD branch.}%
}{%
  \AtlasBoards{A_{N-1}}{%
    \AtlasCell{1}{\AtlasRows{\AtlasRow{1}{N}}}%
    \AtlasCell{2}{\AtlasRows{\AtlasRow{1}{N}}}%
    \AtlasCell{3}{\AtlasRows{\AtlasLong{N-1}{1}\AtlasRow{1}{1}}}%
    \AtlasCell{4}{\AtlasRows{\AtlasLong{N-1}{1}\AtlasRow{1}{1}}}%
  }{D_N^{\mathrm{tw}}}{%
    \AtlasCell{1}{\AtlasRows{\AtlasLong{N-1}{2}\AtlasRow{1}{2}}}%
    \AtlasCell{2}{\AtlasRows{\AtlasRow{1}{2N-2}}}%
    \AtlasCell{3}{\AtlasRows{\AtlasLong{N-1}{2}}}%
  }%
}{%
  \AtlasPair{A_3}{4}{%
    \AtlasKey{1}{[1^4]}\AtlasKey{2}{[1^4]}\AtlasKey{3}{[3,1]}\AtlasKey{4}{[3,1]}%
  }{D_4^{\mathrm{tw}}}{3}{%
    \AtlasKey{1}{[3^2,1^2]}\AtlasKey{2}{[1^6]_t}\AtlasKey{3}{[3^2]_t}%
  }%
  \AtlasNote{$N=4$.}%
}{%
  \AtlasBoards{A_3}{%
    \AtlasCell{1}{\AtlasGrid{1,1,1,1}}\AtlasCell{2}{\AtlasGrid{1,1,1,1}}%
    \AtlasCell{3}{\AtlasGrid{3,1}}\AtlasCell{4}{\AtlasGrid{3,1}}%
  }{D_4^{\mathrm{tw}}}{%
    \AtlasCell{1}{\AtlasGrid{3,3,1,1}}\AtlasCell{2}{\AtlasGrid{1,1,1,1,1,1}}%
    \AtlasCell{3}{\AtlasGrid{3,3}}%
  }%
}

\AtlasDefine{15}{%
  \AtlasPair{E_8}{3}{%
    \AtlasKey{1}{E_7}\AtlasKey{2}{A_5}\AtlasKey{3}{E_8(a_7)}%
  }{E_8}{3}{%
    \AtlasKey{1}{E_7}\AtlasKey{2}{E_6(a_3)}\AtlasKey{3}{E_7(a_5)}%
  }%
  \AtlasNote{Fixed $E_7$ puncture in the $E_8$ exchange.}%
}{%
  \AtlasBoards{E_8}{%
    \AtlasCell{1}{\AtlasBadge{E_7}}\AtlasCell{2}{\AtlasBadge{A_5}}\AtlasCell{3}{\AtlasBadge{E_8(a_7)}}%
  }{E_8}{%
    \AtlasCell{1}{\AtlasBadge{E_7}}\AtlasCell{2}{\AtlasBadge{E_6(a_3)}}\AtlasCell{3}{\AtlasBadge{E_7(a_5)}}%
  }%
}{%
  \AtlasPair{E_8}{3}{%
    \AtlasKey{1}{E_7}\AtlasKey{2}{A_5}\AtlasKey{3}{E_8(a_7)}%
  }{E_8}{3}{%
    \AtlasKey{1}{E_7}\AtlasKey{2}{E_6(a_3)}\AtlasKey{3}{E_7(a_5)}%
  }%
  \AtlasNote{Checked finite exchange.}%
}{%
  \AtlasBoards{E_8}{%
    \AtlasCell{1}{\AtlasOrbit{E_7}{E8}{21012221}}%
    \AtlasCell{2}{\AtlasOrbit{A_5}{E8}{20001010}}%
    \AtlasCell{3}{\AtlasOrbit{E_8(a_7)}{E8}{00020000}}%
  }{E_8}{%
    \AtlasCell{1}{\AtlasOrbit{E_7}{E8}{21012221}}%
    \AtlasCell{2}{\AtlasOrbit{E_6(a_3)}{E8}{20000200}}%
    \AtlasCell{3}{\AtlasOrbit{E_7(a_5)}{E8}{00101000}}%
  }%
}

\AtlasDefine{16}{%
  \AtlasPair{D_6}{3}{%
    \AtlasKey{1}{J_e}\AtlasKey{2}{K_f}\AtlasKey{3}{V_i}%
  }{E_7}{3}{%
    \AtlasKey{1}{C_{e+f}}\AtlasKey{2}{A_6}\AtlasKey{3}{W_i}%
  }%
  \AtlasNote{$D_6$ branch. $J=([2^6]_{\mathrm{blue}},[2^4,1^4])$, $K=([5,3^2,1],[4^2,3,1])$, $V=([4^2,3,1],[4^2,2^2]_{\mathrm{blue}})$, $C=(A_3+A_2+A_1,A_3+A_2,D_4(a_1)+A_1)$, and $W=(D_5+A_1,D_5)$.}%
}{%
  \AtlasBoards{D_6}{%
    \AtlasCell{1}{\AtlasRows{\AtlasRow{2}{6-2e}\AtlasRow{1}{4e}}}%
    \AtlasCell{2}{\AtlasRows{\AtlasLong{5-f}{1}\AtlasLong{3+f}{1}\AtlasRow{3}{1}\AtlasRow{1}{1}}}%
    \AtlasCell{3}{\AtlasRows{\AtlasRow{4}{2}\AtlasLong{3-i}{1}\AtlasLong{1+i}{1}}}%
  }{E_7}{%
    \AtlasCell{1}{\AtlasBadge{C_{e+f}}}\AtlasCell{2}{\AtlasBadge{A_6}}\AtlasCell{3}{\AtlasBadge{W_i}}%
  }%
}{%
  \AtlasPair{D_6}{3}{%
    \AtlasKey{1}{[2^6]_{\mathrm{blue}}}\AtlasKey{2}{[5,3^2,1]}\AtlasKey{3}{[4^2,3,1]}%
  }{E_7}{3}{%
    \AtlasKey{1}{A_3+A_2+A_1}\AtlasKey{2}{A_6}\AtlasKey{3}{D_5+A_1}%
  }%
  \AtlasNote{$e=f=i=0$.}%
}{%
  \AtlasBoards{D_6}{%
    \AtlasCell{1}{\AtlasGrid{2,2,2,2,2,2}}\AtlasCell{2}{\AtlasGrid{5,3,3,1}}%
    \AtlasCell{3}{\AtlasGrid{4,4,3,1}}%
  }{E_7}{%
    \AtlasCell{1}{\AtlasOrbit{A_3+A_2+A_1}{E7}{0002000}}%
    \AtlasCell{2}{\AtlasOrbit{A_6}{E7}{0020200}}%
    \AtlasCell{3}{\AtlasOrbit{D_5+A_1}{E7}{2101101}}%
  }%
}

%% file: figures/Family_Diagrams_Additional.tex
\newcommand{\AtlasTorus}[1]{%
  \begingroup\setlength{\unitlength}{1pt}\linethickness{.50pt}%
  \begin{picture}(38,38)
    \color{atlasink}\put(19,19){\oval(36,25)}%
    \qbezier(8,21)(19,10)(30,21)%
    \qbezier(8,21)(19,28)(30,21)%
    \ifnum#1=2\relax
      \AtlasDot{10}{27}{1}\AtlasDot{28}{26}{2}%
    \fi
    \ifnum#1=4
      \AtlasDot{10}{27}{1}\AtlasDot{28}{26}{2}%
      \AtlasDot{30}{12}{3}\AtlasDot{9}{12}{4}%
    \fi
  \end{picture}\endgroup
}
\newcommand{\AtlasTorusSide}[3]{%
  \begin{minipage}[t]{.46\linewidth}\centering$#1$\par
  \AtlasTorus{#2}\par
  \begin{tabular}{@{}r@{\hspace{4pt}}l@{}}#3\end{tabular}
  \end{minipage}%
}
\newcommand{\AtlasTorusPair}[6]{%
  \par\noindent\begingroup\footnotesize
  \AtlasTorusSide{#1}{#2}{#3}%
  \hfill\raisebox{-22pt}[0pt][0pt]{$\leftrightarrow$}\hfill
  \AtlasSide{#4}{#5}{#6}\par\endgroup
}

\AtlasDefine{17}{%
\AtlasTorusPair{A_{n-1}^{g=1}}{4}{%
\AtlasKey{1}{M}
\AtlasKey{2}{M}
\AtlasKey{3}{M}
\AtlasKey{4}{M}%
}{A_{2n-1}^{\mathrm{tw}}}{4}{%
\AtlasKey{1}{L}
\AtlasKey{2}{L}
\AtlasKey{3}{L}
\AtlasKey{4}{Z}%
}
\AtlasNote{$M=[n-1,1]$, $L=[2n-1,1^2]_t$, and $Z=[2n+1]_t$.}
}{%
\AtlasBoards{A_{n-1}^{g=1}}{%
\AtlasCell{1}{\AtlasRows{\AtlasLong{n-1}{1}\AtlasRow{1}{1}}}
\AtlasCell{2}{\AtlasRows{\AtlasLong{n-1}{1}\AtlasRow{1}{1}}}
\AtlasCell{3}{\AtlasRows{\AtlasLong{n-1}{1}\AtlasRow{1}{1}}}
\AtlasCell{4}{\AtlasRows{\AtlasLong{n-1}{1}\AtlasRow{1}{1}}}%
}{A_{2n-1}^{\mathrm{tw}}}{%
\AtlasCell{1}{\AtlasRows{\AtlasLong{2n-1}{1}\AtlasRow{1}{2}}}
\AtlasCell{2}{\AtlasRows{\AtlasLong{2n-1}{1}\AtlasRow{1}{2}}}
\AtlasCell{3}{\AtlasRows{\AtlasLong{2n-1}{1}\AtlasRow{1}{2}}}
\AtlasCell{4}{\AtlasRows{\AtlasLong{2n+1}{1}}}%
}
}{%
\AtlasTorusPair{A_{1} {}^{g=1}}{4}{%
\AtlasKey{1}{[1^{2}]}
\AtlasKey{2}{[1^{2}]}
\AtlasKey{3}{[1^{2}]}
\AtlasKey{4}{[1^{2}]}%
}{A_{3}^{\mathrm{tw}}}{4}{%
\AtlasKey{1}{\bigl([3,1^{2}]\bigr)_{t}}
\AtlasKey{2}{\bigl([3,1^{2}]\bigr)_{t}}
\AtlasKey{3}{\bigl([3,1^{2}]\bigr)_{t}}
\AtlasKey{4}{\bigl([5]\bigr)_{t}}%
}
}{%
\AtlasBoards{A_{1} {}^{g=1}}{%
\AtlasCell{1}{\AtlasGrid{1,1}}
\AtlasCell{2}{\AtlasGrid{1,1}}
\AtlasCell{3}{\AtlasGrid{1,1}}
\AtlasCell{4}{\AtlasGrid{1,1}}%
}{A_{3}^{\mathrm{tw}}}{%
\AtlasCell{1}{\AtlasGrid{3,1,1}}
\AtlasCell{2}{\AtlasGrid{3,1,1}}
\AtlasCell{3}{\AtlasGrid{3,1,1}}
\AtlasCell{4}{\AtlasGrid{5}}%
}
}

\AtlasDefine{18}{%
\AtlasPair{D_N^{\mathrm{tw}}}{4}{%
\AtlasKey{1}{Q_{k_1}}
\AtlasKey{2}{Q_{k_2}}
\AtlasKey{3}{A_e}
\AtlasKey{4}{Z}%
}{D_N^{\mathrm{tw}}}{4}{%
\AtlasKey{1}{R}
\AtlasKey{2}{R}
\AtlasKey{3}{F}
\AtlasKey{4}{A_e}%
}
\AtlasNote{$Q_k=[2^k,1^{2N-2k}]$, $R=[N^2]$, $F=[1^{2N-2}]_t$, $Z=[2N-2]_t$, and $A_e=([2N-4]\cup\rho_e)_t$.}
}{%
\AtlasBoards{D_N^{\mathrm{tw}}}{%
\AtlasCell{1}{\AtlasRows{\AtlasRow{2}{k_1}\AtlasRow{1}{2N-2k_1}}}
\AtlasCell{2}{\AtlasRows{\AtlasRow{2}{k_2}\AtlasRow{1}{2N-2k_2}}}
\AtlasCell{3}{\AtlasRows{\AtlasLong{2N-4}{1}\AtlasTail{\rho_e}}}
\AtlasCell{4}{\AtlasRows{\AtlasLong{2N-2}{1}}}%
}{D_N^{\mathrm{tw}}}{%
\AtlasCell{1}{\AtlasRows{\AtlasLong{N}{2}}}
\AtlasCell{2}{\AtlasRows{\AtlasLong{N}{2}}}
\AtlasCell{3}{\AtlasRows{\AtlasRow{1}{2N-2}}}
\AtlasCell{4}{\AtlasRows{\AtlasLong{2N-4}{1}\AtlasTail{\rho_e}}}%
}
}{%
\AtlasPair{D_{4}^{\mathrm{tw}}}{4}{%
\AtlasKey{1}{[2^{4}]_{\mathrm{blue}}}
\AtlasKey{2}{[2^{4}]_{\mathrm{blue}}}
\AtlasKey{3}{\bigl([4,1^{2}]\bigr)_{t}}
\AtlasKey{4}{\bigl([6]\bigr)_{t}}%
}{D_{4}^{\mathrm{tw}}}{4}{%
\AtlasKey{1}{[4^{2}]_{\mathrm{blue}}}
\AtlasKey{2}{[5,3]}
\AtlasKey{3}{\bigl([1^{6}]\bigr)_{t}}
\AtlasKey{4}{\bigl([4,1^{2}]\bigr)_{t}}%
}
}{%
\AtlasBoards{D_{4}^{\mathrm{tw}}}{%
\AtlasCell{1}{\AtlasGrid{2,2,2,2}}
\AtlasCell{2}{\AtlasGrid{2,2,2,2}}
\AtlasCell{3}{\AtlasGrid{4,1,1}}
\AtlasCell{4}{\AtlasGrid{6}}%
}{D_{4}^{\mathrm{tw}}}{%
\AtlasCell{1}{\AtlasGrid{4,4}}
\AtlasCell{2}{\AtlasGrid{5,3}}
\AtlasCell{3}{\AtlasGrid{1,1,1,1,1,1}}
\AtlasCell{4}{\AtlasGrid{4,1,1}}%
}
}

\AtlasDefine{19}{%
\AtlasPair{A_{2n-1}^{\mathrm{tw}}}{3}{%
\AtlasKey{1}{H_e}
\AtlasKey{2}{H_f}
\AtlasKey{3}{Z}%
}{A_{2n-1}^{\mathrm{tw}}}{3}{%
\AtlasKey{1}{M}
\AtlasKey{2}{M}
\AtlasKey{3}{X_{e+f}}%
}
\AtlasNote{$\rho_0=[1^2]$, $\rho_1=[2]$, and $(\eta_0,\eta_1,\eta_2)=([1^4],[2^2],[3,1])$. The worked example uses the branch that changes the twist sector.}
}{%
\AtlasBoards{A_{2n-1}^{\mathrm{tw}}}{%
\AtlasCell{1}{\AtlasRows{\AtlasLong{2n-2}{1}\AtlasTail{\rho_e}}}
\AtlasCell{2}{\AtlasRows{\AtlasLong{2n-2}{1}\AtlasTail{\rho_f}}}
\AtlasCell{3}{\AtlasRows{\AtlasLong{2n+1}{1}}}%
}{A_{2n-1}^{\mathrm{tw}}}{%
\AtlasCell{1}{\AtlasRows{\AtlasLong{2n-1}{1}\AtlasRow{1}{1}}}
\AtlasCell{2}{\AtlasRows{\AtlasLong{2n-1}{1}\AtlasRow{1}{1}}}
\AtlasCell{3}{\AtlasRows{\AtlasLong{2n-3}{1}\AtlasTail{\eta_{e+f}}}}%
}
}{%
\AtlasPair{A_{3}}{4}{%
\AtlasKey{1}{[1^{4}]}
\AtlasKey{2}{[2^{2}]}
\AtlasKey{3}{[2^{2}]}
\AtlasKey{4}{[3,1]}%
}{A_{3}^{\mathrm{tw}}}{4}{%
\AtlasKey{1}{[1^{4}]}
\AtlasKey{2}{[1^{4}]}
\AtlasKey{3}{\bigl([5]\bigr)_{t}}
\AtlasKey{4}{\bigl([5]\bigr)_{t}}%
}
}{%
\AtlasBoards{A_{3}}{%
\AtlasCell{1}{\AtlasGrid{1,1,1,1}}
\AtlasCell{2}{\AtlasGrid{2,2}}
\AtlasCell{3}{\AtlasGrid{2,2}}
\AtlasCell{4}{\AtlasGrid{3,1}}%
}{A_{3}^{\mathrm{tw}}}{%
\AtlasCell{1}{\AtlasGrid{1,1,1,1}}
\AtlasCell{2}{\AtlasGrid{1,1,1,1}}
\AtlasCell{3}{\AtlasGrid{5}}
\AtlasCell{4}{\AtlasGrid{5}}%
}
}

\AtlasDefine{20}{%
\AtlasPair{A_3^{\mathrm{tw}}}{3}{%
\AtlasKey{1}{P_e}
\AtlasKey{2}{\bigl([1^{5}]\bigr)_{t}}
\AtlasKey{3}{\bigl([1^{5}]\bigr)_{t}}%
}{A_5^{\mathrm{tw}}}{3}{%
\AtlasKey{1}{[3^{2}]}
\AtlasKey{2}{X_e}
\AtlasKey{3}{\bigl([3,1^{4}]\bigr)_{t}}%
}
\AtlasNote{$(P_0,P_1)=([2,1^2],[2^2])$ and $(X_0,X_1)=([3,2^2]_t,[3^2,1]_t)$.}
}{%
\AtlasBoards{A_3^{\mathrm{tw}}}{%
\AtlasCell{1}{\AtlasBadge{P_e}}
\AtlasCell{2}{\AtlasGrid{1,1,1,1,1}}
\AtlasCell{3}{\AtlasGrid{1,1,1,1,1}}%
}{A_5^{\mathrm{tw}}}{%
\AtlasCell{1}{\AtlasGrid{3,3}}
\AtlasCell{2}{\AtlasBadge{X_e}}
\AtlasCell{3}{\AtlasGrid{3,1,1,1,1}}%
}
}{%
\AtlasPair{A_{3}^{\mathrm{tw}}}{3}{%
\AtlasKey{1}{[2^{2}]}
\AtlasKey{2}{\bigl([1^{5}]\bigr)_{t}}
\AtlasKey{3}{\bigl([1^{5}]\bigr)_{t}}%
}{A_{5}^{\mathrm{tw}}}{3}{%
\AtlasKey{1}{[3^{2}]}
\AtlasKey{2}{\bigl([3,1^{4}]\bigr)_{t}}
\AtlasKey{3}{\bigl([3^{2},1]\bigr)_{t}}%
}
}{%
\AtlasBoards{A_{3}^{\mathrm{tw}}}{%
\AtlasCell{1}{\AtlasGrid{2,2}}
\AtlasCell{2}{\AtlasGrid{1,1,1,1,1}}
\AtlasCell{3}{\AtlasGrid{1,1,1,1,1}}%
}{A_{5}^{\mathrm{tw}}}{%
\AtlasCell{1}{\AtlasGrid{3,3}}
\AtlasCell{2}{\AtlasGrid{3,1,1,1,1}}
\AtlasCell{3}{\AtlasGrid{3,3,1}}%
}
}

\AtlasDefine{21}{%
\AtlasPair{A_{2n-2}}{4}{%
\AtlasKey{1}{P}
\AtlasKey{2}{R}
\AtlasKey{3}{R}
\AtlasKey{4}{M}%
}{A_{2n-1}^{\mathrm{tw}}}{3}{%
\AtlasKey{1}{U}
\AtlasKey{2}{F}
\AtlasKey{3}{L}%
}
\AtlasNote{The branch with nearly rectangular partitions has $P=[2^{n-1},1]$, $R=[(n-1)^2,1]$, $M=[2n-2,1]$, $U=[n,n-1,1]$, $F=[1^{2n+1}]_t$, and $L=[2n-1,1^2]_t$.}
}{%
\AtlasBoards{A_{2n-2}}{%
\AtlasCell{1}{\AtlasRows{\AtlasRow{2}{n-1}\AtlasRow{1}{1}}}
\AtlasCell{2}{\AtlasRows{\AtlasLong{n-1}{2}\AtlasRow{1}{1}}}
\AtlasCell{3}{\AtlasRows{\AtlasLong{n-1}{2}\AtlasRow{1}{1}}}
\AtlasCell{4}{\AtlasRows{\AtlasLong{2n-2}{1}\AtlasRow{1}{1}}}%
}{A_{2n-1}^{\mathrm{tw}}}{%
\AtlasCell{1}{\AtlasRows{\AtlasLong{n}{1}\AtlasLong{n-1}{1}\AtlasRow{1}{1}}}
\AtlasCell{2}{\AtlasRows{\AtlasRow{1}{2n+1}}}
\AtlasCell{3}{\AtlasRows{\AtlasLong{2n-1}{1}\AtlasRow{1}{2}}}%
}
}{%
\AtlasPair{A_{2}}{4}{%
\AtlasKey{1}{[2,1]}
\AtlasKey{2}{[1^{3}]}
\AtlasKey{3}{[1^{3}]}
\AtlasKey{4}{[2,1]}%
}{A_{3}^{\mathrm{tw}}}{3}{%
\AtlasKey{1}{[2,1^{2}]}
\AtlasKey{2}{\bigl([1^{5}]\bigr)_{t}}
\AtlasKey{3}{\bigl([3,1^{2}]\bigr)_{t}}%
}
}{%
\AtlasBoards{A_{2}}{%
\AtlasCell{1}{\AtlasGrid{2,1}}
\AtlasCell{2}{\AtlasGrid{1,1,1}}
\AtlasCell{3}{\AtlasGrid{1,1,1}}
\AtlasCell{4}{\AtlasGrid{2,1}}%
}{A_{3}^{\mathrm{tw}}}{%
\AtlasCell{1}{\AtlasGrid{2,1,1}}
\AtlasCell{2}{\AtlasGrid{1,1,1,1,1}}
\AtlasCell{3}{\AtlasGrid{3,1,1}}%
}
}

\AtlasDefine{22}{%
\AtlasPair{D_4}{3}{%
\AtlasKey{1}{[2^{2},1^{4}]}
\AtlasKey{2}{[2^{2},1^{4}]}
\AtlasKey{3}{B_i}%
}{D_5^{\mathrm{tw}}}{3}{%
\AtlasKey{1}{[5^{2}]}
\AtlasKey{2}{\bigl([2^{2},1^{4}]\bigr)_{t}}
\AtlasKey{3}{W_i}%
}
\AtlasNote{$B_i=([2^4]_{\rm blue},[3,2^2,1],[3^2,1^2])$ and $W_i=[2^{i+2},1^{4-2i}]_t$, with $i=0,1,2$.}
}{%
\AtlasBoards{D_4}{%
\AtlasCell{1}{\AtlasGrid{2,2,1,1,1,1}}
\AtlasCell{2}{\AtlasGrid{2,2,1,1,1,1}}
\AtlasCell{3}{\AtlasBadge{B_i}}%
}{D_5^{\mathrm{tw}}}{%
\AtlasCell{1}{\AtlasGrid{5,5}}
\AtlasCell{2}{\AtlasGrid{2,2,1,1,1,1}}
\AtlasCell{3}{\AtlasBadge{W_i}}%
}
}{%
\AtlasPair{D_{4}}{3}{%
\AtlasKey{1}{[2^{2},1^{4}]}
\AtlasKey{2}{[2^{2},1^{4}]}
\AtlasKey{3}{[3,2^{2},1]}%
}{D_{5}^{\mathrm{tw}}}{3}{%
\AtlasKey{1}{[5^{2}]}
\AtlasKey{2}{\bigl([2^{2},1^{4}]\bigr)_{t}}
\AtlasKey{3}{\bigl([2^{3},1^{2}]\bigr)_{t}}%
}
}{%
\AtlasBoards{D_{4}}{%
\AtlasCell{1}{\AtlasGrid{2,2,1,1,1,1}}
\AtlasCell{2}{\AtlasGrid{2,2,1,1,1,1}}
\AtlasCell{3}{\AtlasGrid{3,2,2,1}}%
}{D_{5}^{\mathrm{tw}}}{%
\AtlasCell{1}{\AtlasGrid{5,5}}
\AtlasCell{2}{\AtlasGrid{2,2,1,1,1,1}}
\AtlasCell{3}{\AtlasGrid{2,2,2,1,1}}%
}
}

\AtlasDefine{23}{%
\AtlasPair{D_5}{3}{%
\AtlasKey{1}{[2^{4},1^{2}]}
\AtlasKey{2}{[2^{4},1^{2}]}
\AtlasKey{3}{B_e}%
}{D_6}{3}{%
\AtlasKey{1}{[3^{3},1^{3}]}
\AtlasKey{2}{Q_e}
\AtlasKey{3}{[5^{2},1^{2}]}%
}
\AtlasNote{$(B_0,B_1)=([5,3,1^2],[4^2,1^2])$ and $(Q_0,Q_1)=([3^3,1^3],[3^2,2^2,1^2])$.}
}{%
\AtlasBoards{D_5}{%
\AtlasCell{1}{\AtlasGrid{2,2,2,2,1,1}}
\AtlasCell{2}{\AtlasGrid{2,2,2,2,1,1}}
\AtlasCell{3}{\AtlasBadge{B_e}}%
}{D_6}{%
\AtlasCell{1}{\AtlasGrid{3,3,3,1,1,1}}
\AtlasCell{2}{\AtlasBadge{Q_e}}
\AtlasCell{3}{\AtlasGrid{5,5,1,1}}%
}
}{%
\AtlasPair{D_{5}}{3}{%
\AtlasKey{1}{[2^{4},1^{2}]}
\AtlasKey{2}{[2^{4},1^{2}]}
\AtlasKey{3}{[5,3,1^{2}]}%
}{D_{6}}{3}{%
\AtlasKey{1}{[3^{3},1^{3}]}
\AtlasKey{2}{[3^{3},1^{3}]}
\AtlasKey{3}{[5^{2},1^{2}]}%
}
}{%
\AtlasBoards{D_{5}}{%
\AtlasCell{1}{\AtlasGrid{2,2,2,2,1,1}}
\AtlasCell{2}{\AtlasGrid{2,2,2,2,1,1}}
\AtlasCell{3}{\AtlasGrid{5,3,1,1}}%
}{D_{6}}{%
\AtlasCell{1}{\AtlasGrid{3,3,3,1,1,1}}
\AtlasCell{2}{\AtlasGrid{3,3,3,1,1,1}}
\AtlasCell{3}{\AtlasGrid{5,5,1,1}}%
}
}

\AtlasDefine{24}{%
\AtlasPair{A_{2n-1}}{4}{%
\AtlasKey{1}{P}
\AtlasKey{2}{P}
\AtlasKey{3}{P}
\AtlasKey{4}{P}%
}{D_{2n}}{4}{%
\AtlasKey{1}{U}
\AtlasKey{2}{U}
\AtlasKey{3}{U}
\AtlasKey{4}{V}%
}
\AtlasNote{$P=[n,n-1,1]$, $U=[(2n-1)^2,1^2]$, and $V=[(2n)^2]$. The worked pair uses the finite $A_5/D_4$ closure branch.}
}{%
\AtlasBoards{A_{2n-1}}{%
\AtlasCell{1}{\AtlasRows{\AtlasLong{n}{1}\AtlasLong{n-1}{1}\AtlasRow{1}{1}}}
\AtlasCell{2}{\AtlasRows{\AtlasLong{n}{1}\AtlasLong{n-1}{1}\AtlasRow{1}{1}}}
\AtlasCell{3}{\AtlasRows{\AtlasLong{n}{1}\AtlasLong{n-1}{1}\AtlasRow{1}{1}}}
\AtlasCell{4}{\AtlasRows{\AtlasLong{n}{1}\AtlasLong{n-1}{1}\AtlasRow{1}{1}}}%
}{D_{2n}}{%
\AtlasCell{1}{\AtlasRows{\AtlasLong{2n-1}{2}\AtlasRow{1}{2}}}
\AtlasCell{2}{\AtlasRows{\AtlasLong{2n-1}{2}\AtlasRow{1}{2}}}
\AtlasCell{3}{\AtlasRows{\AtlasLong{2n-1}{2}\AtlasRow{1}{2}}}
\AtlasCell{4}{\AtlasRows{\AtlasLong{2n}{2}}}%
}
}{%
\AtlasPair{A_{5}}{4}{%
\AtlasKey{1}{[2^{3}]}
\AtlasKey{2}{[3^{2}]}
\AtlasKey{3}{[3^{2}]}
\AtlasKey{4}{[4,2]}%
}{D_{4}^{\mathrm{tw}}}{3}{%
\AtlasKey{1}{[1^{8}]}
\AtlasKey{2}{\bigl([2^{3}]\bigr)_{t}}
\AtlasKey{3}{\bigl([4,2]\bigr)_{t}}%
}
}{%
\AtlasBoards{A_{5}}{%
\AtlasCell{1}{\AtlasGrid{2,2,2}}
\AtlasCell{2}{\AtlasGrid{3,3}}
\AtlasCell{3}{\AtlasGrid{3,3}}
\AtlasCell{4}{\AtlasGrid{4,2}}%
}{D_{4}^{\mathrm{tw}}}{%
\AtlasCell{1}{\AtlasGrid{1,1,1,1,1,1,1,1}}
\AtlasCell{2}{\AtlasGrid{2,2,2}}
\AtlasCell{3}{\AtlasGrid{4,2}}%
}
}

\AtlasDefine{25}{%
\AtlasPair{A_3}{4}{%
\AtlasKey{1}{[1^{4}]}
\AtlasKey{2}{P_e}
\AtlasKey{3}{[2^{2}]}
\AtlasKey{4}{[3,1]}%
}{D_4^{\mathrm{tw}}}{3}{%
\AtlasKey{1}{[2^{4}]_{\mathrm{blue}}}
\AtlasKey{2}{H_e}
\AtlasKey{3}{\bigl([3^{2}]\bigr)_{t}}%
}
\AtlasNote{$P_e=[2^{1+e},1^{2-2e}]$ and $H_e=[2^{2+e},1^{2-2e}]_t$, with $e=0,1$.}
}{%
\AtlasBoards{A_3}{%
\AtlasCell{1}{\AtlasGrid{1,1,1,1}}
\AtlasCell{2}{\AtlasBadge{P_e}}
\AtlasCell{3}{\AtlasGrid{2,2}}
\AtlasCell{4}{\AtlasGrid{3,1}}%
}{D_4^{\mathrm{tw}}}{%
\AtlasCell{1}{\AtlasGrid{2,2,2,2}}
\AtlasCell{2}{\AtlasBadge{H_e}}
\AtlasCell{3}{\AtlasGrid{3,3}}%
}
}{%
\AtlasPair{A_{3}}{4}{%
\AtlasKey{1}{[1^{4}]}
\AtlasKey{2}{[2^{2}]}
\AtlasKey{3}{[2^{2}]}
\AtlasKey{4}{[3,1]}%
}{D_{4}^{\mathrm{tw}}}{3}{%
\AtlasKey{1}{[2^{4}]_{\mathrm{blue}}}
\AtlasKey{2}{\bigl([2^{3}]\bigr)_{t}}
\AtlasKey{3}{\bigl([3^{2}]\bigr)_{t}}%
}
}{%
\AtlasBoards{A_{3}}{%
\AtlasCell{1}{\AtlasGrid{1,1,1,1}}
\AtlasCell{2}{\AtlasGrid{2,2}}
\AtlasCell{3}{\AtlasGrid{2,2}}
\AtlasCell{4}{\AtlasGrid{3,1}}%
}{D_{4}^{\mathrm{tw}}}{%
\AtlasCell{1}{\AtlasGrid{2,2,2,2}}
\AtlasCell{2}{\AtlasGrid{2,2,2}}
\AtlasCell{3}{\AtlasGrid{3,3}}%
}
}

%% file: Scan_Counts.tex
\newcommand{\UmbrellaFamilies}{25}
\newcommand{\MainFamilies}{4}
\newcommand{\SubfamilyCount}{145}
\newcommand{\FiniteRuleGroups}{6}
\newcommand{\IsolatedComparisons}{22}
\newcommand{\CandidateDualities}{5{,}278}
\newcommand{\OrganizedCandidatePairs}{5{,}256}
\newcommand{\OrganizedCandidatePercent}{99.58}
\newcommand{\RegularCensus}{167{,}244{,}647}
\newcommand{\ApproxScanCount}[1]{\ensuremath{\sim}\,#1}
\newcommand{\AfterSchurTwelve}{22{,}730{,}000}
\newcommand{\AfterSchurTwenty}{19{,}298{,}580}
\newcommand{\AfterHLTwenty}{19{,}297{,}000}
\newcommand{\AfterDeeperSchur}{3{,}790{,}000}
\newcommand{\AfterDeeperHL}{3{,}630{,}000}
\newcommand{\AfterCoulomb}{3{,}630{,}000}
\newcommand{\AfterSpinors}{3{,}490{,}000}
\newcommand{\AfterFlavor}{3{,}407{,}000}
\newcommand{\AfterIdentical}{2{,}893{,}200}
\newcommand{\AfterAtypical}{90{,}000}
\newcommand{\AfterCollisions}{24{,}400}
\newcommand{\AfterExplanations}{19{,}500}
\newcommand{\CatalogueConstructions}{4{,}759}
\newcommand{\CollisionPairs}{1{,}647{,}489}
\newcommand{\CollisionSharedPairs}{847{,}950}
\newcommand{\CollisionSharedPercent}{51.47\%}
\newcommand{\CollisionNoSharedPairs}{799{,}539}

%% file: figures/Five_Way_Duality.tex
\begin{figure}[!htb]
  \centering
  \newcommand{\FiveWaySphere}[2]{%
    \put(#1,#2){\makebox(0,0){\AtlasSphere{3}}}%
  }
  \newcommand{\FiveWayRay}[1]{%
    \put(116,74){\rotatebox{#1}{%
      \begin{picture}(0,0)
        \put(2,0){\line(1,0){30}}
        \qbezier(27.5,2.8)(29.75,1.4)(32,0)
        \qbezier(27.5,-2.8)(29.75,-1.4)(32,0)
      \end{picture}}}%
  }
  \newcommand{\FiveWayColor}[2]{%
    \begingroup
    \colorlet{atlascell}{#1!22}%
    \AtlasGrid{#2}%
    \endgroup
    \kern2pt\textcolor{#1!65!black}{$\scriptstyle\mathrm{#1}$}%
  }
  \newcommand{\FiveWayTwisted}[1]{\AtlasGrid{#1}\kern2pt$\scriptstyle t$}
  {\setlength{\unitlength}{1pt}%
    \footnotesize
    \scalebox{.93}{\begin{picture}(232,155)
      {\color{atlasink}\linethickness{0.65pt}%
        \FiveWayRay{90}\FiveWayRay{18}\FiveWayRay{-54}%
        \FiveWayRay{-126}\FiveWayRay{162}%
        \put(116,74){\makebox(0,0){$\scriptstyle\bullet$}}%
      }
      \FiveWaySphere{116}{128}
      \FiveWaySphere{167}{91}
      \FiveWaySphere{148}{30}
      \FiveWaySphere{84}{30}
      \FiveWaySphere{65}{91}
      \put(116,152){\makebox(0,0){(a) $D_6$}}
      \put(192,91){\makebox(0,0)[l]{(b) $D_6$}}
      \put(148,5){\makebox(0,0){(c) $D_7$}}
      \put(84,5){\makebox(0,0){(d) $D_7$}}
      \put(40,91){\makebox(0,0)[r]{(e) $D_7^{\mathrm{tw}}$}}
    \end{picture}}\par
  }
  \smallskip
  {\footnotesize
    \begin{tabular*}{232pt}{@{}r@{\extracolsep{\fill}}ccccc@{}}
      & (a) & (b) & (c) & (d) & (e)\\[3pt]
      \AtlasNumber{1}
      & \AtlasGrid{2,2,2,2,1,1,1,1}
      & \AtlasGrid{2,2,2,2,1,1,1,1}
      & \AtlasGrid{3,3,3,1,1,1,1,1}
      & \AtlasGrid{3,3,3,1,1,1,1,1}
      & \AtlasGrid{7,7}\\
      \noalign{\vskip 3pt}
      \AtlasNumber{2}
      & \AtlasGrid{2,2,2,2,1,1,1,1}
      & \FiveWayColor{blue}{2,2,2,2,2,2}
      & \AtlasGrid{3,3,3,1,1,1,1,1}
      & \AtlasGrid{3,3,2,2,2,2}
      & \FiveWayTwisted{2,2,1,1,1,1,1,1,1,1}\\
      \noalign{\vskip 6pt}
      \AtlasNumber{3}
      & \AtlasGrid{7,5}
      & \FiveWayColor{red}{6,6}
      & \AtlasGrid{7,7}
      & \AtlasGrid{7,7}
      & \FiveWayTwisted{4,4,4}
    \end{tabular*}\par
  }
  \caption{Five candidate constructions of one isolated theory. The numbers key the punctures to the Young diagrams below. Red and blue mark the very even choices, and $t$ marks a twist.}
  \label{fig:five-way}
\end{figure}
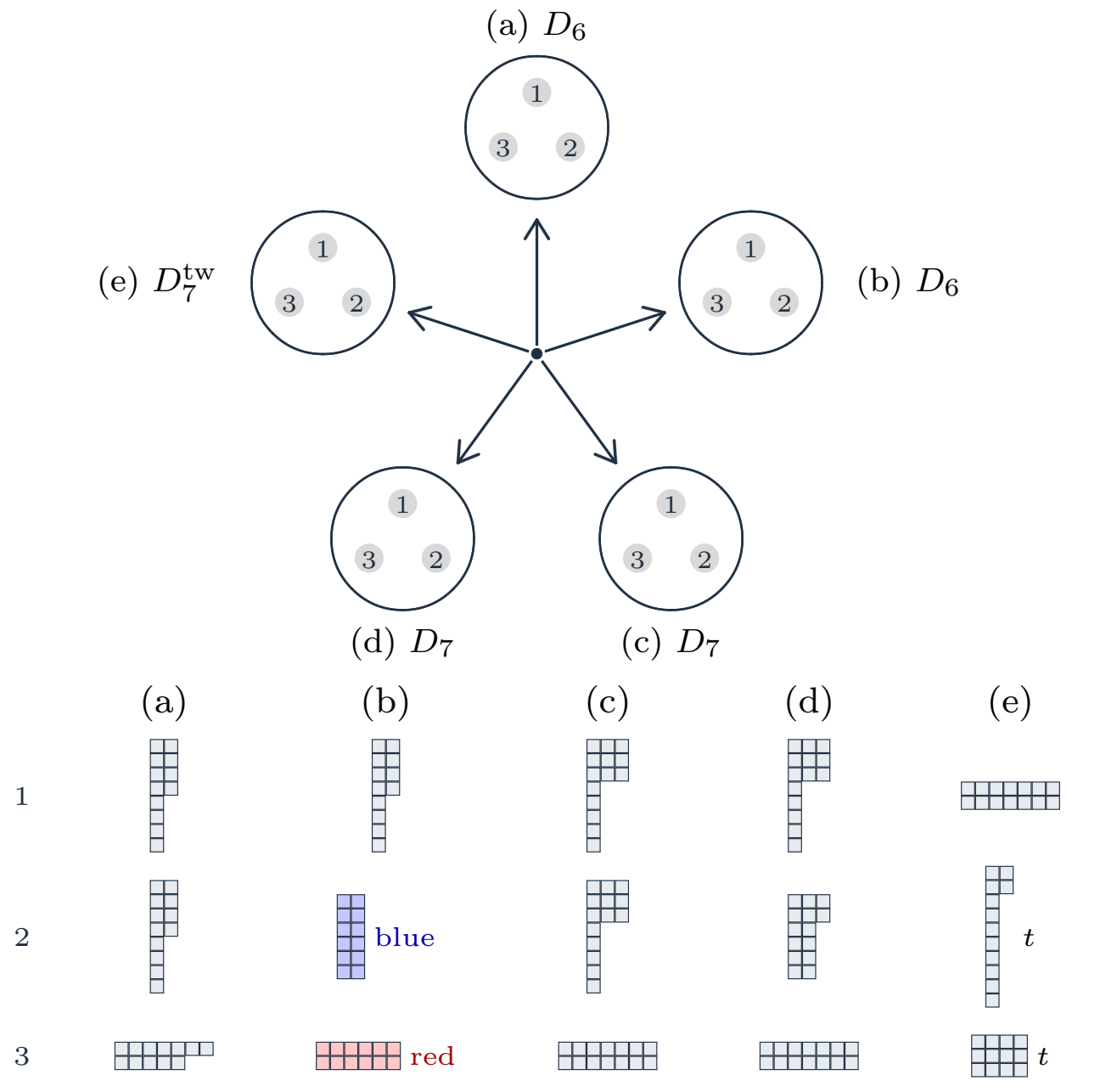

%% file: Supplemental_Material.tex
% Included by Five_Class_S_Conjectures_v4.tex after the references.
\setcounter{secnumdepth}{3}
\setcounter{section}{0}
\setcounter{equation}{0}
\setcounter{table}{0}
\setcounter{figure}{0}
\renewcommand{\thesection}{S\arabic{section}}
\renewcommand{\thesubsection}{\thesection.\arabic{subsection}}
\renewcommand{\thesubsubsection}{\thesubsection.\arabic{subsubsection}}
\makeatletter
\let\p@subsection\@empty
\let\p@subsubsection\@empty
\makeatother
\renewcommand{\theequation}{S\arabic{equation}}
\renewcommand{\thetable}{S\arabic{table}}
\renewcommand{\thefigure}{S\arabic{figure}}
\renewcommand{\theHsection}{supp.\arabic{section}}
\renewcommand{\theHequation}{supp.\arabic{equation}}
\renewcommand{\theHtable}{supp.\arabic{table}}
\renewcommand{\theHfigure}{supp.\arabic{figure}}

\pdfbookmark[0]{Supplemental Material}{supplement.start}
\section*{Supplemental Material}
\phantomsection
\label{app:supplement}

We first report how many constructions survive the screening, then give representative examples for the \UmbrellaFamilies{} families and \FiniteRuleGroups{} finite sets of related examples, followed by the \IsolatedComparisons{} pairs not assigned to a family. A family collects candidate isomorphisms linked by common changes in rank, punctures, or twists. A subfamily specifies one branch and its allowed parameters. The finite sets collect related comparisons at specified ranks and punctures. These labels describe how we organize the catalogue. We then describe the index calculations and give examples of the wavefunctions.

The IndexLab database documents all \SubfamilyCount{} subfamilies and the checks for each pair in the full catalogue. We include the published $A_5/D_5$ pair in Ref.~\cite[Fig.~10(a)]{ZafrirO5} as a reference check.

\phantomsection
\label{app:landscape}

\begin{figure}[!ht]
\centering
\resizebox{0.82\linewidth}{!}{\input{figures/Landscape_Sankey}}
\caption{Flow of constructions through the regular class $\mathcal{S}$ census. The blue ribbons follow the surviving constructions. Each branch reports how many constructions lose their last candidate partner or are set aside at that step. Removal counts are rounded estimates based on the records of the intermediate screening steps. Ribbon widths are rescaled at the dashed links. The final count of \CatalogueConstructions{} refers to distinct constructions surviving the census. The full catalogue contains \CandidateDualities{} candidate pairs, including checked family extensions. Three-way and higher-way candidate equivalences allow one construction to appear in several pairs. Table~\ref{tab:landscape-scan} lists the corresponding checks.}
\label{fig:landscape-scan}
\end{figure}
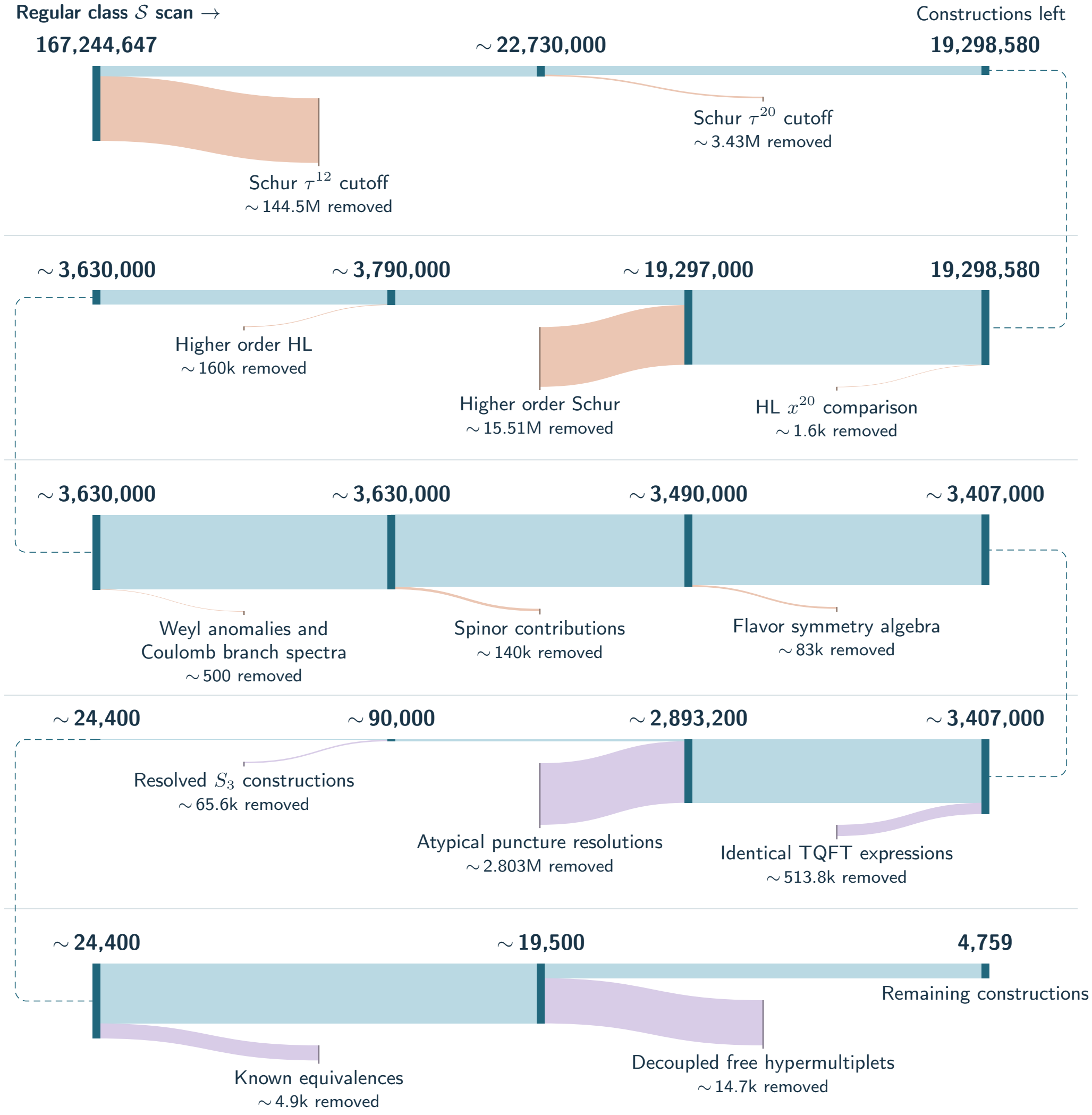

\clearpage

\begin{table}[!htp]
\centering
\caption{Surviving constructions at successive screening steps in Fig.~\ref{fig:landscape-scan}. We count each construction once when it has at least one remaining candidate partner. The final count of \CatalogueConstructions{} refers to distinct constructions surviving the census. The full catalogue contains \CandidateDualities{} candidate pairs, including checked family extensions. Three-way and higher-way candidate equivalences allow one construction to appear in several pairs. Entries marked $\sim$ are rounded where the IndexLab pair records give bounds on the number of distinct surviving constructions.}
\label{tab:landscape-scan}
\small
\renewcommand{\arraystretch}{1.12}
\begin{tabularx}{\linewidth}{@{}>{\raggedright\arraybackslash}p{0.24\linewidth}r>{\raggedright\arraybackslash}X@{}}
\toprule
Screening step & Constructions left & Comparison \\
\midrule
Regular census & \RegularCensus & Enumerated regular punctures and twists \\
Schur $\tau^{12}$ & \ApproxScanCount{\AfterSchurTwelve} & Unrefined Schur coefficients \\
Schur $\tau^{20}$ & \AfterSchurTwenty & Extended Schur cutoff \\
HL $x^{20}$ & \ApproxScanCount{\AfterHLTwenty} & First Hall--Littlewood cutoff \\
Higher order Schur & \ApproxScanCount{\AfterDeeperSchur} & For surfaces differing only by a $\mathbb Z_2$ handle twist, test the first Schur contribution from weights excluded by the twist. Compare through $\tau^{30}$ and to higher orders in selected cases. \\
Higher order HL & \ApproxScanCount{\AfterDeeperHL} & Common $x^{30}$ cutoff and targeted higher orders \\
Weyl anomalies and Coulomb branch spectra & \ApproxScanCount{\AfterCoulomb} & Equality of $a,c$ and Coulomb branch spectra \\
Spinor contributions & \ApproxScanCount{\AfterSpinors} & For $D$-type pairs with matching puncture weights after projection onto the invariant lattice, total puncture $K$ factors, and denominator exponent $2g-2+n$, compare the omitted spinor terms through six powers beyond their onset when this extends the previous Schur cutoff. \\
Flavor symmetry algebra & \ApproxScanCount{\AfterFlavor} & Exclude established mismatches in the flavor symmetry algebra. Keep pairs with an incomplete flavor symmetry comparison. \\
\midrule
Identical formulas & \ApproxScanCount{\AfterIdentical} & Set aside identical Macdonald sums with conjectured full index agreement \\
Resolution of atypical punctures & \ApproxScanCount{\AfterAtypical} & Set aside pairs whose puncture resolutions identify a common conformal family, even when their original coupling loci differ. \\
Resolved $S_3$ constructions & \ApproxScanCount{\AfterCollisions} & Apply the same criterion to resolved $D_4$ constructions, retaining the ordered $S_3$ holonomies. \\
Known equivalences & \ApproxScanCount{\AfterExplanations} & Set aside established Higgsing, gauging, and sewing explanations \\
Decoupled hypermultiplets & \textbf{\CatalogueConstructions} & Require zero decoupled free hypermultiplets on both sides \\
\bottomrule
\end{tabularx}
\par\smallskip
\end{table}

\begin{table}[!htp]
\centering
\caption{Puncture labels and twist data used in the screening.}
\label{tab:screening-notation}
\small
\renewcommand{\arraystretch}{1.12}
\begin{tabularx}{\linewidth}{@{}>{\raggedright\arraybackslash}p{0.24\linewidth}>{\raggedright\arraybackslash}X@{}}
\toprule
Notation & Meaning \\
\midrule
$G\{P_1,\ldots,P_n\}$, $g=1$ & A sphere of six dimensional parent type $G$ with the listed punctures. A superscript $g=1$ specifies a torus. A power outside a puncture label counts copies of that puncture. \\
Untwisted labels & Nahm orbits given by partitions in types $A,D$ and Bala--Carter labels in type $E$. The orbit label $0$ denotes the full puncture~\cite{CDTDefects}. \\
$\mathrm{tw}$, $t$, $t_3$ & The superscript $\mathrm{tw}$ marks a twisted surface. The puncture subscripts $t$ and $t_3$ mark order two and order three monodromy, respectively. \\
Order two punctures & The twisted Nahm labels are $B$-partitions of $2N+1$ for $A_{2N-1}$, $C$-partitions of $2N$ for $A_{2N}$, $C$-partitions of $2N-2$ for $D_N$, and $F_4$ orbits for $E_6$. In a $B$-partition each even part has even multiplicity, and in a $C$-partition each odd part has even multiplicity~\cite{CDTTwistedA,BPTwisted,CDTTwistedD,CDTTwistedE6}. \\
Order three $D_4$ punctures & Nahm orbits of $G_2$~\cite{CDTZ3D4}. \\
$D_4^{S_3}$, $P_{t,\alpha}$ & The subscript $\alpha$ in $P_{t,\alpha}$ specifies the actual order two monodromy in $S_3$. For a fixed choice of generators, the ordered puncture holonomies $m_j$ and handle holonomies $A_i,B_i$ obey $\prod_i[A_i,B_i]\prod_j m_j=1$ and are identified under simultaneous conjugation~\cite{DESD4}. In U8, $\alpha,\beta$ are distinct transpositions and the third puncture carries $(\alpha\beta)^{-1}$. \\
Red and blue & A very even $D_N$ partition labels two orbits~\cite{CDTypeD}. In IndexLab's orthogonal Cartan coordinates for the Nahm embedding, the last coordinate is positive for red and negative for blue. Relative color compares these signs between punctures. \\
\bottomrule
\end{tabularx}
\end{table}

\raggedbottom
\input{Supplemental_Simplest_Representatives}

\input{Supplemental_Isolated_Cases}

\section{\texorpdfstring{\texttt{IndexLab}}{IndexLab}}
\label{app:index-calculations}

\subsection{Indices}
\label{sec:index-methods}

The class $\mathcal{S}$ TQFT builds the index from wavefunctions, one for each regular puncture~\cite{GPRR,GRRYMacdonald,MSTTwistedIndex}. To glue pieces of the surface, we sum over their shared representation labels. The wavefunctions depend on which limit of the index we want to compute. We write $p,q,t$ for the superconformal fugacities and $\mathbf a$ for flavor fugacities. For the expansions, we use $x=\sqrt t$, with $q=t=\tau^2$ in the Schur limit.

\begin{samepage}
IndexLab computes or displays the following indices and theory data.
\begin{itemize}
\setlength{\itemsep}{0pt}
\item Schur, Hall--Littlewood, Macdonald, and full superconformal indices, with flavor refinement where available.
\item Weyl anomaly coefficients $a$ and $c$.
\item Coulomb branch rank and spectrum.
\item Flavor symmetry algebras and levels where available.
\item Counts of decoupled free hypermultiplets.
\end{itemize}
\end{samepage}

Table~\ref{tab:indexlab-coverage} lists the regular class $\mathcal{S}$ parent algebras supported by IndexLab. Protected index calculations extend to genus 20, depending on the puncture and twist data, and the truncation order we can reach. All four indices are also available for selected Argyres--Douglas theories and $SU$, $SO$, and $\mathrm{USp}$ quivers (Sec.~\ref{sec:additional-theories}).

\begin{table}[!htbp]
\centering
\caption{Regular class $\mathcal{S}$ coverage in IndexLab by six dimensional parent rank. Full index coverage includes calculations conditional on proposed elliptic wavefunctions and reconstructions from protected data.}
\label{tab:indexlab-coverage}
\small
\setlength{\tabcolsep}{5pt}
\renewcommand{\arraystretch}{1.14}
\begin{tabularx}{\linewidth}{@{}l>{\raggedright\arraybackslash}p{0.34\linewidth}>{\raggedright\arraybackslash}X@{}}
\toprule
Index & Untwisted parents & Twisted parents \\
\midrule
Hall--Littlewood & $A_1$--$A_{32}$, $D_4$--$D_{32}$, $E_6,E_7,E_8$ & $A_2$--$A_{32}$, $D_4$--$D_{32}$, $E_6$ with $\mathbb Z_2$ twists. $D_4$ with $\mathbb Z_3$ or $S_3$ twists. \\
Schur & $A_1$--$A_{32}$, $D_4$--$D_{32}$, $E_6,E_7,E_8$ & $A_2$--$A_{32}$, $D_4$--$D_{32}$, $E_6$ with $\mathbb Z_2$ twists. $D_4$ with $\mathbb Z_3$ or $S_3$ twists. \\
Macdonald & $A_1$--$A_{32}$, $D_4$--$D_{32}$, $E_6,E_7,E_8$ & $A_2$--$A_{32}$, $D_4$--$D_{32}$, $E_6$ with $\mathbb Z_2$ twists. $D_4$ with $\mathbb Z_3$ or $S_3$ twists. \\
\addlinespace[2pt]
Full SCI &
$A_1$--$A_5$, $D_4,D_5,E_6,E_7,E_8$ for supported surfaces of genus at most three with up to five punctures. Limited coverage at low orders for $A_6$--$A_{10}$ and $D_6$--$D_{10}$. &
$A_2$--$A_5$, $D_4,D_5$, and $E_6$ in supported $\mathbb Z_2$ sectors of genus at most three with up to five punctures. Selected $\mathbb Z_3$ fixtures of $D_4$. Higher orders for selected inputs. \\
\bottomrule
\end{tabularx}
\end{table}

\paragraph{Check against published results}
The \texttt{IndexLab} database (\emph{Database}, \emph{Browse}, \emph{Typos (17)}) lists 17 discrepancies with literature results and gives the corresponding coefficient comparisons.

\subsubsection{Full superconformal index}
\label{sec:full-index}

The full superconformal index (SCI) is the graded trace
\[
\Ia(p,q,t;\mathbf a)=
\operatorname{Tr}_{\delta=0}(-1)^F
p^{j_2-j_1-r}q^{j_2+j_1-r}t^{R+r}\mathbf a^{\mathbf f},
\qquad
\delta=E-2j_2-2R+r .
\]
Here $E$ is the scaling dimension, $j_1,j_2$ are Cartan charges of the rotation group, $R,r$ are Cartan charges of the R symmetry, and $\mathbf f$ denotes flavor charges~\cite[Sec.~2]{RRReview}.

For regular untwisted type $A$, we start at $p=0$, where the wavefunctions are built from Macdonald polynomials. We then solve the eigenvalue equation order by order in $p$ to find the elliptic correction. In the $GL(N)$ polynomial convention, the first Ruijsenaars operator is
\[
D(p)=\sum_{i=1}^{N}
\left(\prod_{j\ne i}\frac{\theta(tz_i/z_j;p)}{\theta(z_i/z_j;p)}\right)
T_{q,z_i},
\qquad
T_{q,z_i}f(\mathbf z)=f(z_1,\ldots,qz_i,\ldots,z_N).
\]
Here $\theta(u;p)=(u;p)_\infty(p/u;p)_\infty$. At $p=0$, the eigenvalue on $P_\lambda$ is $E_\lambda^{(0)}=\sum_i q^{\lambda_i}t^{N-i}$. Write $D(p)=D_0+pD_1+\cdots$ and $F_\lambda=P_\lambda+pF_\lambda^{(1)}+\cdots$, with the $P_\lambda$ component of $F_\lambda^{(1)}$ set to zero. For generic parameters the first correction is
\[
F_\lambda^{(1)}
=\sum_{\mu\ne\lambda}
\frac{[P_\mu](D_1P_\lambda)}{E_\lambda^{(0)}-E_\mu^{(0)}}\,P_\mu ,
\]
where $[P_\mu]$ extracts the coefficient of a Macdonald basis polynomial. We specialize $F_\lambda$ at the puncture and multiply it by the elliptic gamma factors associated with the puncture. Before sewing, we recompute the norm using the physical vector multiplet measure~\cite{GRRDefects,LNSElliptic}.

At first order, the exact mixing coefficients connect labels that differ by a root. On stable regular untwisted $A_1$ through $A_5$ surfaces, the summands have positive leading powers, so we can bound the representation sum at each cutoff. For $A_1,A_2$, we also compute the second order correction within finite bounds on the powers of $p,q,x$.

For the other Lie types, we start from the $p=0$ wavefunctions and propose first order corrections using root or adjoint Pieri coefficients~\cite{KomoriElliptic,VanDiejenEmsiz}. In twisted sectors, we must also choose a folded representation lattice and the corresponding puncture and cap factors. Our type $D$ candidates cover $D_4$ and $D_5$. After sewing the $D_4$ wavefunctions, we checked the Macdonald limit and $p\leftrightarrow q$ symmetry to finite order.

As a separate check, the elliptic integral calculation reproduces the published unrefined series for the rank one $E_6$ theory through radial degree 11. It also reproduces the series for $R_{0,4}$ and the rank one $E_7$ theory through radial degrees 8 and 10, respectively~\cite[Sec.~3.2]{GRRYE6}\cite[Sec.~3]{AMSE7}. Section~\ref{app:elliptic-wavefunctions} gives examples of the first order wavefunctions.

The type $A$ calculation uses the surface defect TQFT and elliptic eigenfunction construction~\cite{GRRDefects,LNSElliptic}, and the integral benchmarks use elliptic inversion and Higgsing~\cite{GRRYE6,AMSE7}.

\subsubsection{Macdonald index}

The Macdonald index is $\Ia_{\rm Mac}(q,t;\mathbf a)=\Ia(0,q,t;\mathbf a)$. The trace keeps the separate weights $q^{2j_1}t^{R+r}$ on the surviving states. A regular puncture $\rho$ specifies a Nahm embedding of $SU(2)$ in the parent group. Write $H_\rho$ for its Cartan element and decompose the adjoint as $\operatorname{adj}G=\bigoplus_d V_d\otimes R_{\rho,d}$, where $V_d$ is $d$ dimensional and $R_{\rho,d}$ carries the flavor action. For an untwisted regular puncture,
\begin{align*}
\psi_{\rho,\lambda}^{\rm Mac}(\mathbf a;q,t)
  &=K_\rho^{\rm Mac}(\mathbf a;q,t)\,
    P_\lambda(\mathbf a\,t^{H_\rho/2};q,t),\\
K_\rho^{\rm Mac}(\mathbf a;q,t)
  &=\PE\left[\frac{\sum_d t^{(d+1)/2}\chi_{R_{\rho,d}}(\mathbf a)}{1-q}\right].
\end{align*}
Here $P_\lambda$ is a monic Macdonald polynomial and $\PE$ denotes the plethystic exponential. For a surface of genus $g$ with $n$ punctures, our diagonal sewing convention is
\[
\Ia_{g,n}=\sum_\lambda
\mathcal N_\lambda^{\,g-1}
\frac{\displaystyle\prod_{i=1}^{n}\psi_{\rho_i,\lambda}(\mathbf a_i)}{\psi_{\mathrm{pr},\lambda}^{\,2g-2+n}} ,
\]
where $\mathcal N_\lambda$ is the squared norm before orthonormalization and $\psi_{\mathrm{pr},\lambda}$ is the wavefunction evaluated at the principal embedding (the TQFT cap). In twisted sectors the holonomy selects the allowed labels and the twisted puncture and cap factors enter the sewing sum~\cite{GRRYMacdonald,LPRTypeD,BPTwisted}. IndexLab accepts higher genus inputs in supported sectors. For example, the Schur index calculation allows $0\le g\le20$.

At a fixed cutoff, we use the leading powers for the fundamental weights to determine which representations can contribute, provided those powers are positive. We evaluate principal specializations and norms with root product formulas, and general untwisted polynomials by triangular orthogonalization. In the $\Spin(8)$ comparisons, weighted degree 20 means keeping $q^a x^b$ with $2a+b\le20$. In the Schur specialization $t=q=\tau^2$, we have $x=\sqrt t=\tau$, so $q^a x^b$ becomes $\tau^{2a+b}$. The weighted degree is therefore the power of $\tau$ in that limit. Setting $q=0$ leaves the Hall--Littlewood terms with $a=0$. The Macdonald calculation uses root system polynomials and their evaluation formulas~\cite{MacdonaldRoots,CherednikEvaluation}, together with the untwisted and twisted sewing prescriptions~\cite{GRRYMacdonald,LPRTypeD,BPTwisted}.

\subsubsection{Hall--Littlewood index}

The Hall--Littlewood (HL) index is $\Ia_{\rm HL}(t;\mathbf a)=\Ia_{\rm Mac}(0,t;\mathbf a)$. It keeps the $j_1=0$ states of the Macdonald trace with weight $t^{R+r}$~\cite[Sec.~5]{RRReview}. In the untwisted wavefunction, $P_\lambda(\cdot;0,t)$ is the Hall--Littlewood polynomial, and $K_\rho^{\rm HL}=\PE[\sum_d t^{(d+1)/2}\chi_{R_{\rho,d}}]$. For untwisted punctures, we evaluate these polynomials by Weyl symmetrization in the Hecke representation. In twisted sectors, we take the Hall--Littlewood limit of that sector's sewing data. Examples with four twisted punctures show why additional conditions are needed to identify this index with a Higgs branch Hilbert series~\cite{KLLSSHL}. The calculation uses the $q=0$ root system polynomials~\cite{MacdonaldRoots} and the corresponding untwisted and twisted sewing prescriptions~\cite{GRRYMacdonald,LPRTypeD,BPTwisted,KLLSSHL}.

\subsubsection{Schur index}

The Schur index sets $t=q=\tau^2$ and is independent of $p$. It is the trace with weight $\tau^{2(E-R)}$ over Schur states. In the untwisted sewing sum the monic Macdonald polynomial becomes a Weyl character, so the wavefunction is
\[
\psi_{\rho,\lambda}^{\rm Schur}
=K_\rho^{\rm Mac}(\mathbf a;\tau^2,\tau^2)\,
\chi_\lambda(\mathbf a\,\tau^{H_\rho}) .
\]
The highest weight evaluations at the punctures and at the principal specialization determine the leading power of each summand. If the leading powers for the fundamental weights are positive, we discard representations whose terms start beyond the cutoff. We then expand the remaining characters and $K$ factors exactly. For twisted fixtures, we use the corresponding folded or twining characters. We keep very even colors in the puncture data because changing one color can change the characters in the sum. The Schur calculation uses $q$ deformed Yang--Mills sewing and its Macdonald formulation, together with the type $D$ and twisted $A$ character prescriptions~\cite{GRRYSchur,GRRYMacdonald,LPRTypeD,BPTwisted}. An independent class $\mathcal{S}$ Schur index database \cite{ChoSchurTool} appeared while we were developing \texttt{IndexLab}.

\subsection{Argyres--Douglas and Lagrangian theories}
\label{sec:additional-theories}

Regular punctures give us one way to compute an index. For Argyres--Douglas theories, we also use irregular puncture formulas and, where available, gauge descriptions that flow to the fixed point. For Lagrangian theories, we project directly onto gauge singlets. When a theory also has a regular class $\mathcal{S}$ construction, these calculations provide separate checks.

\subsubsection{Argyres--Douglas examples}

The rank one Argyres--Douglas (AD) examples $H_0=(A_1,A_2)$, $H_1=(A_1,A_3)\simeq(A_1,D_3)$, and $H_2=(A_1,D_4)$ admit full index calculations using ultraviolet $\mathcal N=1$ gauge theories that flow to the interacting $\mathcal N=2$ fixed points~\cite{MSMinimalAD,MSADFlows,AMSADFlows}. We remove the decoupled chiral factors and expand the elliptic gamma integrand before projection onto gauge singlets with the Haar measure. The radial expansion uses
\[
 p=s^3y,\qquad q=s^3/y,\qquad t=s^4/v .
\]
We set flavor fugacities to one and keep the $y,v$ dependence. The resulting series reproduce the published coefficients through $s^9$ for $H_0$, $s^8$ for $H_1$, and $s^9$ for $H_2$~\cite{MSMinimalAD,MSADFlows,AMSADFlows}. For $H_1$, the separate $A_3$ and $D_3$ flow integrals give the same 23 nonzero monomials through $s^8$~\cite{MSADFlows,AMSADFlows}.

For the protected limits, we use vacuum characters of vertex operator algebras or irregular wavefunctions for the Schur index, rational Hall--Littlewood formulas, and proposed Macdonald formulas with two fugacities~\cite{BNADSchur,SXYADVOA,BNADMacdonald,SongADHL,KKSADMacdonald}. We also use published identifications of two regular twisted $A_2$ fixtures to evaluate their full indices as $\Ia_{H_2}\Ia_{\rm hyper}$ and $\Ia_{H_2}^2$~\cite[Secs.~3.4 and 3.5]{BPTwisted}.

\subsubsection{Lagrangian theories}

For a conformal $\mathcal N=2$ gauge theory with compact gauge group $G$, we compute
\begin{align*}
\Ia_G&=\int_G d\mu_G(\mathbf z)\,
\PE\bigl[f_V\,\chi_{\rm adj}(\mathbf z)+f_H\,\chi_{\mathcal R}(\mathbf z)\bigr],\\
f_H&=\frac{\sqrt t-pq/\sqrt t}{(1-p)(1-q)},\\
f_V&=-\frac{p}{1-p}-\frac{q}{1-q}
      +\frac{pq/t-t}{(1-p)(1-q)} .
\end{align*}
Here $d\mu_G$ is normalized Haar measure and $\mathcal R$ contains one copy of the matter representation for a half hypermultiplet and $R\oplus\overline R$ for a full hypermultiplet in $R$~\cite[Sec.~2]{GRRYMacdonald}. We construct the characters of the gauge group representations for products of $SU$, $SO$, and $USp$ factors, check the vanishing of the one loop beta functions and the conditions for half hypermultiplets, and evaluate the integral as a constant term in the gauge fugacities with the Weyl integration measure. 

\section{Elliptic wavefunctions for the full superconformal index}
\label{app:elliptic-wavefunctions}

The examples below give the first order data used in Sec.~\ref{sec:full-index}. We write $F_\lambda$ for the eigenfunction before puncture specialization. After specialization and multiplication by the puncture factor, we write it as $\psi_{\rho,\lambda}$. The $D_4$, $E_6$, and $E_8$ wavefunctions and their sewing prescriptions remain conjectural.

\subsection{\texorpdfstring{The elliptic $A_1$ vacuum wavefunction}{The elliptic A1 vacuum wavefunction}}

For a maximal $A_1$ puncture write $z=z_1/z_2$. The monic Macdonald polynomial for the adjoint representation is
\[
P_{(1,-1)}=z+z^{-1}+\frac{(1+q)(1-t)}{1-qt}.
\]
To first order in $p$, the vacuum eigenfunction is
\[
F_0=1+p\,\frac{q(1-t)(1-t^2)}{t(1-q)(1-qt)}P_{(1,-1)}
       +\OO(p^2).
\]
The coefficient of $P_{(1,-1)}$ follows from the elliptic expansion in Ref.~\cite[Sec.~2.2]{LNSElliptic} at $n=2$, $\lambda=0$.
To obtain the puncture wavefunction in this convention, we multiply by the maximal puncture factor
\[
K_{\rm full}^{\Gamma}=\prod_{w\in\operatorname{wt}(\operatorname{adj} A_1)}
\Gamma(t\mathbf a^w;p,q).
\]
In the similarity transformation, the nonzero weights give the factor that depends on flavor fugacities. The zero weight gives $\Gamma(t;p,q)$, which is independent of $\mathbf a$ but depends on $p,q,t$~\cite[Sec.~5]{RRReview}. At $p=0$, the product reduces to the Macdonald maximal $K$ factor.

\subsection{\texorpdfstring{A candidate elliptic vacuum wavefunction for $D_4$}{A candidate elliptic vacuum wavefunction for D4}}

At first order, the candidate $D_4$ vacuum wavefunction contains an adjoint contribution. In the monic Macdonald convention, the eigenfunction starts at
\[
F_0^{D_4}=1+
pq\,\frac{(1-t)(1-t^6)}{t}\,P_{\omega_2}^{D_4}
+\OO(pq^2,p^2),
\]
where $\omega_2$ labels the adjoint and $t$ is held fixed. The elliptic dressing and sewing for this candidate remain conjectural.

\subsection{\texorpdfstring{First order mixing in the candidate $E_6$ wavefunction}{First order mixing in the candidate E6 wavefunction}}

For a root $\alpha$ with $\mu=\lambda+\alpha$ dominant, define $m=1+\langle\lambda,\alpha^\vee\rangle$, $u=q^m t^{\operatorname{ht}(\alpha)}$, and $M_{\lambda\mu}=[P_\mu](\chi_{\rm adj}P_\lambda)$. Here $[P_\mu](\chi_{\rm adj}P_\lambda)$ means the coefficient of $P_\mu$ when the product $\chi_{\rm adj}P_\lambda$ is expanded in the Macdonald polynomial basis. Also, $\operatorname{ht}(\alpha)=\langle\rho^\vee,\alpha\rangle$, with $\rho^\vee$ the half sum of positive coroots. In the convention used by the $E_6$ and $E_8$ calculations in IndexLab, the coefficient of $pP_\mu$ in $F_\lambda^{(-)}$ is
\[
 c^{(-)}_{\lambda\mu}=-M_{\lambda\mu}
 \frac{(1-t)(1-q/t)}{1-q}\,\frac{u}{1-u}.
\]
The coefficient of $P_\lambda$ in $F_\lambda^{(-)}$ is fixed to one. At the orders displayed in the $A_1$ and $D_4$ examples, the vacuum coefficients instead take the form
\[
 c^{(+)}_{\lambda\mu}=M_{\lambda\mu}\,\beta
 \frac{1-(t/q)u}{1-u},\qquad
 \beta=\frac{q(1-t)}{t(1-q)}.
\]
For each off diagonal coefficient this gives $c^{(+)}_{\lambda\mu}-c^{(-)}_{\lambda\mu}=\beta M_{\lambda\mu}$. To compare the two conventions within a fixed parent, let $C(z)=1-p\beta\chi_{\rm adj}(z)$ and $\sigma_\lambda=1+p\beta M_{\lambda\lambda}$. Multiplication by the common factor and restoration of the monic normalization give $F_\lambda^{(-)}(z)=\sigma_\lambda C(z)F_\lambda^{(+)}(z)+\OO(p^2)$.

To change conventions while preserving the dressed wavefunctions up to $\sigma_\lambda$, we need $K_\rho^{(-)}=K_\rho^{(+)}/C(z_\rho)$ through first order, where $z_\rho=\mathbf a\,t^{H_\rho/2}$. We make the same change at the principal cap. In the norm integral, the effective density becomes $W^{(-)}(z)=W^{(+)}(z)/[C(z)C(z^{-1})]$, giving $\mathcal N_\lambda^{(-)}=\sigma_\lambda^2\mathcal N_\lambda^{(+)}+\OO(p^2)$. In the diagonal sewing formula, the norm contributes $\sigma_\lambda^{2(g-1)}$, the $n$ punctures contribute $\sigma_\lambda^n$, and the cap denominator contributes $\sigma_\lambda^{-(2g-2+n)}$. Their exponents sum to zero, so the sewn index is unchanged through first order in $p$.

The implementation for exceptional Lie algebras uses $c^{(-)}$ with the unmodified puncture and cap elliptic gamma factors. Relative to the Macdonald norm, its first order norm correction is $p M_{\lambda\lambda}(t-1-q+q/t)/(1-q)$. The type $A$ code uses the corresponding factor $(t-1+q-q/t)/(1-q)$. The two factors differ by $2\beta$, as required by the change in effective norm density. In the implementation for exceptional Lie algebras, the elliptic gamma factors and norm together form a candidate prescription. To treat the coefficient shift as a change of convention, we must also change the dressing and norm as described above.

The ansatz uses quasi-minuscule Pieri coefficients and the affine elliptic operator framework~\cite{VanDiejenEmsiz,KomoriElliptic}. The truncated index expansions depend on the candidate wavefunction and its sewing prescription.

\subsection{\texorpdfstring{A candidate elliptic vacuum wavefunction for $E_8$}{A candidate elliptic vacuum wavefunction for E8}}

For a maximal $E_8$ puncture, the same ansatz gives an explicit vacuum eigenfunction before multiplication by the puncture factor. Let $\vartheta$ be the highest root of $E_8$, which labels the adjoint $\mathbf{248}$. It is the unique dominant root, so only the adjoint appears in the first order vacuum correction. Here $\operatorname{ht}(\vartheta)=29$, $m=1$, and $M_{0,\vartheta}=1$ by monicity. We use the $(-)$ convention above. Thus
\[
F_0^{E_8}(\mathbf a)=1-
p\,\frac{q t^{29}(1-t)(1-q/t)}{(1-q)(1-q t^{29})}
P_\vartheta^{E_8}(\mathbf a;q,t)+\OO(p^2).
\]
At unit flavor fugacities the monic Macdonald polynomial in this expression is
\[
P_\vartheta^{E_8}(\mathbf 1;q,t)=248+
\frac{q-t}{1-q t^{29}}
\bigl(1+t^6+t^{10}+t^{12}+t^{16}+t^{18}+t^{22}+t^{28}\bigr).
\]
The second term follows from Ref.~\cite[Thm.~4.1 and Table~4.1]{Stembridge}. It vanishes at $q=t$, where the polynomial gives the adjoint dimension $248$.

\begin{samepage}
To check the operator, write $D(p)=D_0+pD_1+\OO(p^2)$ and denote the $E_8$ root system by $\Phi$. The character matrix used in IndexLab gives the following zeroth order eigenvalues and action of $D_1$ on the vacuum:
\begin{align*}
 D_0P_\lambda&=\varepsilon_\lambda P_\lambda,\qquad
 \varepsilon_\lambda=\sum_{\alpha\in\Phi}
 q^{\langle\lambda,\alpha^\vee\rangle}t^{29+\operatorname{ht}(\alpha)},\\
 D_1\cdot 1&=B(q,t)\chi_{\rm adj},\\
 B(q,t)&=(1-t)(1-q/t)\bigl[(1+q)t^{58}\\
 &\hspace{3em}+(1+q^{-1})t^{29}+S_1(t)\bigr],\\
 S_1(t)&=\sum_{\substack{\alpha\in\Phi\\\langle\vartheta,\alpha^\vee\rangle=1}}
 t^{29+\operatorname{ht}(\alpha)}.
\end{align*}
Since $[P_\vartheta]\chi_{\rm adj}=1$, the first order eigenfunction coefficient is $B(q,t)/(\varepsilon_0-\varepsilon_\vartheta)$. Summing over the $E_8$ roots gives exactly the coefficient of $pP_\vartheta$ shown above, providing a check of the displayed vacuum coefficient.
\end{samepage}

It remains conjectural that dressing and sewing this candidate gives the physical index. We take the regular puncture data from the $E_8$ fixture catalogue~\cite[Sec.~2.2]{CDTZTypeE8}.

%% file: figures/Landscape_Sankey.tex
% Included in the Supplemental Material. Counts come from Scan_Counts.tex.
\begingroup
\definecolor{sankeyink}{HTML}{193447}
\definecolor{sankeybar}{HTML}{21677C}
\definecolor{sankeyflow}{HTML}{BAD9E3}
\definecolor{sankeyveto}{HTML}{EBC6B3}
\definecolor{sankeyscope}{HTML}{D8CCE7}
\definecolor{sankeyrule}{HTML}{D7E0E4}
% Arguments are start x, end x, top y, incoming and continuing widths,
% branch color, mechanism label, and number removed. Each row has its own width scale.
\newcommand{\SankeyFlow}[8]{%
  \pgfmathsetmacro{\scanmid}{(#1+#2)/2}%
  \pgfmathsetmacro{\scancontrol}{(#1+\scanmid)/2}%
  \pgfmathsetmacro{\scantop}{#3+#5}%
  \pgfmathsetmacro{\scanbottom}{#3+#4}%
  \pgfmathsetmacro{\scanshift}{10}%
  \path[fill=sankeyflow] (#1,#3) rectangle (#2,\scantop);
  \path[fill=#6] (#1,\scantop)
    .. controls (\scancontrol,\scantop) and (\scancontrol,\scantop+\scanshift)
    .. (\scanmid,\scantop+\scanshift)
    -- (\scanmid,\scanbottom+\scanshift)
    .. controls (\scancontrol,\scanbottom+\scanshift) and (\scancontrol,\scanbottom)
    .. (#1,\scanbottom) -- cycle;
  \draw[#6!60!black,line width=0.65pt]
    (\scanmid,\scantop+\scanshift) -- (\scanmid,\scanbottom+\scanshift+1.5);
  \node[sankeylabel,anchor=north] at (\scanmid,\scanbottom+\scanshift+4) {#7\\{\footnotesize #8}};
}
\newcommand{\SankeyStage}[4]{%
  \path[fill=sankeybar] (#1-1.8,#2) rectangle (#1+1.8,#2+#3);
  \node[sankeycount,anchor=south] at (#1,#2-5) {#4};
}
\newcommand{\SankeyLink}[4]{%
  \draw[sankeylink] (#1,#2) -- (#3,#2) -- (#3,#4) -- (#1,#4);
}

\begin{tikzpicture}[x=1pt,y=-1pt,
  sankeylabel/.style={align=center,font=\sffamily\small,text=sankeyink,inner sep=0pt},
  sankeycount/.style={font=\sffamily\bfseries\normalsize,text=sankeyink,inner sep=0pt},
  sankeylink/.style={draw=sankeybar,densely dashed,rounded corners=5pt,->}]
\path[use as bounding box] (0,0) rectangle (506,519);
\foreach \scanrow in {109,211,318,415} {
  \draw[sankeyrule,line width=0.5pt] (9,\scanrow) -- (497,\scanrow);
}
\node[sankeylabel,anchor=west,font=\sffamily\small\bfseries] at (14,8)
  {Regular class $\mathcal{S}$ scan $\rightarrow$};
\node[sankeylabel,anchor=east] at (492,8) {Constructions left};

% Branch labels give differences between successive construction counts.
% Schur screening
\SankeyFlow{51}{253}{32}{34}{4.62}{sankeyveto}{Schur $\tau^{12}$ cutoff}{\ApproxScanCount{144.5M} removed}
\SankeyFlow{253}{455}{32}{4.62}{3.92}{sankeyveto}{Schur $\tau^{20}$ cutoff}{\ApproxScanCount{3.43M} removed}
\SankeyStage{51}{32}{34}{\RegularCensus}
\SankeyStage{253}{32}{4.62}{\ApproxScanCount{\AfterSchurTwelve}}
\SankeyStage{455}{32}{3.92}{\AfterSchurTwenty}
\SankeyLink{455}{34}{492}{151}

% Hall--Littlewood and deeper index comparisons
\SankeyFlow{455}{320}{134}{34}{33.8}{sankeyveto}{HL $x^{20}$ comparison}{\ApproxScanCount{1.6k} removed}
\SankeyFlow{320}{185}{134}{33.8}{6.64}{sankeyveto}{Higher order Schur}{\ApproxScanCount{15.51M} removed}
\SankeyFlow{185}{51}{134}{6.64}{6.36}{sankeyveto}{Higher order HL}{\ApproxScanCount{160k} removed}
\SankeyStage{455}{134}{34}{\AfterSchurTwenty}
\SankeyStage{320}{134}{33.8}{\ApproxScanCount{\AfterHLTwenty}}
\SankeyStage{185}{134}{6.64}{\ApproxScanCount{\AfterDeeperSchur}}
\SankeyStage{51}{134}{6.36}{\ApproxScanCount{\AfterDeeperHL}}
\SankeyLink{51}{137.2}{14}{253}

% Anomalies, Coulomb branch spectra, spinor contributions, and flavor
\SankeyFlow{51}{185}{236}{34}{33.8}{sankeyveto}{Weyl anomalies and\\Coulomb branch spectra}{\ApproxScanCount{500} removed}
\SankeyFlow{185}{320}{236}{33.8}{32.7}{sankeyveto}{Spinor contributions}{\ApproxScanCount{140k} removed}
\SankeyFlow{320}{455}{236}{32.7}{31.9}{sankeyveto}{Flavor symmetry algebra}{\ApproxScanCount{83k} removed}
\SankeyStage{51}{236}{34}{\ApproxScanCount{\AfterDeeperHL}}
\SankeyStage{185}{236}{33.8}{\ApproxScanCount{\AfterCoulomb}}
\SankeyStage{320}{236}{32.7}{\ApproxScanCount{\AfterSpinors}}
\SankeyStage{455}{236}{31.9}{\ApproxScanCount{\AfterFlavor}}
\SankeyLink{455}{252}{492}{355}

% Catalogue selection by structural explanations
\SankeyFlow{455}{320}{338}{34}{28.9}{sankeyscope}{Identical TQFT expressions}{\ApproxScanCount{513.8k} removed}
\SankeyFlow{320}{185}{338}{28.9}{0.9}{sankeyscope}{Atypical puncture resolutions}{\ApproxScanCount{2.803M} removed}
\SankeyFlow{185}{51}{338}{0.9}{0.24}{sankeyscope}{Resolved $S_3$ constructions}{\ApproxScanCount{65.6k} removed}
\SankeyStage{455}{338}{34}{\ApproxScanCount{\AfterFlavor}}
\SankeyStage{320}{338}{28.9}{\ApproxScanCount{\AfterIdentical}}
\SankeyStage{185}{338}{0.9}{\ApproxScanCount{\AfterAtypical}}
\SankeyStage{51}{338}{0.24}{\ApproxScanCount{\AfterCollisions}}
\SankeyLink{51}{338.12}{14}{457}

% Residual worklist and the zero-free-hypermultiplet requirement
\SankeyFlow{51}{253}{440}{34}{27.17}{sankeyscope}{Known equivalences}{\ApproxScanCount{4.9k} removed}
\SankeyFlow{253}{455}{440}{27.17}{6.63}{sankeyscope}{Decoupled free hypermultiplets}{\ApproxScanCount{14.7k} removed}
\SankeyStage{51}{440}{34}{\ApproxScanCount{\AfterCollisions}}
\SankeyStage{253}{440}{27.17}{\ApproxScanCount{\AfterExplanations}}
\SankeyStage{455}{440}{6.63}{\CatalogueConstructions}
\node[sankeylabel,anchor=north] at (455,450) {Remaining constructions};
\end{tikzpicture}
\endgroup

%% file: Supplemental_Simplest_Representatives.tex
\section{\texorpdfstring{Representative comparisons}{Representative comparisons}}
\label{supp:simplest-representatives}

We give one pair for each family or finite set. Some families share a pair. Here $r_{\rm CB}$ is the common Coulomb branch rank, and $\Delta_{\rm CB}$ lists the Coulomb branch spectrum, including repeated scaling dimensions.

\subsection{\texorpdfstring{U1. Eight-box partitions and $\Spin(8)$}{U1. Eight-box partitions and Spin(8)}}
\label{supp:minimum-u1}
\[
\begin{gathered}
D_{4}\bigl\{[2^{4}]_{\mathrm{blue}},\,[2^{4}]_{\mathrm{red}},\,[3,1^{5}]\bigr\}\\
\longleftrightarrow D_{5}\bigl\{[3^{2},1^{4}],\,\bigl([3^{3},1]\bigr)^{2}\bigr\}.
\end{gathered}
\]
$r_{\mathrm{CB}}=2$, $\Delta_{\mathrm{CB}}=\{4,6\}$, and $(n_v,n_h)=(18,40)$. 

\subsection{\texorpdfstring{U2. Relations between different ranks in type $D$}{U2. Relations between different ranks in type D}}
\label{supp:minimum-u2}
\[
\begin{gathered}
D_{5}\bigl\{[3,1^{7}],\,[3^{3},1],\,[5^{2}],\,[7,3]\bigr\}\\
\longleftrightarrow D_{6}\bigl\{[3^{3},1^{3}],\,[4^{2},3,1],\,[6^{2}]_{\mathrm{blue}},\,[9,3]\bigr\}.
\end{gathered}
\]
$r_{\mathrm{CB}}=5$, $\Delta_{\mathrm{CB}}=\{2,4,6^{2},8\}$, and $(n_v,n_h)=(47,72)$.

\subsection{\texorpdfstring{U3. $A/D$ comparisons with rectangular punctures}{U3. A/D comparisons with rectangular punctures}}
\label{supp:minimum-u3}
\[
\begin{gathered}
A_{5}\bigl\{[1^{6}],\,[2^{3}],\,[3^{2}]\bigr\}\\
\longleftrightarrow D_{4}\bigl\{\bigl([1^{8}]\bigr)^{2},\,[5,3]\bigr\}.
\end{gathered}
\]
$r_{\mathrm{CB}}=1$, $\Delta_{\mathrm{CB}}=\{6\}$, and $(n_v,n_h)=(11,40)$.

\subsection{\texorpdfstring{U4. Nilpotent orbit matching in $B_2/C_2$}{U4. Nilpotent orbit matching in B2/C2}}
\label{supp:minimum-u4}
\[
\begin{gathered}
D_{4}\bigl\{[2^{4}]_{\mathrm{blue}},\,[2^{4}]_{\mathrm{red}},\,[3,1^{5}]\bigr\}\\
\longleftrightarrow D_{4}^{\mathrm{tw}}\bigl\{[5,3],\,\bigl(\bigl([1^{6}]\bigr)_{t}\bigr)^{2}\bigr\}.
\end{gathered}
\]
$r_{\mathrm{CB}}=2$, $\Delta_{\mathrm{CB}}=\{4,6\}$, and $(n_v,n_h)=(18,40)$. 

\subsection{\texorpdfstring{U5. Complementary columns in $A$-type puncture partitions}{U5. Complementary columns in A-type puncture partitions}}
\label{supp:minimum-u5}
\[
\begin{gathered}
A_{2}\bigl\{\bigl([1^{3}]\bigr)^{2},\,\bigl([2,1]\bigr)^{2}\bigr\}\\
\longleftrightarrow A_{3}^{\mathrm{tw}}\bigl\{[2,1^{2}],\,\bigl([1^{5}]\bigr)_{t},\,\bigl([3,1^{2}]\bigr)_{t}\bigr\}.
\end{gathered}
\]
$r_{\mathrm{CB}}=2$, $\Delta_{\mathrm{CB}}=\{2,3\}$, and $(n_v,n_h)=(8,18)$.

\subsection{\texorpdfstring{U6. Exceptional puncture replacements}{U6. Exceptional puncture replacements}}
\label{supp:minimum-u6}
\[
\begin{gathered}
E_{6}\bigl\{0,\,2A_{2},\,D_{5}\bigr\}\\
\longleftrightarrow E_{7}\bigl\{(3A_{1})'',\,A_{4}+A_{2},\,E_{6}\bigr\}.
\end{gathered}
\]
$r_{\mathrm{CB}}=4$, $\Delta_{\mathrm{CB}}=\{4,6,8,12\}$, and $(n_v,n_h)=(56,112)$.

\subsection{\texorpdfstring{U7. Puncture closures involving $E_6$}{U7. Puncture closures involving E6}}
\label{supp:minimum-u7}
\[
\begin{gathered}
D_{7}\bigl\{[1^{14}],\,[5,4^{2},1],\,[7^{2}]\bigr\}\\
\longleftrightarrow E_{6}\bigl\{0,\,3A_{1},\,E_{6}(a_{1})\bigr\}.
\end{gathered}
\]
$r_{\mathrm{CB}}=2$, $\Delta_{\mathrm{CB}}=\{6,12\}$, and $(n_v,n_h)=(34,93)$.

\subsection{\texorpdfstring{U8. $D_4$ triality and ordered twist holonomies}{U8. D4 triality and ordered twist holonomies}}
\label{supp:minimum-u8}
\[
\begin{gathered}
A_{3}^{\mathrm{tw}}\bigl\{[1^{4}],\,\bigl(\bigl([2^{2},1]\bigr)_{t}\bigr)^{2}\bigr\}\\
\longleftrightarrow D_{4}^{S_3}\bigl\{\bigl([4,1^{2}]\bigr)_{t,\alpha},\,\bigl([4,1^{2}]\bigr)_{t,\beta},\,\bigl(0\bigr)_{t_3,(\alpha\beta)^{-1}}\bigr\}.
\end{gathered}
\]
$r_{\mathrm{CB}}=2$, $\Delta_{\mathrm{CB}}=\{4^{2}\}$, and $(n_v,n_h)=(14,26)$.
Here $\alpha=(12)$, $\beta=(23)$, and the order three holonomy is $(\alpha\beta)^{-1}$, so the ordered product is the identity. The $A_3$ twisted punctures use their recorded common order two sector.

\subsection{\texorpdfstring{U9. From tori to spheres in type $A$}{U9. From tori to spheres in type A}}
\label{supp:minimum-u9}
\[
\begin{gathered}
A_{1} {}^{g=1}\bigl\{\bigl([1^{2}]\bigr)^{4}\bigr\}\\
\longleftrightarrow A_{3}^{\mathrm{tw}}\bigl\{\bigl(\bigl([3,1^{2}]\bigr)_{t}\bigr)^{3},\,\bigl([5]\bigr)_{t}\bigr\}.
\end{gathered}
\]
$r_{\mathrm{CB}}=4$, $\Delta_{\mathrm{CB}}=\{2^{4}\}$, and $(n_v,n_h)=(12,16)$.

\subsection{\texorpdfstring{U10. Two-column punctures and twisted puncture replacements in type $D$}{U10. Two-column punctures and twisted puncture replacements in type D}}
\label{supp:minimum-u10}
\[
\begin{gathered}
D_{4}^{\mathrm{tw}}\bigl\{\bigl([2^{4}]_{\mathrm{blue}}\bigr)^{2},\,\bigl([4,1^{2}]\bigr)_{t},\,\bigl([6]\bigr)_{t}\bigr\}\\
\longleftrightarrow D_{4}^{\mathrm{tw}}\bigl\{[4^{2}]_{\mathrm{blue}},\,[5,3],\,\bigl([1^{6}]\bigr)_{t},\,\bigl([4,1^{2}]\bigr)_{t}\bigr\}.
\end{gathered}
\]
$r_{\mathrm{CB}}=4$, $\Delta_{\mathrm{CB}}=\{2,4^{2},6\}$, and $(n_v,n_h)=(28,45)$.

\subsection{\texorpdfstring{U11. Related puncture closures in type $D$}{U11. Related puncture closures in type D}}
\label{supp:minimum-u11}
\[
\begin{gathered}
D_{4}\bigl\{[2^{4}]_{\mathrm{blue}},\,[2^{4}]_{\mathrm{red}},\,[3,1^{5}]\bigr\}\\
\longleftrightarrow D_{4}^{\mathrm{tw}}\bigl\{[5,3],\,\bigl(\bigl([1^{6}]\bigr)_{t}\bigr)^{2}\bigr\}.
\end{gathered}
\]
$r_{\mathrm{CB}}=2$, $\Delta_{\mathrm{CB}}=\{4,6\}$, and $(n_v,n_h)=(18,40)$.

\subsection{\texorpdfstring{U12. Short rows in paired $A/D$ puncture partitions}{U12. Short rows in paired A/D puncture partitions}}
\label{supp:minimum-u12}
\[
\begin{gathered}
A_{5}^{\mathrm{tw}}\bigl\{[1^{6}],\,\bigl([2^{2},1^{3}]\bigr)_{t},\,\bigl([7]\bigr)_{t}\bigr\}\\
\longleftrightarrow D_{4}\bigl\{[1^{8}],\,[3,2^{2},1],\,[3^{2},1^{2}]\bigr\}.
\end{gathered}
\]
$r_{\mathrm{CB}}=2$, $\Delta_{\mathrm{CB}}=\{3,6\}$, and $(n_v,n_h)=(16,39)$.

\subsection{\texorpdfstring{U13. Puncture replacements in twisted $A/D$ constructions}{U13. Puncture replacements in twisted A/D constructions}}
\label{supp:minimum-u13}
\[
\begin{gathered}
A_{3}^{\mathrm{tw}}\bigl\{[2^{2}],\,\bigl(\bigl([1^{5}]\bigr)_{t}\bigr)^{2}\bigr\}\\
\longleftrightarrow D_{4}^{\mathrm{tw}}\bigl\{[3^{2},1^{2}],\,\bigl(\bigl([2^{3}]\bigr)_{t}\bigr)^{2}\bigr\}.
\end{gathered}
\]
$r_{\mathrm{CB}}=2$, $\Delta_{\mathrm{CB}}=\{3,4\}$, and $(n_v,n_h)=(12,24)$.

\subsection{\texorpdfstring{U14. Two $A$ presentations of one $D$ trinion}{U14. Two A presentations of one D trinion}}
\label{supp:minimum-u14}
\[
\begin{gathered}
A_{5}\bigl\{\bigl([2^{2},1^{2}]\bigr)^{2},\,[2^{3}]\bigr\}\\
\longleftrightarrow D_{6}^{\mathrm{tw}}\bigl\{[6^{2}]_{\mathrm{blue}},\,\bigl([2^{5}]\bigr)_{t},\,\bigl([3^{2},2^{2}]\bigr)_{t}\bigr\}.
\end{gathered}
\]
$r_{\mathrm{CB}}=3$, $\Delta_{\mathrm{CB}}=\{3,4,6\}$, and $(n_v,n_h)=(23,48)$.

\subsection{\texorpdfstring{U15. Full punctures in $D_4$}{U15. Full punctures in D4}}
\label{supp:minimum-u15}
\[
\begin{gathered}
D_{4}\bigl\{[1^{8}],\,[2^{4}]_{\mathrm{blue}},\,[3,2^{2},1]\bigr\}\\
\longleftrightarrow D_{6}^{\mathrm{tw}}\bigl\{[6^{2}]_{\mathrm{blue}},\,\bigl([2^{5}]\bigr)_{t},\,\bigl([3^{2},2,1^{2}]\bigr)_{t}\bigr\}.
\end{gathered}
\]
$r_{\mathrm{CB}}=3$, $\Delta_{\mathrm{CB}}=\{4,6^{2}\}$, and $(n_v,n_h)=(29,55)$.

\subsection{\texorpdfstring{U16. $R_{0,N}$ theories and conformal SQCD}{U16. R0,N theories and conformal SQCD}}
\label{supp:minimum-u16}
\[
\begin{gathered}
A_{3}\bigl\{\bigl([1^{4}]\bigr)^{2},\,[2^{2}]\bigr\}\\
\longleftrightarrow D_{4}^{\mathrm{tw}}\bigl\{[4^{2}]_{\mathrm{red}},\,\bigl([1^{6}]\bigr)_{t},\,\bigl([2^{3}]\bigr)_{t}\bigr\}.
\end{gathered}
\]
$r_{\mathrm{CB}}=1$, $\Delta_{\mathrm{CB}}=\{4\}$, and $(n_v,n_h)=(7,24)$.

\subsection{\texorpdfstring{U17. Exceptional puncture changes with a shared puncture}{U17. Exceptional puncture changes with a shared puncture}}
\label{supp:minimum-u17}
\[
\begin{gathered}
E_{6}\bigl\{A_{2},\,2A_{2}+A_{1},\,E_{6}(a_{3})\bigr\}\\
\longleftrightarrow E_{6}\bigl\{2A_{2},\,2A_{2}+A_{1},\,D_{4}\bigr\}.
\end{gathered}
\]
$r_{\mathrm{CB}}=5$, $\Delta_{\mathrm{CB}}=\{3,6^{2},8,12\}$, and $(n_v,n_h)=(65,98)$.

\subsection{\texorpdfstring{U18. Comparisons between classical Lie types and $E_7$}{U18. Comparisons between classical Lie types and E7}}
\label{supp:minimum-u18}
\[
\begin{gathered}
D_{6}\bigl\{[3,1^{9}],\,[2^{6}]_{\mathrm{blue}},\,[6^{2}]_{\mathrm{blue}}\bigr\}\\
\longleftrightarrow E_{7}\bigl\{2A_{2},\,A_{3}+A_{2}+A_{1},\,E_{6}\bigr\}.
\end{gathered}
\]
$r_{\mathrm{CB}}=4$, $\Delta_{\mathrm{CB}}=\{4,6,8,10\}$, and $(n_v,n_h)=(52,96)$.

\subsection{\texorpdfstring{U19. Puncture replacements at fixed $A$-type parent}{U19. Puncture replacements at fixed A-type parent}}
\label{supp:minimum-u19}
\[
\begin{gathered}
A_{3}\bigl\{[1^{4}],\,\bigl([2^{2}]\bigr)^{2},\,[3,1]\bigr\}\\
\longleftrightarrow A_{3}^{\mathrm{tw}}\bigl\{\bigl([1^{4}]\bigr)^{2},\,\bigl(\bigl([5]\bigr)_{t}\bigr)^{2}\bigr\}.
\end{gathered}
\]
$r_{\mathrm{CB}}=2$, $\Delta_{\mathrm{CB}}=\{2,4\}$, and $(n_v,n_h)=(10,24)$.

\subsection{\texorpdfstring{U20. Rectangular punctures in twisted type $A$ at different ranks}{U20. Rectangular punctures in twisted type A at different ranks}}
\label{supp:minimum-u20}
\[
\begin{gathered}
A_{3}^{\mathrm{tw}}\bigl\{[2^{2}],\,\bigl(\bigl([1^{5}]\bigr)_{t}\bigr)^{2}\bigr\}\\
\longleftrightarrow A_{5}^{\mathrm{tw}}\bigl\{[3^{2}],\,\bigl([3,1^{4}]\bigr)_{t},\,\bigl([3^{2},1]\bigr)_{t}\bigr\}.
\end{gathered}
\]
$r_{\mathrm{CB}}=2$, $\Delta_{\mathrm{CB}}=\{3,4\}$, and $(n_v,n_h)=(12,24)$.

\subsection{\texorpdfstring{U21. $A$-type comparisons with a full twisted puncture}{U21. A-type comparisons with a full twisted puncture}}
\label{supp:minimum-u21}
\[
\begin{gathered}
A_{2}\bigl\{\bigl([2,1]\bigr)^{2},\,\bigl([1^{3}]\bigr)^{2}\bigr\}\\
\longleftrightarrow A_{3}^{\mathrm{tw}}\bigl\{[2,1^{2}],\,\bigl([1^{5}]\bigr)_{t},\,\bigl([3,1^{2}]\bigr)_{t}\bigr\}.
\end{gathered}
\]
$r_{\mathrm{CB}}=2$, $\Delta_{\mathrm{CB}}=\{2,3\}$, and $(n_v,n_h)=(8,18)$.

\subsection{\texorpdfstring{U22. Rectangular punctures and partial closures in $D_4/D_5$}{U22. Rectangular punctures and partial closures in D4/D5}}
\label{supp:minimum-u22}
\[
\begin{gathered}
D_{4}\bigl\{\bigl([2^{2},1^{4}]\bigr)^{2},\,[3,2^{2},1]\bigr\}\\
\longleftrightarrow D_{5}^{\mathrm{tw}}\bigl\{[5^{2}],\,\bigl([2^{2},1^{4}]\bigr)_{t},\,\bigl([2^{3},1^{2}]\bigr)_{t}\bigr\}.
\end{gathered}
\]
$r_{\mathrm{CB}}=3$, $\Delta_{\mathrm{CB}}=\{4^{2},6\}$, and $(n_v,n_h)=(25,47)$.

\subsection{\texorpdfstring{U23. Related puncture partitions at $D_5,D_6,D_7$}{U23. Related puncture partitions at D5,D6,D7}}
\label{supp:minimum-u23}
\[
\begin{gathered}
D_{5}\bigl\{\bigl([2^{4},1^{2}]\bigr)^{2},\,[5,3,1^{2}]\bigr\}\\
\longleftrightarrow D_{6}\bigl\{\bigl([3^{3},1^{3}]\bigr)^{2},\,[5^{2},1^{2}]\bigr\}.
\end{gathered}
\]
$r_{\mathrm{CB}}=4$, $\Delta_{\mathrm{CB}}=\{3,4,5,8\}$, and $(n_v,n_h)=(36,64)$.

\subsection{\texorpdfstring{U24. Repeated rectangular and atypical punctures in $A/D$}{U24. Repeated rectangular and atypical punctures in A/D}}
\label{supp:minimum-u24}
\[
\begin{gathered}
A_{5}\bigl\{[2^{3}],\,\bigl([3^{2}]\bigr)^{2},\,[4,2]\bigr\}\\
\longleftrightarrow D_{4}^{\mathrm{tw}}\bigl\{[1^{8}],\,\bigl([2^{3}]\bigr)_{t},\,\bigl([4,2]\bigr)_{t}\bigr\}.
\end{gathered}
\]
$r_{\mathrm{CB}}=3$, $\Delta_{\mathrm{CB}}=\{2,4,6\}$, and $(n_v,n_h)=(21,40)$.

\subsection{\texorpdfstring{U25. $A_3$ gluing and $D_4/D_5$ presentations}{U25. A3 gluing and D4/D5 presentations}}
\label{supp:minimum-u25}
\[
\begin{gathered}
A_{3}\bigl\{[1^{4}],\,\bigl([2^{2}]\bigr)^{2},\,[3,1]\bigr\}\\
\longleftrightarrow D_{4}^{\mathrm{tw}}\bigl\{[2^{4}]_{\mathrm{blue}},\,\bigl([2^{3}]\bigr)_{t},\,\bigl([3^{2}]\bigr)_{t}\bigr\}.
\end{gathered}
\]
$r_{\mathrm{CB}}=2$, $\Delta_{\mathrm{CB}}=\{2,4\}$, and $(n_v,n_h)=(10,24)$.

\subsection{\texorpdfstring{F1. Two full $A$ punctures and untwisted $D$ constructions}{F1. Two full A punctures and untwisted D constructions}}
\label{supp:minimum-f1}
\[
\begin{gathered}
A_{4}\bigl\{\bigl([1^{5}]\bigr)^{2},\,[3,1^{2}]\bigr\}\\
\longleftrightarrow D_{6}\bigl\{[3^{3},1^{3}],\,[3^{4}],\,[5^{2},1^{2}]\bigr\}.
\end{gathered}
\]
$r_{\mathrm{CB}}=3$, $\Delta_{\mathrm{CB}}=\{3,4,5\}$, and $(n_v,n_h)=(21,48)$.

\subsection{\texorpdfstring{F2. Puncture replacements in $A_4/D_5$}{F2. Puncture replacements in A4/D5}}
\label{supp:minimum-f2}
\[
\begin{gathered}
A_{4}\bigl\{[1^{5}],\,[2,1^{3}],\,[3,2],\,[4,1]\bigr\}\\
\longleftrightarrow D_{5}\bigl\{[2^{4},1^{2}],\,[4^{2},1^{2}],\,[5^{2}],\,[7,1^{3}]\bigr\}.
\end{gathered}
\]
$r_{\mathrm{CB}}=5$, $\Delta_{\mathrm{CB}}=\{2,3,4^{2},5\}$, and $(n_v,n_h)=(31,54)$.

\subsection{\texorpdfstring{F3. $A_5/D_6$ closures that depend on color}{F3. A5/D6 closures that depend on color}}
\label{supp:minimum-f3}
\[
\begin{gathered}
A_{5}\bigl\{[1^{6}],\,[2^{3}],\,[3,1^{3}]\bigr\}\\
\longleftrightarrow D_{6}\bigl\{[2^{6}]_{\mathrm{blue}},\,[4^{2},2^{2}]_{\mathrm{red}},\,[5^{2},1^{2}]\bigr\}.
\end{gathered}
\]
$r_{\mathrm{CB}}=4$, $\Delta_{\mathrm{CB}}=\{3,4,5,6\}$, and $(n_v,n_h)=(32,64)$.

\subsection{\texorpdfstring{F4. Puncture replacements in $A_7/D_5$}{F4. Puncture replacements in A7/D5}}
\label{supp:minimum-f4}
\[
\begin{gathered}
A_{7}^{\mathrm{tw}}\bigl\{[1^{8}],\,\bigl([4^{2},1]\bigr)_{t},\,\bigl([9]\bigr)_{t}\bigr\}\\
\longleftrightarrow D_{5}\bigl\{[1^{10}],\,[3,2^{2},1^{3}],\,[5^{2}]\bigr\}.
\end{gathered}
\]
$r_{\mathrm{CB}}=2$, $\Delta_{\mathrm{CB}}=\{4,8\}$, and $(n_v,n_h)=(22,57)$.

\subsection{\texorpdfstring{F5. Repeated $A$-type punctures and untwisted $D$ rectangular partitions}{F5. Repeated A-type punctures and untwisted D rectangular partitions}}
\label{supp:minimum-f5}
\[
\begin{gathered}
A_{5}\bigl\{\bigl([2,1^{4}]\bigr)^{2},\,[2^{3}]\bigr\}\\
\longleftrightarrow D_{6}\bigl\{[3^{2},1^{6}],\,[3^{4}],\,[5^{2},1^{2}]\bigr\}.
\end{gathered}
\]
$r_{\mathrm{CB}}=5$, $\Delta_{\mathrm{CB}}=\{3,4,5^{2},6\}$, and $(n_v,n_h)=(41,72)$.

\subsection{\texorpdfstring{F6. Two-column punctures in $A_5$ comparisons}{F6. Two-column punctures in A5 comparisons}}
\label{supp:minimum-f6}
\[
\begin{gathered}
A_{5}^{\mathrm{tw}}\bigl\{[2,1^{4}],\,[2^{2},1^{2}],\,\bigl(\bigl([7]\bigr)_{t}\bigr)^{2}\bigr\}\\
\longleftrightarrow D_{5}\bigl\{[3^{2},2^{2}],\,\bigl([4^{2},1^{2}]\bigr)^{2},\,[7,3]\bigr\}.
\end{gathered}
\]
$r_{\mathrm{CB}}=6$, $\Delta_{\mathrm{CB}}=\{2,3^{2},4,5,6\}$, and $(n_v,n_h)=(40,60)$.

%% file: Supplemental_Isolated_Cases.tex
\section{\texorpdfstring{Pairs not assigned to a family}{Pairs not assigned to a family}}
\label{supp:isolated-comparisons}

The catalogue contains \IsolatedComparisons{} candidate pairs that have not been assigned to a family. Some share a construction with other pairs in the catalogue.

\paragraph{I1.}
\label{supp:isolated-1}
\[
\begin{gathered}
D_{4}^{\mathrm{tw}}\bigl\{[4^{2}]_{\mathrm{blue}},\,\bigl([1^{6}]\bigr)_{t},\,\bigl([2,1^{4}]\bigr)_{t}\bigr\}\\
\longleftrightarrow D_{5}^{\mathrm{tw}}\bigl\{[5^{2}],\,\bigl(\bigl([2^{3},1^{2}]\bigr)_{t}\bigr)^{2}\bigr\}.
\end{gathered}
\]
$r_{\mathrm{CB}}=2$, $\Delta_{\mathrm{CB}}=\{4,6\}$, and $(n_v,n_h)=(18,38)$.

\paragraph{I2.}
\label{supp:isolated-2}
\[
\begin{gathered}
A_{4}\bigl\{\bigl([1^{5}]\bigr)^{2},\,[3,1^{2}]\bigr\}\\
\longleftrightarrow A_{9}^{\mathrm{tw}}\bigl\{[3^{3},1],\,\bigl([3^{2},1^{5}]\bigr)_{t},\,\bigl([11]\bigr)_{t}\bigr\}.
\end{gathered}
\]
$r_{\mathrm{CB}}=3$, $\Delta_{\mathrm{CB}}=\{3,4,5\}$, and $(n_v,n_h)=(21,48)$.

\paragraph{I3.}
\label{supp:isolated-3}
\[
\begin{gathered}
A_{4}\bigl\{\bigl([1^{5}]\bigr)^{2},\,[2^{2},1]\bigr\}\\
\longleftrightarrow A_{9}^{\mathrm{tw}}\bigl\{[3^{3},1],\,\bigl([3,2^{2},1^{4}]\bigr)_{t},\,\bigl([11]\bigr)_{t}\bigr\}.
\end{gathered}
\]
$r_{\mathrm{CB}}=4$, $\Delta_{\mathrm{CB}}=\{3,4,5^{2}\}$, and $(n_v,n_h)=(30,58)$.

\paragraph{I4.}
\label{supp:isolated-4}
\[
\begin{gathered}
A_{4}\bigl\{\bigl([2,1^{3}]\bigr)^{2},\,[1^{5}]\bigr\}\\
\longleftrightarrow A_{9}^{\mathrm{tw}}\bigl\{[3^{2},2,1^{2}],\,\bigl([3^{3},1^{2}]\bigr)_{t},\,\bigl([11]\bigr)_{t}\bigr\}.
\end{gathered}
\]
$r_{\mathrm{CB}}=4$, $\Delta_{\mathrm{CB}}=\{3,4^{2},5\}$, and $(n_v,n_h)=(28,54)$.

\paragraph{I5.}
\label{supp:isolated-5}
\[
\begin{gathered}
A_{5}\bigl\{[2^{2},1^{2}],\,[1^{6}],\,[3,2,1]\bigr\}\\
\longleftrightarrow A_{11}^{\mathrm{tw}}\bigl\{[3^{3},2,1],\,\bigl([5^{2},1^{3}]\bigr)_{t},\,\bigl([13]\bigr)_{t}\bigr\}.
\end{gathered}
\]
$r_{\mathrm{CB}}=4$, $\Delta_{\mathrm{CB}}=\{3,4,5,6\}$, and $(n_v,n_h)=(32,63)$.

\paragraph{I6.}
\label{supp:isolated-6}
\[
\begin{gathered}
A_{5}\bigl\{[2^{3}],\,[1^{6}],\,[3,1^{3}]\bigr\}\\
\longleftrightarrow A_{11}^{\mathrm{tw}}\bigl\{[3^{3},1^{3}],\,\bigl([5^{2},3]\bigr)_{t},\,\bigl([13]\bigr)_{t}\bigr\}.
\end{gathered}
\]
$r_{\mathrm{CB}}=4$, $\Delta_{\mathrm{CB}}=\{3,4,5,6\}$, and $(n_v,n_h)=(32,64)$.

\paragraph{I7.}
\label{supp:isolated-7}
\[
\begin{gathered}
A_{6}\bigl\{[1^{7}],\,[2,1^{5}],\,[4,3]\bigr\}\\
\longleftrightarrow D_{7}\bigl\{[2^{6},1^{2}],\,[4^{2},1^{6}],\,[7^{2}]\bigr\}.
\end{gathered}
\]
$r_{\mathrm{CB}}=5$, $\Delta_{\mathrm{CB}}=\{4,5,6^{2},7\}$, and $(n_v,n_h)=(51,96)$.

\paragraph{I8.}
\label{supp:isolated-8}
\[
\begin{gathered}
A_{7}^{\mathrm{tw}}\bigl\{[6,1^{2}],\,\bigl([1^{9}]\bigr)_{t},\,\bigl([3^{3}]\bigr)_{t}\bigr\}\\
\longleftrightarrow D_{7}^{\mathrm{tw}}\bigl\{[6^{2},1^{2}],\,\bigl([3^{4}]\bigr)_{t},\,\bigl([4^{3}]\bigr)_{t}\bigr\}.
\end{gathered}
\]
$r_{\mathrm{CB}}=5$, $\Delta_{\mathrm{CB}}=\{3^{2},4,5,7\}$, and $(n_v,n_h)=(39,66)$.

\paragraph{I9.}
\label{supp:isolated-9}
\[
\begin{gathered}
E_{6}\bigl\{\bigl(0\bigr)^{2},\,E_{6}(a_{1})\bigr\}\\
\longleftrightarrow E_{8}\bigl\{2A_{2},\,D_{4}(a_{1})+A_{2},\,E_{8}(a_{1})\bigr\}.
\end{gathered}
\]
$r_{\mathrm{CB}}=5$, $\Delta_{\mathrm{CB}}=\{6,8,9,12^{2}\}$, and $(n_v,n_h)=(89,168)$.

\paragraph{I10.}
\label{supp:isolated-10}
\[
\begin{gathered}
A_{5}\bigl\{\bigl([2^{2},1^{2}]\bigr)^{2},\,[2,1^{4}]\bigr\}\\
\longleftrightarrow A_{9}^{\mathrm{tw}}\bigl\{[3^{2},1^{4}],\,\bigl([3^{3},1^{2}]\bigr)_{t},\,\bigl([11]\bigr)_{t}\bigr\}.
\end{gathered}
\]
$r_{\mathrm{CB}}=5$, $\Delta_{\mathrm{CB}}=\{3,4^{2},5,6\}$, and $(n_v,n_h)=(39,68)$.

\paragraph{I11.}
\label{supp:isolated-11}
\[
\begin{gathered}
A_{6}\bigl\{\bigl([2^{2},1^{3}]\bigr)^{2},\,[3^{2},1]\bigr\}\\
\longleftrightarrow A_{9}^{\mathrm{tw}}\bigl\{[3^{3},1],\,\bigl([3,1^{8}]\bigr)_{t},\,\bigl([11]\bigr)_{t}\bigr\}.
\end{gathered}
\]
$r_{\mathrm{CB}}=5$, $\Delta_{\mathrm{CB}}=\{3,4,5^{2},7\}$, and $(n_v,n_h)=(43,76)$.

\paragraph{I12.}
\label{supp:isolated-12}
\[
\begin{gathered}
A_{5}\bigl\{[2^{2},1^{2}],\,[1^{6}],\,[2^{3}]\bigr\}\\
\longleftrightarrow A_{11}^{\mathrm{tw}}\bigl\{[3^{3},1^{3}],\,\bigl([5^{2},1^{3}]\bigr)_{t},\,\bigl([13]\bigr)_{t}\bigr\}.
\end{gathered}
\]
$r_{\mathrm{CB}}=5$, $\Delta_{\mathrm{CB}}=\{3,4,5,6^{2}\}$, and $(n_v,n_h)=(43,76)$.

\paragraph{I13.}
\label{supp:isolated-13}
\[
\begin{gathered}
A_{5}\bigl\{[2^{2},1^{2}],\,[1^{6}],\,[2^{3}]\bigr\}\\
\longleftrightarrow A_{11}^{\mathrm{tw}}\bigl\{[3^{3},1^{3}],\,\bigl([5,4^{2}]\bigr)_{t},\,\bigl([13]\bigr)_{t}\bigr\}.
\end{gathered}
\]
$r_{\mathrm{CB}}=5$, $\Delta_{\mathrm{CB}}=\{3,4,5,6^{2}\}$, and $(n_v,n_h)=(43,76)$.

\paragraph{I14.}
\label{supp:isolated-14}
\[
\begin{gathered}
A_{5}\bigl\{\bigl([2,1^{4}]\bigr)^{2},\,[2^{3}]\bigr\}\\
\longleftrightarrow A_{11}^{\mathrm{tw}}\bigl\{[4^{2},3,1],\,\bigl([3^{3},1^{4}]\bigr)_{t},\,\bigl([13]\bigr)_{t}\bigr\}.
\end{gathered}
\]
$r_{\mathrm{CB}}=5$, $\Delta_{\mathrm{CB}}=\{3,4,5^{2},6\}$, and $(n_v,n_h)=(41,72)$.

\paragraph{I15.}
\label{supp:isolated-15}
\[
\begin{gathered}
E_{7}\bigl\{\bigl(2A_{1}\bigr)^{2},\,E_{7}(a_{1})\bigr\}\\
\longleftrightarrow E_{8}\bigl\{2A_{2},\,A_{3}+A_{2},\,E_{8}(a_{1})\bigr\}.
\end{gathered}
\]
$r_{\mathrm{CB}}=6$, $\Delta_{\mathrm{CB}}=\{6,8,10,12^{2},18\}$, and $(n_v,n_h)=(126,208)$.

\paragraph{I16.}
\label{supp:isolated-16}
\[
\begin{gathered}
D_{5}\bigl\{[3,1^{7}],\,\bigl([4^{2},1^{2}]\bigr)^{2},\,[7,3]\bigr\}\\
\longleftrightarrow E_{6}^{\mathrm{tw}}\bigl\{2A_{2},\,A_{3},\,\bigl(\bigl(F_{4}\bigr)_{t}\bigr)^{2}\bigr\}.
\end{gathered}
\]
$r_{\mathrm{CB}}=8$, $\Delta_{\mathrm{CB}}=\{2,3^{2},4,5,6^{2},8\}$, and $(n_v,n_h)=(66,92)$.

\paragraph{I17.}
\label{supp:isolated-17}
\[
\begin{gathered}
D_{6}\bigl\{[2^{2},1^{8}],\,\bigl([6^{2}]_{\mathrm{blue}}\bigr)^{3}\bigr\}\\
\longleftrightarrow D_{7}\bigl\{[5,1^{9}],\,\bigl([7^{2}]\bigr)^{3}\bigr\}.
\end{gathered}
\]
$r_{\mathrm{CB}}=8$, $\Delta_{\mathrm{CB}}=\{2,4^{2},6^{3},8,10\}$, and $(n_v,n_h)=(84,120)$. The recorded Schur check reaches $\tau^{40}$. The full flavor comparison remains pending.

\paragraph{I18.}
\label{supp:isolated-18}
\[
\begin{gathered}
D_{6}\bigl\{[3,1^{9}],\,[4^{2},1^{4}],\,[4^{2},3,1]\bigr\}\\
\longleftrightarrow D_{8}\bigl\{[3^{5},1],\,\bigl([6^{2},2^{2}]_{\mathrm{blue}}\bigr)^{2}\bigr\}.
\end{gathered}
\]
$r_{\mathrm{CB}}=8$, $\Delta_{\mathrm{CB}}=\{4^{2},6^{3},8^{2},10\}$, and $(n_v,n_h)=(96,136)$.

\paragraph{I19.}
\label{supp:isolated-19}
\[
\begin{gathered}
E_{7}\bigl\{2A_{2},\,E_{7}(a_{4}),\,E_{6},\,E_{7}(a_{1})\bigr\}\\
\longleftrightarrow E_{7}\bigl\{(A_{3}+A_{1})'',\,A_{6},\,E_{6},\,E_{7}(a_{1})\bigr\}.
\end{gathered}
\]
$r_{\mathrm{CB}}=9$, $\Delta_{\mathrm{CB}}=\{2,4,6^{3},8^{2},10,12\}$, and $(n_v,n_h)=(115,152)$.

\paragraph{I20.}
\label{supp:isolated-20}
\[
\begin{gathered}
E_{6}^{\mathrm{tw}}\bigl\{\bigl(A_{2}+2A_{1}\bigr)^{2},\,\bigl(\bigl(F_{4}\bigr)_{t}\bigr)^{2}\bigr\}\\
\longleftrightarrow E_{7}\bigl\{A_{4}+A_{2},\,\bigl(E_{6}(a_{1})\bigr)^{2},\,E_{7}(a_{1})\bigr\}.
\end{gathered}
\]
$r_{\mathrm{CB}}=10$, $\Delta_{\mathrm{CB}}=\{2,3^{2},5^{2},6,8,9^{2},12\}$, and $(n_v,n_h)=(114,140)$.

\paragraph{I21.}
\label{supp:isolated-21}
\[
\begin{gathered}
D_{5}\bigl\{\bigl([4^{2},1^{2}]\bigr)^{4}\bigr\}\\
\longleftrightarrow D_{6}\bigl\{\bigl([5^{2},1^{2}]\bigr)^{3},\,[7,3,1^{2}]\bigr\}.
\end{gathered}
\]
$r_{\mathrm{CB}}=11$, $\Delta_{\mathrm{CB}}=\{2,3^{4},4,5^{3},6,8\}$, and $(n_v,n_h)=(83,104)$.

\paragraph{I22.}
\label{supp:isolated-22}
\[
\begin{gathered}
E_{6}\bigl\{\bigl(D_{5}(a_{1})\bigr)^{4}\bigr\}\\
\longleftrightarrow E_{7}\bigl\{\bigl(E_{6}(a_{1})\bigr)^{3},\,E_{7}(a_{3})\bigr\}.
\end{gathered}
\]
$r_{\mathrm{CB}}=14$, $\Delta_{\mathrm{CB}}=\{2,3^{4},5^{3},6,8,9^{3},12\}$, and $(n_v,n_h)=(150,172)$.